\documentclass[12pt]{iopart}
\usepackage{amssymb}
\expandafter\let\csname equation*\endcsname\relax
\expandafter\let\csname endequation*\endcsname\relax
\usepackage{amsmath}
\usepackage{mathtools}

\usepackage[colorinlistoftodos]{todonotes}

\usepackage[nameinlink, noabbrev, capitalize]{cleveref}
\usepackage{subfig}
\usepackage{graphicx,cite}
\usepackage[normalem]{ulem}
\usepackage{booktabs}       

\newcommand{\be}{\begin{equation}}
\newcommand{\ee}{\end{equation}}
\newcommand{\ba}{\begin{eqnarray}}
\newcommand{\ea}{\end{eqnarray}}
\renewcommand{\tt}{\texttt}

\usepackage{annotate-equations}

\begin{document}

\title[Plasma Surrogate Modelling]{Plasma Surrogate Modelling using Fourier Neural Operators}

\author{Vignesh Gopakumar\textsuperscript{1,2}, Stanislas Pamela\textsuperscript{1}, Lorenzo Zanisi\textsuperscript{1}, Zongyi Li\textsuperscript{4}, Ander Gray\textsuperscript{1}, Daniel Brennand\textsuperscript{1}, Nitesh Bhatia\textsuperscript{1}, Gregory Stathopoulos\textsuperscript{3}, Matt Kusner\textsuperscript{2}, Marc Peter Deisenroth\textsuperscript{2}, Anima Anandkumar\textsuperscript{4}, JOREK Team\footnote{See the author list of Hoelzl et al., Nucl. Fusion 61, 065001 (2021)} and MAST Team\footnote{See the author list of Meyer et al., Nucl. Fusion 57, 102014 (2017)}}


\renewcommand\footnoterule{ %
  \kern-3pt \hrule width 12cm \kern 5.6pt \textwidth 0.8pt
}
\renewcommand*\thefootnote{\alph{footnote}}

\address{\textsuperscript{1}UK Atomic Energy Authority, Oxford, OX14 3EB, UK}
\address{\textsuperscript{2}University College London, London, WC1E 6BT, UK}
\address{\textsuperscript{4}University of Liverpool, Liverpool, L69 3BX, UK}
\address{\textsuperscript{4}Caltech, Pasadena, 91125, USA}

\ead{vignesh.gopakumar@ukaea.uk}


\begin{abstract}
    Predicting plasma evolution within a Tokamak reactor is crucial to realizing the goal of sustainable fusion. Capabilities in forecasting the spatio-temporal evolution of plasma rapidly and accurately allow us to quickly iterate over design and control strategies on current Tokamak devices and future reactors. Modelling plasma evolution using numerical solvers is often expensive, consuming many hours on supercomputers,  and hence, we need alternative inexpensive surrogate models. We demonstrate accurate predictions of plasma evolution both in simulation and experimental domains using deep learning-based surrogate modelling tools, viz., Fourier Neural Operators (FNO). We show that FNO has a speedup of six orders of magnitude over traditional solvers in predicting the plasma dynamics simulated from magnetohydrodynamic models, while maintaining a high accuracy (MSE in the normalised domain $\approx$ $10^{-5}$). Our modified version of the FNO is capable of solving multi-variable Partial Differential Equations (PDE), and can capture the dependence among the different variables in a single model. FNOs can also predict plasma evolution on real-world experimental data observed by the cameras positioned within the MAST Tokamak, i.e., cameras looking across the central solenoid and the divertor in the Tokamak. We show that FNOs are able to accurately forecast the evolution of plasma and have the potential to be deployed for real-time monitoring. We also illustrate their capability in forecasting the plasma shape, the locations of interactions of the plasma with the central solenoid and the divertor for the full (available) duration of the plasma shot within MAST. The FNO offers a viable alternative for surrogate modelling as it is quick to train and infer, and requires fewer data points, while being able to do zero-shot super-resolution and getting high-fidelity solutions.
    
\end{abstract}


%
%
%
%
%


\section{Introduction}\label{introduction}

\begin{figure}[t]
    \centering
    \subfloat[Experiment]{\includegraphics[width=3.7cm]{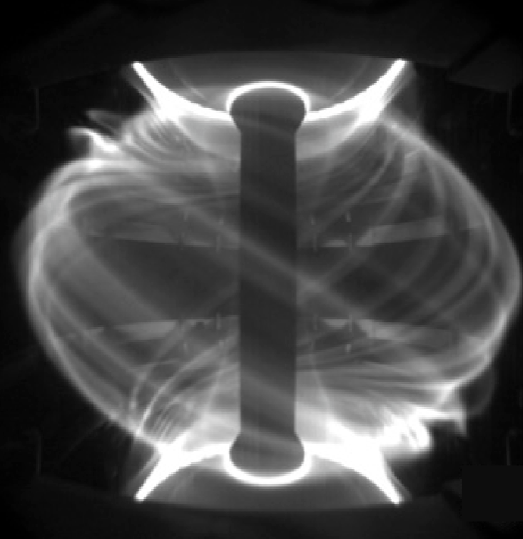}}
    \label{fig: mast_fastcam}
    \centering
    \subfloat[Simulation]{\includegraphics[width=8.0cm]{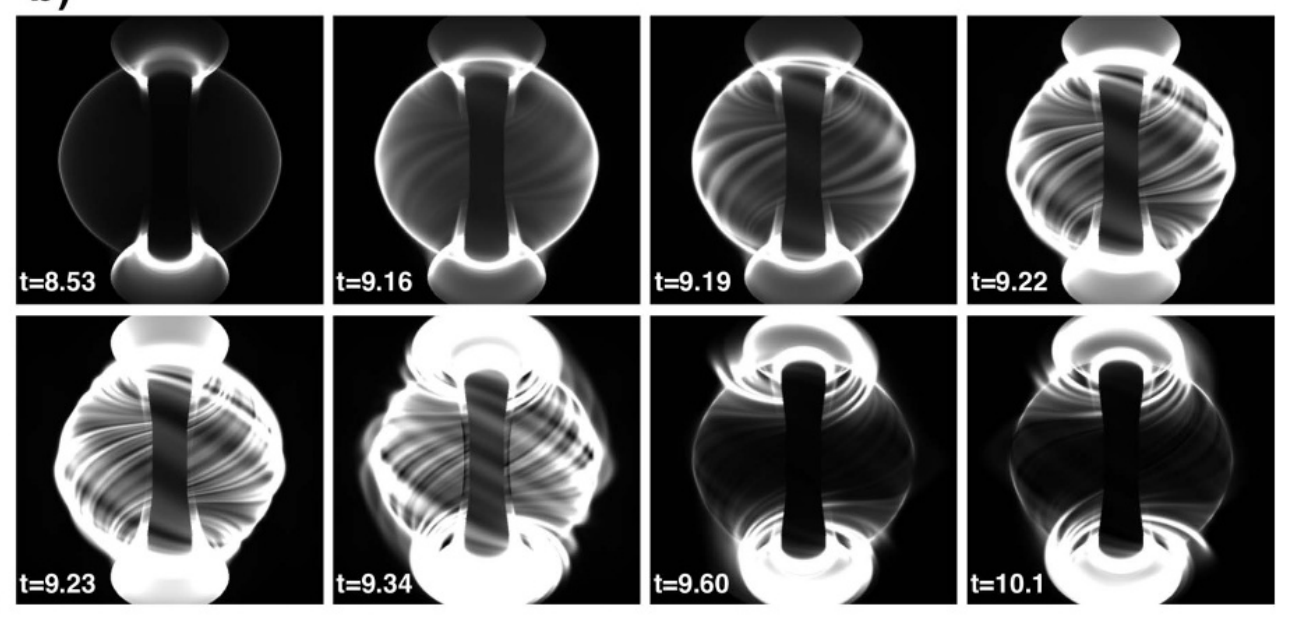}}
    \label{fig: elms_siobhan}%
    \caption{(a): Fast camera image of a MAST plasma during an Edge-Localised-Mode, where plasma filaments are ejected outwards due to MHD instabilities at the plasma edge (figure reproduced from \cite{McArdle_MAST_2010}). (b): The evolution of the filamentary structures during the multi-mode ELM simulation (Figure reproduced from \cite{Smith_2020}) imaged with a synthetic fast camera diagnostic (time given in ms). Full-scale MHD simulations as shown in this figure demonstrate the structural similarities of MHD observed in simulations as well as within the experiment diagnostically captured using the visible camera.}
    \label{fig: elms}
\end{figure}

Predicting the evolution of plasma dynamics is crucial to building sustainable fusion devices as it allows us to anticipate plasma behaviour and make real-time decisions over the input control parameters and alter the trajectory of plasma evolution. Such predictions also allow us to improve our understanding of the physics governing plasma interactions. In particular, Magnetohydrodynamic (MHD) modelling is of critical importance as it improves our understanding of the dynamical evolution of filaments that burst out of the plasma edge and reach the Tokamak’s first wall, as shown in \cref{fig: elms}(a) \cite{Smith_2020}. These plasma filaments, called Edge-Localised Modes (ELMs), are described by non-linear MHD, and can be modelled numerically, as shown in \cref{fig: elms}(b).

In fusion research, prediction of future plasma states is mostly undertaken using physics-based numerical modelling codes, such as BOUT++ \cite{Riva2019}, JOREK \cite{Hoelzl2021jorek}, JINTRAC \cite{jintrac} and SOLPS \cite{WIESEN2015480}. However, these simulations are computationally intensive, taking weeks for effective simulations to be run on high-performance computers \cite{Riva2019}. They are also often complex in terms of software infrastructure and advanced solver library dependencies, making them difficult to deploy for rapid iterative modelling. With numerical solvers, we also encounter cases where the simulations do not converge due to model mismatch or other issues.

Machine learning (ML) offers an alternative data-driven path for obtaining quick, inexpensive approximations to numerical simulations, allowing for rapid iterative modelling \cite{simintelligence}. Within plasma physics, ML-based surrogate models have been used to provide quick approximations for a range of solutions, from emulating turbulent transport \cite{van_de_Plassche_2020, Ho_2021, Meneghini_2017}, estimating the Tritium breeding ratio across Tokamak designs \cite{Manek_2023} to modelling edge plasma \cite{Gopakumar_2020, DASBACH2023101396}. 

Recently, a new class of data-driven ML models, known as neural operators (NO), has recently been at the forefront of surrogate modelling. Unlike other surrogate models, they directly learn the solution operator of the partial differential equation (PDE) in a discretization-convergent manner \cite{kovachki2021neural,li2022pino}. Specifically, neural operators can predict at any resolution, unlike standard neural networks like convolutional neural networks. In other words, they learn the underlying continuum, making them suitable for learning solutions of PDEs. The output of neural operators converges to a unique function upon mesh refinement~\cite{bartolucci2023neural}. NO approaches are more data-efficient compared to other surrogate models \cite{de2022cost}, converge faster than traditional surrogate models, and can generalise better to various PDE settings. Considering these advantages, NO-based surrogate models have found application in emulating the plasma evolution across the domain of a Tokamak \cite{poels2023fast, gopakumar2023fourier}.

{\bf Our Contributions: } 
\begin{itemize}
   \item We demonstrate trained FNOs that can be deployed as an emulator to perform real-time forecasting of the evolution of plasma as characterised by fast cameras on the MAST Tokamak. Our studies show that the FNO is capable of modelling the plasma across the full (available) shot duration ranging from the ramp-up to the L and H modes of operation.
    \item We devise a modified version of the Fourier Neural Operator, capable of handling multiple variables that govern a family of PDEs. The multi-variable FNO allows us to pass field information of all the interested variables to the model, allowing for learning and encoding the mutual dependence of variables, which is essential for capturing the non-linear effects. 
    \item We demonstrate the utility of the multi-variable FNO in (surrogate) modelling the evolution of plasma across a range of reduced MHD cases with increasing complexity. We conduct an extensive analysis of the various features of the FNO, performing ablation studies that evaluate the impact of various choices of the hyperparameters across the model and training. 
  
\end{itemize}

In this paper, we undertake an extensive study of the FNO due to its previously established superior cost-accuracy trade-off in both training and testing \cite{de2022cost}. We use neural operators to model plasma within fusion devices, in both the simulation space and within the experimental domain. For the simulation domain, we consider the reduced MHD models to provide data to train the neural operator \cite{Hoelzl2021jorek}. We study the solution of MHD equations in two spatial dimensions with multiple variables of interest. The simulations that we are interested in modelling represent the diffusion of a blob of plasma within a specified domain. The simulations presented here describe a simplified scenario to that of a full ELM simulation as given in \cref{fig: elms}(b). We iterate over increasingly complex cases of fluid models that govern the diffusion of a plasma blob in toroidal geometry. 

In addition to numerical modelling, we demonstrate that FNOs can be trained to predict the evolution of experimental plasma as observed by high-speed cameras on the MAST spherical Tokamak. The ability to predict future plasma states based on camera imaging, as well as other plasma diagnostics, could allow us to design efficient control feedback loops for future fusion reactors. We build emulators of the fast camera that views across the central solenoid and at the divertor. We find these models to be capable of accurately predicting plasma evolution in real-time.  

In this study, we use FNOs for both MHD simulations and experimental camera data due to the structural similarities of the plasma observed in both. The visual similarity of the MHD modelled in the simulation, and that captured by camera images in a fusion device is elucidated in \cref{fig: elms}(a). The fast camera images capture the MHD behaviour that is observed experimentally. Through this work, the FNO’s utility is demonstrated in predicting both experimental and simulation dynamics. Although this study considers a simplified simulation setup, with filament cross-sections, in a 2D poloidal plane, instead of the full 3D representation, this paper lays the groundwork for future studies where surrogates could be trained from both simulation and experimental data simultaneously. However, in \cref{challenges_scale}, arguments of the challenges and potential methods of extending to a 3D modelling case are discussed. 

A preliminary version of this work was presented at the AI for Science workshop during NeurIPS 2022, where we demonstrated, as a proof-of-concept, the utility of the FNO in modelling MHD scenarios \cite{gopakumar2023fourier}. In the previous work, we conducted a benchmark of the performance of the FNO against more traditional surrogate model architectures such as Convolutional-LSTM \cite{ConvLSTM} and U-Net \cite{UNet}. Our findings established that the FNO takes significantly less time to train, requires fewer parameters, and outperforms the Convolutional-LSTM and the U-Net in emulating the diffusion of a blob of plasma described by reduced MHD \cite{gopakumar2023fourier}. 

The paper is organised as follows: \cref{section: methods} produces a full description of the datasets and methods involved in the paper, ranging from the physics description of the simulations, the diagnostic setup of the experimental camera data and a formal definition of the FNO, its training configuration and the extension to the multi-variable FNO. \cref{section: results} outlines the results of our experiments with using the FNO to learn and predict results from various simulation settings and camera configurations. The section comprises an in-depth analysis of the multi-variable FNO trained to solve different PDEs, exploring the impact of hyperparameters, traini§ng regime, and the training data. We also include a performance comparison of the individual FNO with that of the multi-variable FNO. \cref{section: results} is concluded with the experimental results from the FNO trained to model the plasma evolution as captured by the visible camera looking at the central solenoid and the divertor. We summarise our key results in \cref{section: conclusion} and conclude with a discussion on future extensions of the work. 

\section{Datasets and Methods}
\label{section: methods}

\subsection{MHD Simulations of Plasma Blobs}\label{simulation Data}

The FNO as a Neural-PDE solver is tested on three datasets of simplified Magnetohydrodynamic models, simulating the radial convection of plasma blobs in toroidal geometry, with coordinates system ($R,Z,\phi$). These three cases are designed with increasing complexity to test the utility, versatility, and capabilities of the FNO to emulate MHD models. 
The simplified MHD model is similar to the so-called Reduced-MHD model from \cite{Hoelzl2021jorek}, but with all magnetic fields and currents kept to zero, which is similar to the Euler system, and can be summarised by the continuity, momentum, and pressure equations:

\begin{equation}
      \frac{\partial\rho}{\partial t} =
	- \nabla \cdot \left( \rho \vec{v} \right)
	+ D \nabla^2 \rho 
 \label{CONTINUITY}
\end{equation}

\begin{equation}
    \rho \frac{\partial \vec{v}}{\partial t} =
    - \rho \vec{v} \cdot \nabla \vec{v}
    - \nabla p + \mu \nabla^2\vec{v} 
    \label{MOMENTUM}
\end{equation}

\begin{equation}
 \frac{\partial p}{\partial t} =
    - \vec{v}\cdot\nabla p - \gamma p \nabla \cdot \vec{v} + \kappa \nabla^2 T    
    \label{ENERGY}
\end{equation}

%

    where $\rho$ is the density, $p$ the pressure, $T$ the temperature, and $\vec{v}$ the velocity. $D$ is the diffusion coefficient, $\mu$ the viscosity, and $\kappa$ the thermal conductivity. The ratio of specific heats $\gamma$ is taken to be that of a monatomic gas, $\frac{5}{3}$.

To obtain the Reduced-MHD model, the momentum \cref{MOMENTUM} is projected in the poloidal plane using the operation $\nabla\phi\cdot\nabla\times(R^2...)$, while assuming the poloidal velocity takes the form $\vec{v}=R^2\nabla\phi\times\nabla\Phi$, where the electric potential $\Phi$ is equivalent to the velocity stream function in toroidal coordinates. This projection results in a system of three equations for the three variables $\rho$, $\Phi$, $T$. A pseudo-variable is also used for the vorticity to reduce the level of derivatives in the system: $\tt{w} = \nabla^2\Phi$. We refer the reader to \cite{Hoelzl2021jorek} and references therein for the full system of Reduced-MHD equations.

The simulations are all run in a square slab geometry of $1m$ in width and height, centred at major radius $R=10m$. The runs are performed with toroidally axisymmetric plasma blobs initialised on top of a low background density and temperature. In the absence of a plasma current equilibrium to hold the density blobs in place, the pressure gradient term in the momentum equation generates a buoyancy effect, causing the blobs to move outwards. The simulations are run with the JOREK code, which is routinely used to model non-linear MHD instabilities in realistic geometries \cite{Hoelzl2021jorek}. In this paper, we restrict ourselves to two simplified MHD models.

A further reduction of the model can be obtained by using constant temperature and removing \cref{ENERGY} from the system, therefore evolving only $\rho$ and $\Phi$. With this isothermal model, the pressure gradient term in the momentum \cref{MOMENTUM}, which is proportional to $\rho$, acts as a buoyancy force on the blob, affecting its momentum $\rho\vec{v}$, which is also proportional to $\rho$. In other words, the velocity of the blob will be the same no matter what its density is. To have blobs moving at different speeds, the temperature inside the blob is needed: the higher the temperature, the faster the blob will move.

The values of the diffusive parameters are kept constant inside the entire domain, for all simulations, at $D = 3.4m^2s^{-1}$, $\mu = 2.25\times 10^{-6}kg.m^{-1}s^{-1}$, and $\kappa = 2.25\times 10^{-7}kg.m^{-1}s^{-1}$. The background density level is set to $10^{18}m^{-3}$, and the background temperature is set to $1eV$ (for simulations that evolve temperature). 

Three datasets are considered: with a single isothermal blob, with a single blob with non-uniform temperature, and with multiple blobs with non-uniform temperature. To create the datasets, the blob’s initial conditions are varied. The sampled parameters are the number of blobs, their positions, their width, and their density and temperature amplitudes. This is summarised in Table-\ref{table: data_generation}. For datasets with a single blob, the initial position is fixed at the centre of the domain, $(R,Z) = (10.0,0.0)$. The smaller single-blob datasets are composed of 120 simulations (100 for training, 20 for testing), while the main dataset with multiple blobs contains 2000 simulations. 

\begin{table}
\caption{Domain range and sampling strategies across the initial condition parameters. The case numbers represent the simulation settings in which that parameter was sampled. 1 - single isothermal blob, 2 - single blob with non-uniform temperature, 3 - multiple blobs with non-uniform temperature. $U\;[a, b]$ refers to a uniform distribution bounded by $a$ and $b$.   } 
  \centering
  \begin{tabular}{llll}
  \br
  Parameter & Distribution & Nature & Case\\
    \mr
    Number of Blobs & $U\;[1, \;10]$ & Discrete & 3\\
    R - Position of Blobs & $U\;[9.4, \;10.4]m$ & Continuous & 3 \\
    Z - Position of Blobs & $U\;[-0.4, \;0.4]m$ & Continuous & 3\\
    Width & $U\;[0.02,0.1]m$ & Continuous & 1,2,3  \\
    Density Amplitude of Blobs & $U\;[1.0,\; 4.0]\times 10^{19}m^{-3}$ & Continuous & 1,2,3 \\
    Temperature Amplitude of Blobs & $U\;[12,\; 72]eV$ & Continuous & 2,3\\
    \br
    \end{tabular}
  \label{table: data_generation}
\end{table}

All cases are run with the same time-step size of $0.15\mu s$, for 2000 time-steps in total. This is long enough for the blobs to evolve radially until they reach the wall (where Dirichlet boundary conditions are applied) and disappear through both convective mixing and diffusion. Since a small time-step size is necessary in the simulations to ensure numerical stability, the datasets are down-sampled to 200 slices for the training set. Likewise, the 2D poloidal grid resolution, which is 200$\times$200 bi-cubic finite-elements in the simulations, is down-sampled to a regular grid of 100$\times$100. The smaller single-blob datasets were run on the EOSC-hub cloud using 16 cores per run, and the larger dataset of 2000 multi-blob simulations was run on CINECA's Marconi HPC cluster using 48 cores per run. Each simulation requires approximately 350 core hours. Though each simulation is down-sampled to 200 time instances, we keep the focus of our emulator on short-term evolution up to the first 50 time steps.

\begin{figure}[ht]
    \centering
    \includegraphics[width=1.0\textwidth]{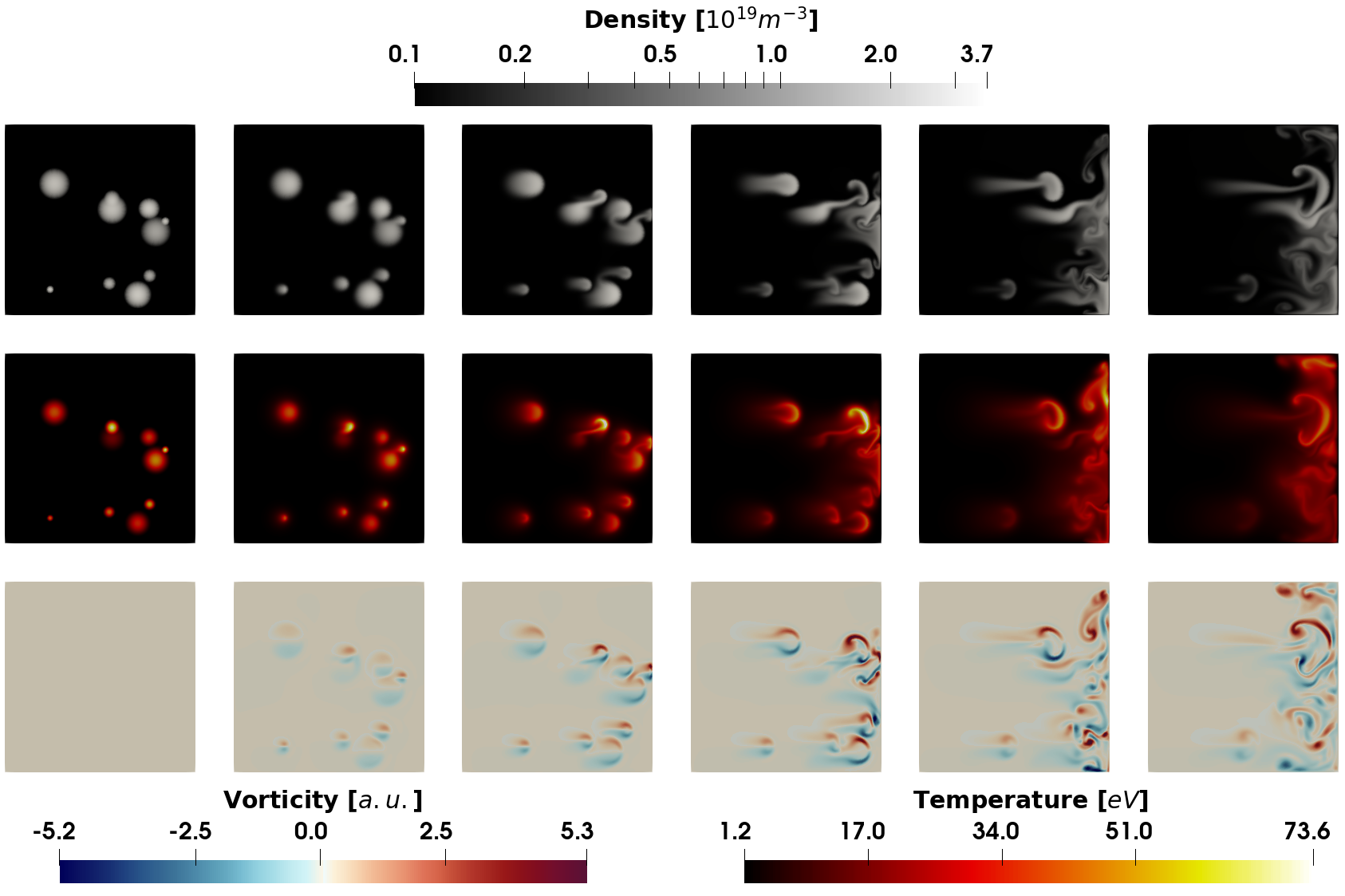}
    \caption{Evolution of multiple blobs inside a slab for the first 500 time-steps of the numerical simulation (the full run is 2000 steps), taken at intervals of 100 steps (about 15$\mu s$). The top row shows the density, the middle row the temperature, and the bottom row the vorticity.}
    \label{fig: multi_blob_example}
\end{figure}

Simulations of blobs in slab geometry are related to turbulence and MHD instabilities at the plasma edge of a Tokamak by assuming that they represent a poloidal cross-section of a filament. As seen in \cref{fig: elms}, filament structures are aligned to the magnetic field, they are elongated in the toroidal direction, but have a nearly circular cross section in the poloidal plane. Such slab simulations represent the dynamic evolution of these filaments in the plane perpendicular to the magnetic field. Extensive modelling of blobs in slab geometry can be found in \cite{EASY_2016,MILITELLO_2017} and references therein. \cref{fig: multi_blob_example} shows an example of a numerical simulation of multiple blobs evolving for the first 500 steps (out of 2000), with the density, temperature, and vorticity. Filaments with higher temperatures can be seen to move at a faster velocity.

\subsection{Fast-Camera Images on MAST Tokamak}
\label{experimental_data}

MAST was equipped with fast camera imaging diagnostics at several points across the Tokamak for capturing images within the visible spectrum, capturing the plasma evolution in real-time \footnote{https://ccfe.ukaea.uk/research/mast-upgrade/}.

These Photron cameras
\footnote{https://photron.com/} are essential diagnostics that contributed to fundamental understandings of key plasma phenomena in Tokamaks in recent years \cite{Kirk2006}. While most of the camera data has been used for qualitative analysis in the past, it can be exploited using more advanced methods to provide statistical insight into plasma turbulence \cite{Walkden2022}, as well as global instabilities (disruptions) \cite{Ham2022}. Within the scope of this work, we focus on two camera views. The first looks at the central solenoid and the plasma confined around it, extending to the walls and providing a holistic view of the plasma (\textit{rbb camera}). The second camera configuration under study is that providing the divertor-view, looking at the bottom of the Tokamak, where energy travels from the plasma to the exhaust region (\textit{rba camera}). \cref{fig: elms} shows a view of the main plasma as seen by the camera, together with a synthetic rendering (right) of the camera, simulating the MHD physics. The cameras have a temporal resolution of 1.2ms, with spatial resolution varying according to the desired calibration setup for each experiment. They are tuned to visible wavelength and, therefore, capture mostly the Balmer D$_{\alpha}$ light emitted mostly from the plasma edge. As seen in \cref{fig: elms}, the fast-visible cameras are able to capture complex dynamics of plasma filamentary eruptions at the plasma edge, the Edge-Localised Modes (ELMs) \cite{Kirk2006}, hence being able to predict real-time evolution allows us to anticipate such plasma instabilities and other confinement related characteristics. Within the scope of this project, our focus relies on modelling the plasma confinement and its interactions with wall accurately, leaving the modelling of ELMs as the next potential research avenue. 


\begin{figure}[h!]
    \centering
    \subfloat[View of the \textit{rbb} camera.]{\includegraphics[width=7.0cm]{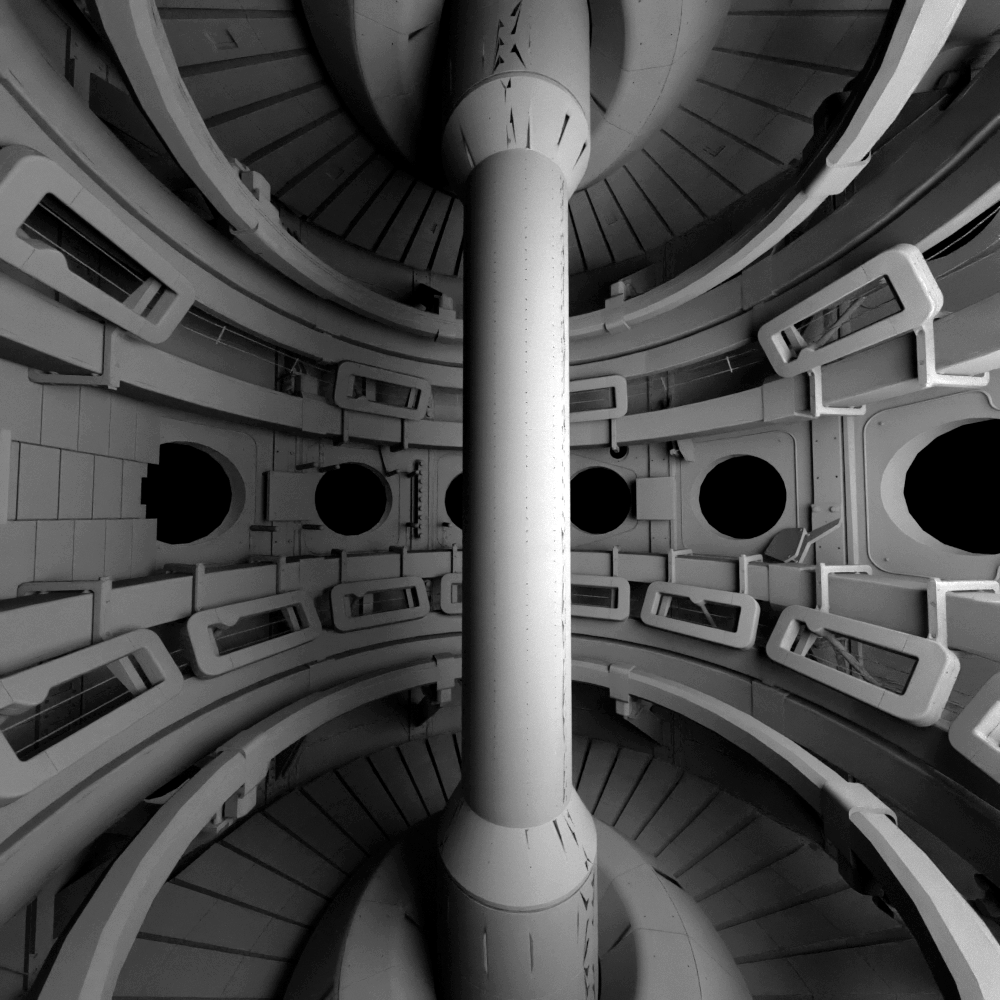}}
    \label{fig:rbb_view}
    \centering
    \hspace{10mm}
    \subfloat[View of the \textit{rba} camera.]{\includegraphics[width=7.0cm]{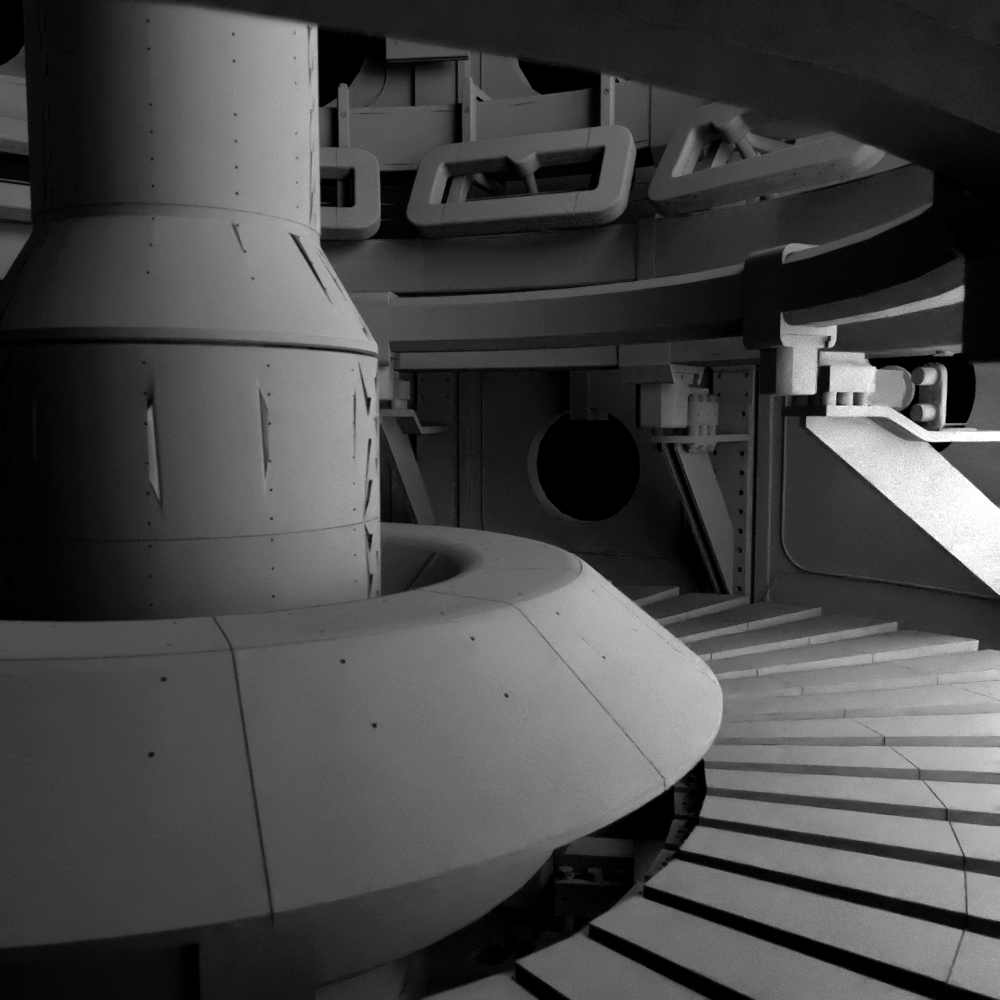}}
    \label{fig:rba_view}
    \caption{Views of the fast cameras on MAST in the absence of plasma, obtained using illuminations on the 3D CAD model of MAST. Camera shown in figure (a), looks across the central solenoid, while the camera in figure (b) shows the view at the divertor.}
    \label{fig: MAST_CAD}
\end{figure}

\subsection{Fourier Neural Operator}\label{fno}

The Fourier Neural Operator (FNO) \cite{li2021fourier} is a neural network-based operator learning method \cite{kovachki2021neural} that learns a mapping between two function spaces from a finite collection of observed input-output pairs. In the Neural Operator (NO) setting, we learn the mapping from the input function space $A$ to the output function space $U$. The input values are given by the function $a \in A$, defined on a bounded domain $D \subset \mathbb{R}^{d_a}$. Similarly, the output values are given by the function $u$, valued and defined on the bounded domain $D' \subset \mathbb{R}^{d_u}$. The NO can be written as a parameterised representation of this operator across function spaces: 
\begin{equation}
    NO: G_\theta : A \rightarrow U 
\end{equation}
$G_\theta$ is a neural network configuration parameterised by $\theta$ and is composed of three elements: 
\begin{enumerate}
    \item \textbf{Lifting}: A fully local, point-wise operation that projects the input domain to a higher dimensional latent representation $a\; \in \; \mathbb{R}^{d_a} \rightarrow \nu_0 \; \in \; \mathbb{R}^{d_{\nu_0}}$. 
    \item \textbf{Iterative Kernel Integration}: Expressed as a sum of a local, linear operator and a non-local integral kernel operator, that maps from $\nu_t \rightarrow \nu_{t+1}$, iterated from $\nu_0$ to $\nu_{T}$. The kernel integration is followed by a point-wise non-linearity. 
    \begin{equation}
    \label{eqn: IKI}
        \nu_{t+1} = \sigma\bigg(W\nu_t(x) + \kappa(a;\phi)\nu_t(x)\bigg)
    \end{equation}

    \item \textbf{Projection}: Similar to lifting, characterised by a fully local point-wise operation that maps the latent representation to that of the output domain $\nu_T\; \in \; \mathbb{R}^{d_{\nu_T}} \rightarrow u \; \in \; \mathbb{R}^{d_u}$.
\end{enumerate}

In \cref{eqn: IKI}, $\sigma$ denotes the nonlinear activation, $W$ represents the local, linear operator, and $\kappa$ represents the chosen kernel which is parameterised by $\phi$. $x$ represents the input that is fed into that layer. 

Conventionally, the lifting and projection within an NO are characterised by linear layers acting on the respective feature spaces. The kernel integration operator is characterised as the neural layer that evaluates a chosen kernel over the higher dimensional latent representation fed into it:
\begin{equation}
\big(\kappa(a;\phi)\nu_t\big)(x) = \int_D \kappa\big(x, y, a(x), a(y) ; \phi \big)\nu_t(y)dy.
\end{equation}

As demonstrated in \cite{li2021fourier}, the FNO is devised by characterising the kernel integral operator as an operation within the Fourier space. Replacing $\kappa$ with the Fourier integral operator $\mathcal{F(\kappa_\phi)}$ we obtain    
\begin{equation}
\label{fourier integral transform}
    \big(\kappa(a;\phi)\nu_t\big)(x) = \mathcal{F}^{-1}\big(\mathcal{F}(\kappa_\phi) . \mathcal{F}(\nu_t)\big)(x).
\end{equation}
Here, $\mathcal{F}$ and $\mathcal{F}^{-1}$ represent the Fourier and inverse Fourier transform, respectively. $\kappa$, the kernel is parameterised by learnable weights $\phi \; \in \; \theta_\kappa$. Within the FNO, the kernel integrator reduces to the inverse Fourier transform of the inner product between the Fourier integral operator $\mathcal{F}(\kappa_\phi)$ and the Fourier transform of the input to that Fourier layer $\mathcal{F}(\nu_t)$.   

\begin{figure}
    \centering
    \includegraphics[width=0.75\textwidth]{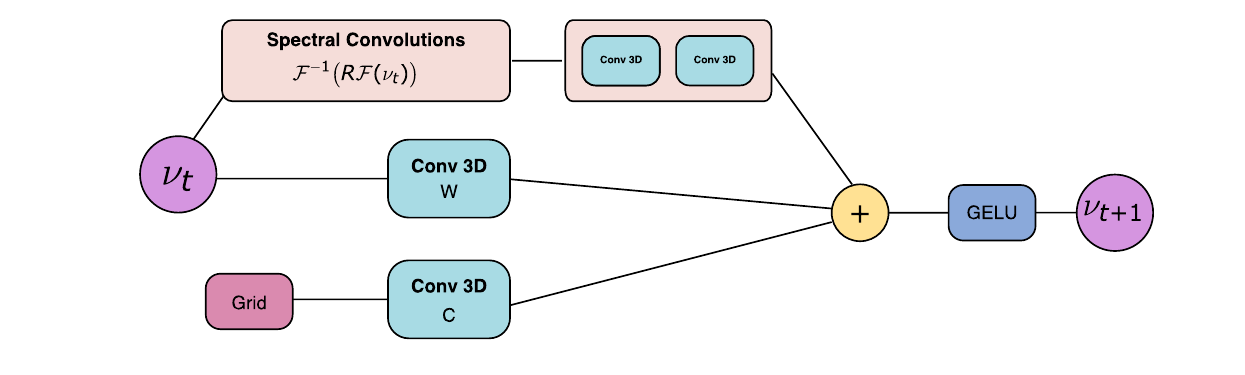}
    \caption{Modified architecture of a single Fourier Layer. FNO is constructed by stacking multiple Fourier layers on top of each other, sandwiched by a lifting and projection operation.}
    \label{fig: fno_arch}
\end{figure}

In practice, the iterative kernel integral transform with Fourier transform is characterised by two learnable weight matrices $R$ and $W$ as shown in \cref{fig: fno_arch}. $R$ learns the mapping behaviour within the Fourier space, while $W$ learns the mapping required within the input Euclidean space. $R$ characterises the non-local integral kernel operator, while $W$ represents the local linear operators. The weight matrix $R$ within the Fourier layer enables convolution within the Fourier space, allowing the network to be parameterised directly in it. The grid discretisation is also fed in at each layer to help the network learn the discretised Fourier representation of the data and undergoes point-wise linear transformations, characterised by $C$, its operation is similar to that of $W$. The output of a Fourier layer can be expressed as
\begin{equation}
    y = \sigma\bigg(\mathcal{F}^{-1}\big(R\mathcal{F}(x)\big) + Wx + Cg + b\bigg),
    \label{fno_eqn}
\end{equation}
where $x$ is the input to that layer, $y$ the output, $g$ the grid discretisation, $b$ the bias and $\sigma$ is the nonlinear activation function. 

\cref{fno_eqn} resembles a single Fourier layer within the FNO. Multiple Fourier layers are stacked on top of each other to complete the iterative kernel integral operation. The stacked Fourier layers are sandwiched between the lifting and projection operations to complete the FNO architecture. 

\subsubsection{Comparison to Convolutions}~\\
The spectral convolutions characterised by $\kappa$ in \cref{fourier integral transform} have an analogous relationship to the convolutional operation. Convolution, being a product of functions across a certain range, is an integral transform. Expressed across an infinite range, it takes the mathematical form
\begin{equation}
\label{convolution_operation}
    (f*g)(t) = \int_{-\infty}^{+\infty}f(\tau).g(t-\tau)d\tau,
\end{equation}
where $f$ and $g$ are functions over which the convolution operation is performed. By replacing the convolution operation with the Fourier transform in \cref{convolution_operation}, we obtain
\begin{equation}
    \label{fourier_operation}
    (\mathcal{F}f)(k) = \int_{-\infty}^{+\infty} f(x)e^{-i2\pi \tau k}d\tau.
\end{equation}
Comparing \cref{convolution_operation} and \cref{fourier_operation}, it becomes evident how \cref{fourier integral transform} is a special form of the convolutional operation, one that takes place in the Fourier space. In practice, both operations are applied across a discretisation, characterised by the data that is fed to the model, but since we decompose the data into Fourier modes within the FNO and apply learning across the decomposed components, we can learn continuous representations of the required mapping as opposed to learning discrete representations with conventional convolutions. 

Named spectral convolutions in \cite{li2021fourier}, allow for several advantages over a standard convolutional neural network, where filters are directly learned without a spectral transform.
Firstly, by learning within the Fourier space, we allow the network to identify global features that are embedded within the data, highlighting the dominant modes in the process. By performing the Fourier decomposition over the data as it parses through the network, the weights are tuned to identify and learn the modes that possess a majority of the information, allowing the network to generalise better. Secondly, the spectral convolutions being in the function space allows for learning the mapping in a discretisation convergent manner, allowing for learning mesh-free solutions. Lastly, the Fourier transform in \cref{fourier_operation} is implemented using a Fast Fourier Transform (FFT) \cite{FFT} which has a computational complexity of $\mathcal{O}(n\log{}n)$, whereas the standard convolution operation without spectral transform has a complexity of $\mathcal{O}(kn^2)$, where $k$ is the size of the kernel and $n$ is the representation dimension. However, within the FNO, we only deal with a complexity of $\mathcal{O}(nk_{max}$) since we truncate the Fourier modes up to a desired hyperparameter $k_{max}$.

\subsubsection{Training Configuration}~\\
\label{training_config}
For emulating the simulations, we use the \textbf{FNO-2d} configuration\cite{li2021fourier}, where a 2-d Fourier Neural Operator which only convolves in space, is deployed in an autoregressive manner \cite{dalal2019autoregressive}. The FNO takes in as its input the field values across a 2D grid for an initial set of time instances (\textit{T\_in}) along with the grid discretisation in both dimensions. Similar to spectral numerical solvers, the grid discretisation, fed in with the field data, provides better-structured information for discrete Fourier transform utilised within the FFT. The FNO takes the initial fields as the input and outputs a set of later time instances (step) of the field value across the same grid. The output of the FNO, coupled with the later time steps from the input, are then fed back in an autoregressive manner to estimate further field evolution in time, until the desired time instance (\textit{T\_out}) is reached. Within the autoregressive setting, we roll out the evolution of the fields up until the desired time step and then perform the backpropagation through time to enable learning of the time evolution of the PDE. For further information regarding the theory behind operator learning as ascribed by the FNO and its implementation in various case settings, we invite the reader to refer to the original work in \cite{li2021fourier}. Unlike the FNO 3D configuration, we only perform Fourier transformations across the spatial dimensions and do not include the temporal dimension. We have chosen the 2D configuration of the FNO as it is lighter, quicker, and gives us the flexibility in deciding the length of the temporal evolution.

\begin{figure}[h!]
    \centering
    \includegraphics[width=0.85\textwidth]{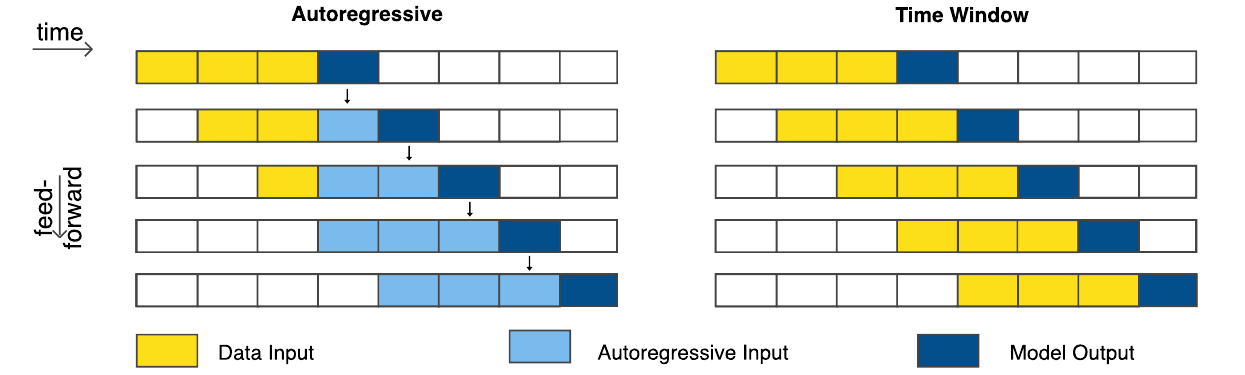}
    \caption{Schematic representation of the two kinds of training configurations deployed within this study. For the simulations, we use the autoregressive configuration (left). We only use the initial set of field values to propagate towards the desired time step, where the intermediary model output is fed back into the model as inputs to further the time evolution. For the experimental data, we use a sequential time window-based configuration. Here, the model output is not factored back into the model, but we perform a fixed window mapping across time. As we go across the time duration of the data, the window is slid further by one-time instance (right). In the example above, we have \textit{T\_in}=3, step=1, \textit{T\_out}=8, the time focus of each forward propagation of the model is given along the $x$-axis while the evolution across the full-time domain is given along the $y$-axis. Each cell block represents the 2D field value at a certain instance of time. The input and output sizes shown in this diagram are representative only and vary across each experiment.}
    \label{fig:ar_tw}
\end{figure}

For the case of the FNO for camera, instead of performing an autoregressive time roll out, we perform a fixed time window mapping, where the FNO takes in the camera information for fixed time steps (\textit{T\_in}) and outputs the next set of frames characterising the future plasma state (\textit{step}). The next set of inputs is obtained by sliding the interested time window by one, as shown in \cref{fig:ar_tw}. 

\subsubsection{Multi-variable FNO}~\\
\label{mv_fno}

As depicted in Equations (\ref{CONTINUITY}), (\ref{MOMENTUM}) and (\ref{ENERGY}), the physics model represents a highly correlated and interdependent system where density, temperature, and electric potential factor into the time evolution equations of each other. Taking this family of PDEs into account, it is vital to construct a surrogate model that is capable of emulating all variables together, allowing for information exchange within the model. The original FNO \cite{li2021fourier} was designed to model for a single variable and needed to be modified to handle multiple variables. The work done in \cite{gopakumar2023fourier} demonstrates the impact of the FNO on predicting MHD evolution but modelled each variable individually. Considering the simplicity of the reduced MHD case used in \cite{gopakumar2023fourier}, it was permissible to model the variables independently. However, for the case under consideration, it was imperative that for physical consistency, we model the variables together. As we move into more complex MHD cases, we introduce more correlated variables such as density and temperature, which are both required to evolve together to fully grasp the propagation of the plasma blob. 

To account for the multi-variable setting, we designed the FNO to take in the variables as a separate channel. Initial construction of the FNO would consider data in the format \textit{[batch size, temporal projection, spatial along x-axis, spatial along y-axis]}. We modified that model to consider data in the format \textit{[batch size, variables, temporal projection, spatial along x-axis, spatial along y-axis]}. This modification allows us to account for all the variables within a single model, which facilitates information exchange across them. By having the multiple variables of the physics model as a channel, we allow for the Fourier transform to be applied across the spatial domain for all variables of interest. The $R$ matrix that allows for learning within the Fourier space in \cref{fno_eqn} is modified to account for the new variables, and the local, linear operator $W$ acts across the variable and spatial domain. By providing the variables as a separate channel across the FNO, we introduce the possibility of performing the operations of lifting and projection defined in \cref{fno} not just across the temporal features but across the fields as well. Thus, within this configuration of a multi-variable FNO, the feature space becomes two-fold, accounting for the time as well as the variables. We also modify the FNO with skip connections to reduce the information loss as the data progresses through the network \cite{srivastava2015highway}. 


\section{Results}
\label{section: results}

Prior to elaborating on the finer details of the results from our experiments, we provide an overview of the performance obtained by the FNO surrogates across all the interested modelling tasks in \cref{table: overview}. Across our range of experiments, emulating the plasma, we find that the FNO achieves considerable accuracy (Mean Squared Error (MSE) in the normalised domain $\sim 10^{-4}$). Despite being heavily parameterised, the FFT within the network helps the FNO achieve quick training and execution times for surrogate models\cite{gopakumar2023fourier}. The Fourier features extracted help provide the model with implicit bias found in the data, allowing it to converge quicker on the spatial evolution of the plasma. By transforming the data to the Fourier space, we allow the network to easily extract the dominant features found in the data as they would be highlighted across the various Fourier modes.  As mentioned in \cref{introduction}, we focus our experiments solely on the FNO as the chosen surrogate. For a detailed benchmark of MHD emulators across various models, we refer to our prior work \cite{gopakumar2023fourier}.

\begin{table}[h!]
    \centering
    \scalebox{0.75}{
    \begin{tabular}{llllll}
    Case & Variables & Parameters & Train Time (m) & Training Size & MSE  \\
    \mr
    Single blob -- isothermal & 2 & 6,321,000 & 48 & 100 & $7.58\times 10^{-5}$ \\
    Single blob -- non-uniform temperature & 3 & 9,467,000 & 97 & 160 & $2.56\times 10^{-5}$ \\ 
    Multiple blobs -- non-uniform temperature & 3 & 9,467,000 & 682 & 1500 & $1.33\times 10^{-4}$  \\
    Camera -- Solenoid (\textit{rbb}) & 1 & 201,900 & 1568 & 7661 & $1.97\times 10^{-3}$ \\
    Camera -- Divertor (\textit{rba}) & 1 & 201,900 & 935  & 5851 & $3.80\times 10^{-3}$ \\
    \br
  \end{tabular}
  }
  \caption{Overview of the performance obtained by various FNO-based surrogate models built to emulate the plasma evolution in both simulations and experiments. The case represents the various simulation settings of JOREK or the camera diagnostic on MAST, for which we are training an emulator. The variables mention the number of variables from the PDE/diagnostic that we model together. Parameters refer to the number of parameters in the trained FNO, and MSE represents the mean squared error in the normalised domain. The Training Size represented the simulation/experimental input-output pairs. All models are trained on a single Nvidia A100 chip.}   \label{table: overview}
\end{table}

\label{results}
\subsection{Multi-variable FNO over MHD Simulations} \label{fno_over_mhd}
We design and train multi-variable FNOs to model the three MHD cases under study: Single isothermal blob, single blob in a non-uniform temperature field, and multiple blobs in a non-uniform temperature field, as described in \cref{simulation Data}. Each multi-variable FNO is constructed as described in \cref{mv_fno} and trained using the autoregressive framework outlined in \cref{training_config}.  

For each multi-variable FNO, we chose an input feed of 10 time steps of the field information (\textit{T\_in} = 10), and an output size of 5 (\textit{step} = 5). Each time step within the FNO corresponds to 1.5 $\mu s$ of simulation time. Each FNO was autoregressively rolled out until the $50^{th}$ time instance, allowing the neural operator to learn the time evolution of the function till then (\textit{T\_out}=50). Each Fourier layer within the multi-variable FNO has 16 modes and a temporal channel width of 32 (unless mentioned otherwise). The temporal channel width in the FNO is akin to the channel width in a CNN, where the temporal data is scaled in the FNO channels instead of the RGB color space in the CNN channels \cite{Lenet}. The channel variable width was maintained at the number of variables that the FNO learned. The grid discretisation associated with the field data is added to the training data within each model's forward call as an addition to the temporal channel, providing further field information to the model. We utilise the discretisation used within JOREK to generate the training data as the grid features within the FNO. We deploy a two-fold normalisation strategy considering the nature of the dataset. The physical field information represented within the MHD cases is in different scales, with densities ranging from 0 to $10^{20}$ $m^{-3}$ and temperature ranging up to $100$ $eV$. Since we consider the gradual diffusion of an in-homogeneous density blob(s), the data distribution within the spatial domain is severely imbalanced. Taking these aspects of the training data into consideration, initially the field values are scaled down by a factor, indicative of the range of operation found within the simulation itself. The physics scaling would divide the data by the order of magnitude around which the plasma operates for each variable. We perform this by scaling the field value with the order of magnitude of its maximum found in the training data. This is followed up by a linear range scaling, allowing the field values to lie between -1 and 1. The normalisation pipeline is implemented separately for each of the variables that the multi-variable FNO accounts for.  Each model is trained for up to 500 epochs using the Adam optimiser \cite{adam} with a step-decaying learning rate. The learning rate is initially set to 0.005 and scheduled to decrease by half after every 100 epochs.The input, output dimensions and the model architecture used for each case is given in \ref{appendix_fno_arch}.

\begin{figure}[h!]
    \centering
    \includegraphics[width=10.0cm]{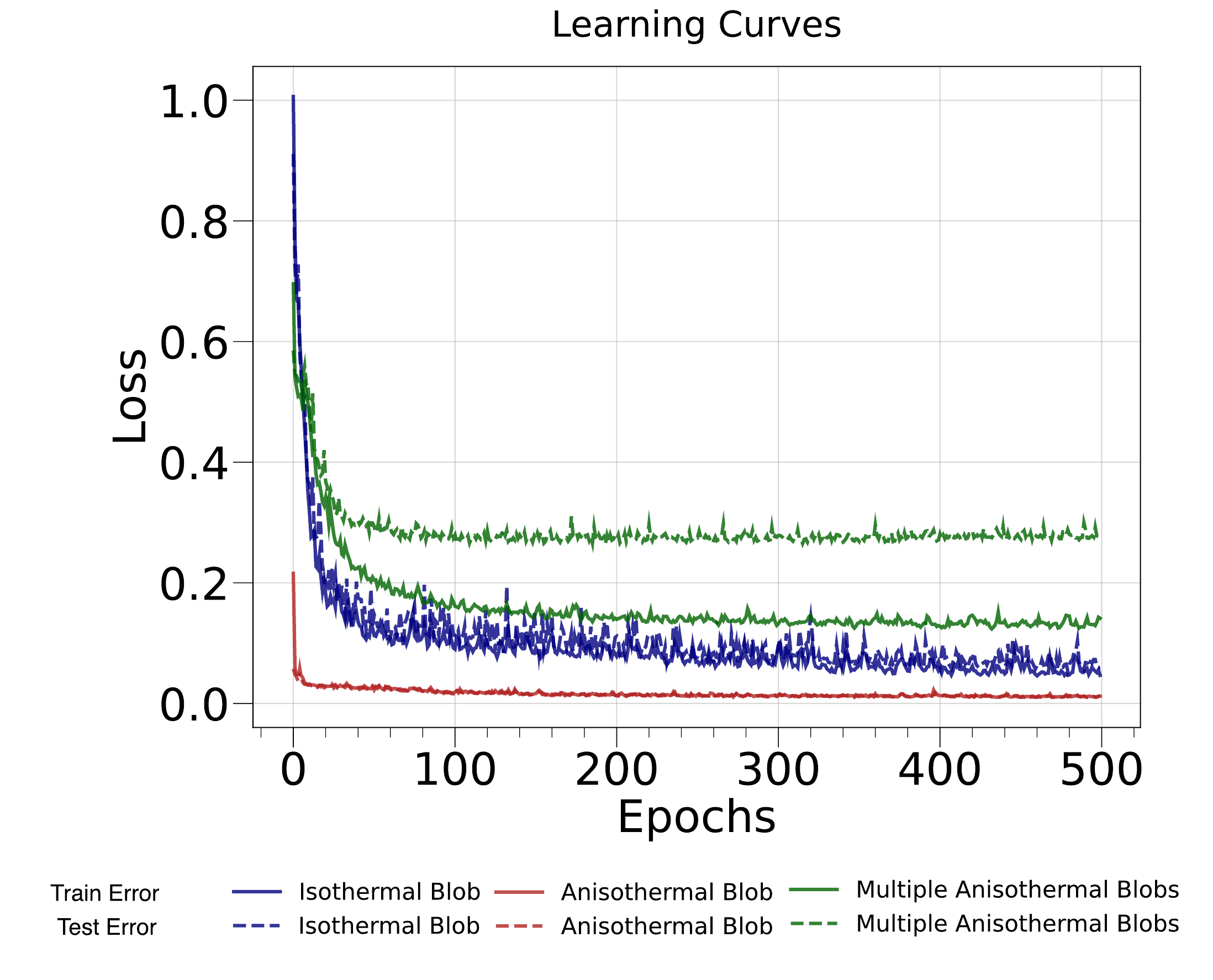}
    \caption{Learning curves associated with training the multi-variable FNO across the various the various Reduced MHD Cases. Isothermal blob refers to the single blob diffusion under uniform temperature, Anisothermal blob refers to the single blob under a non-uniform temperature, and Anisothermal multiple Blobs extend that to the case with several density blobs under a non-uniform temperature distribution.}
    \label{fig: learning_curves}
\end{figure}

The loss function is expressed as an equally weighted linear combination of two elements: the reconstruction error and the difference error, expressed as \\[3ex]
%
%
\begin{equation}
\label{eqn: loss_function}
        L = \frac{1}{N} \sum_{i=1}^{N} \Biggl ( \eqnmarkbox[red]{recon} {\sum_{t=1}^{T} \frac{\| \tilde{y}_t - y_t \|_2^2}{\| y_t \|_2^2} } + \eqnmarkbox[blue]{diff}{ \sum_{t=1}^{T} \frac{\big \| \bigl ((\tilde{y}_{t+step} - \tilde{y_t}) - (y_{t+step} - y_t)\bigr) \big \|_2^2}{\| (y_{t+step} - y_t) \|_2^2} }  \Biggr ),
\end{equation} \\
\annotate[yshift=1em,xshift=2ex]{above,left}{recon}{\textbf{Reconstruction Error}}
\annotate[yshift=-1.em]{below,right}{diff}{\textbf{Difference Error}}
\\[2ex]
where $\tilde{y}_t$ represents the FNO output at the time step t, $y_t$ represents the labelled (actual) data at  time frame $t$, $step$ represents the step size of the FNO output, $T$ represents the full length of the autoregressive time roll-out and N, the batch size or the number of simulations we are learning at a time over which we sum over. 

The reconstruction error helps the FNO to learn to minimise the prediction error for a particular instance in time, whereas the difference error helps learn the time evolution of the plasma better as it minimises the temporal differences that arise in the autoregressive roll-out of the FNO.



The focus of this research relies on the short-term evolution of up to 50 time steps from a total of 200 time steps, seeing how well the multi-variable FNO can model the evolution of all the physical variables in an integrated manner. We restrict ourselves to the short-term due to the errors arising from the autoregressive time rollouts, which are further explored in \cref{ind_vs_multi}, \ref{step_size} and \ref{mode_ablations}. To see the performance of the FNO deployed in an autoregressive setting for the full time range of the simulation, see \ref{appendix_long_rollouts}.

\subsubsection{Single Blob -- Isothermal}~\\

For the first case, that of a single blob in an isothermal setting (hereby referred to as an isothermal blob), the multi-variable FNO models the evolution of the two variables: density $\rho$ and electric potential $\Phi$. Considering the simple structure of the physics, the FNO performs well in (surrogate) modelling the evolution of both the variables arriving at an MSE (in the normalised domain) of $7.58\times 10^{-5}$ in 48 minutes on a single Nvidia A100 chip. The multi-variable FNO is capable of emulating the evolution of the plasma up to the $50th$ time step in 0.025 seconds, offering a speedup of 6 orders of magnitude. Further plots showing the contrast of the FNO output with that of JOREK by way of discrepancy plots can be found in \ref{appendix_errorplots}.

\begin{figure}[h!]
    \centering
    \subfloat[Density]{\includegraphics[width=7.5cm]{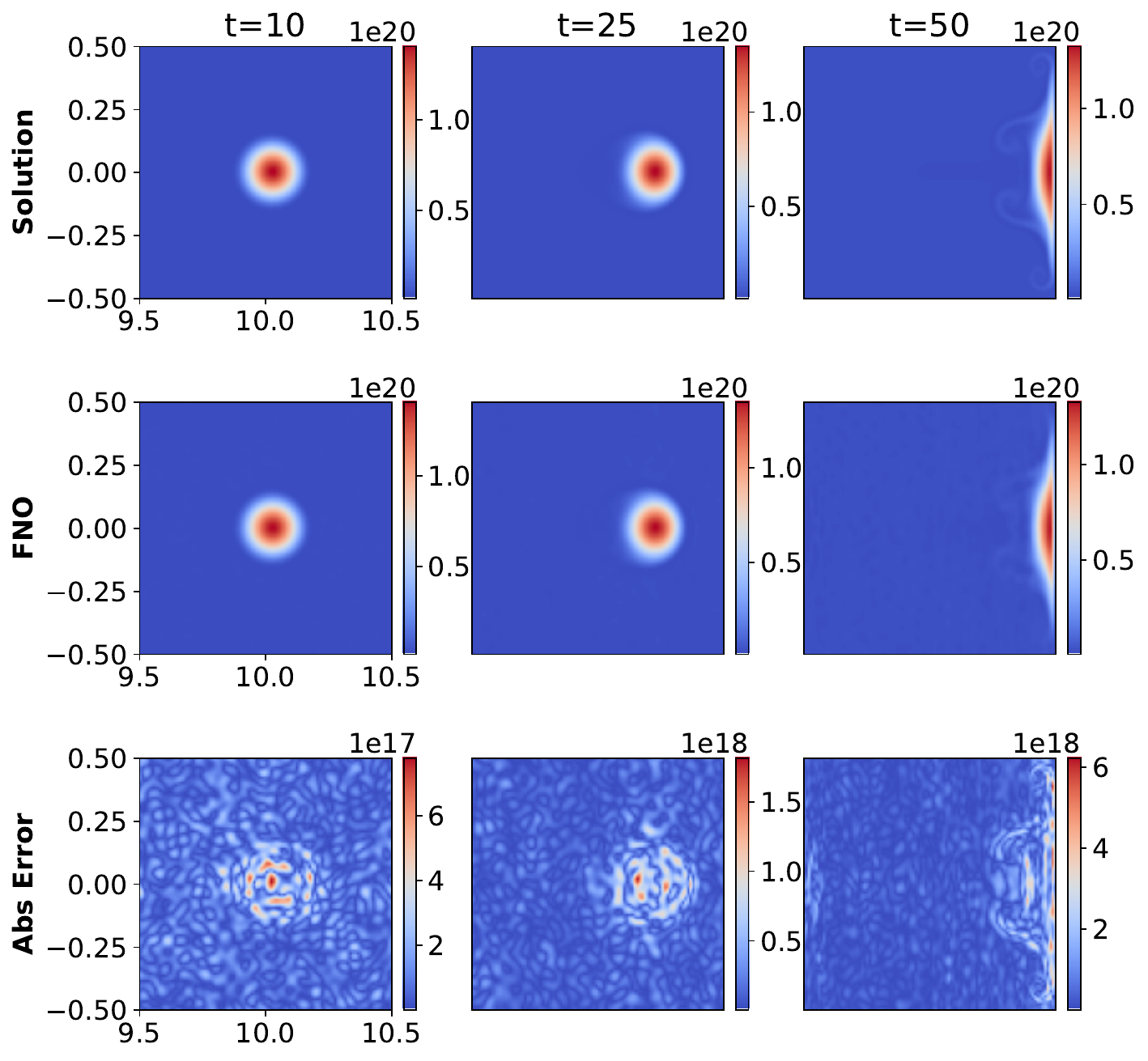}}
    \label{fig: iso_rho}
    \centering
    \subfloat[Electric Potential]{\includegraphics[width=7.7cm]{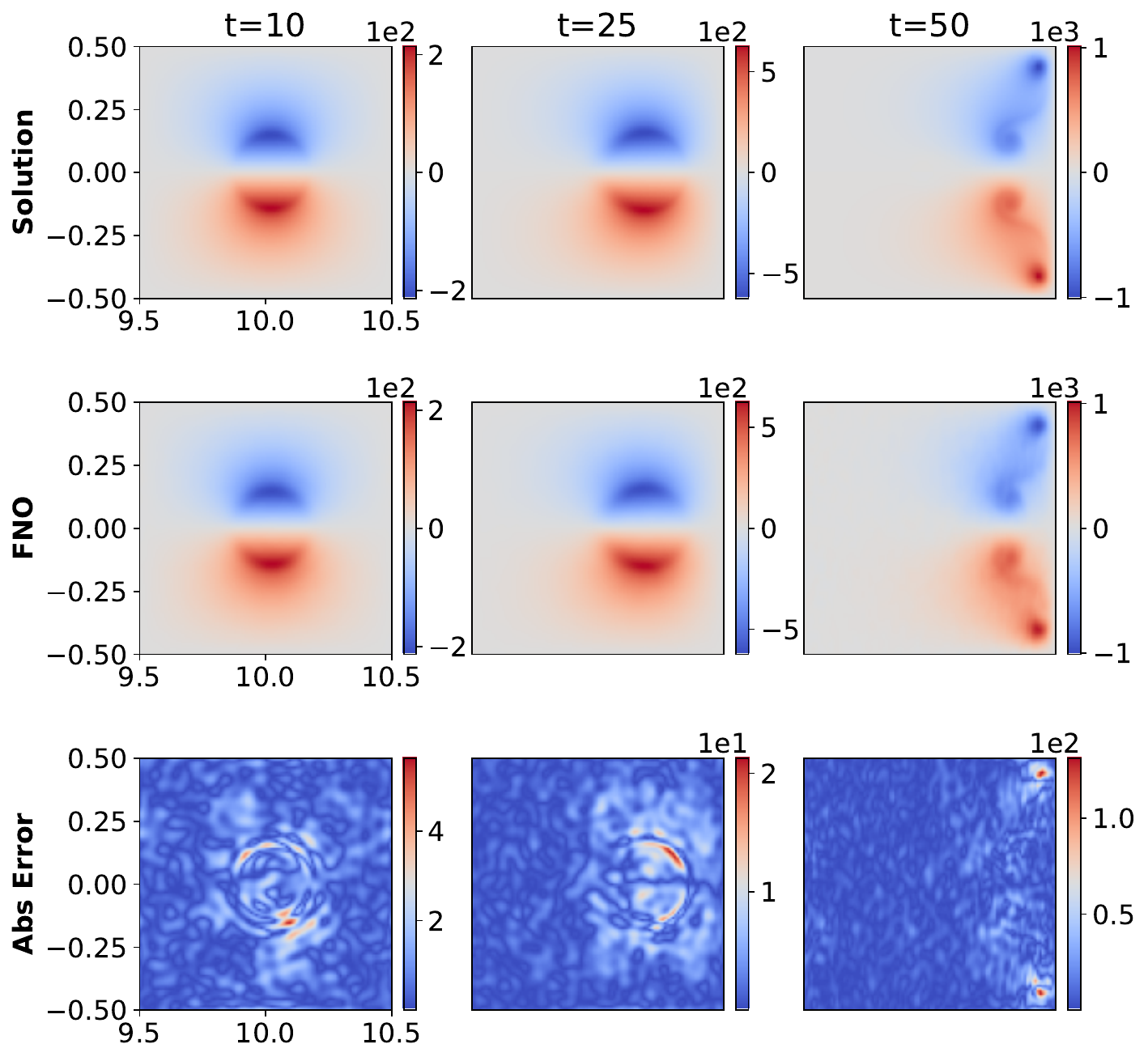}}
    \label{fig: iso_phi}
    \caption{Isothermal Blob: Temporal evolution of (a) density and (b) electric potential of the plasma evolution as obtained using the JOREK code (top of each image), that of the trained multi-variable FNO (middle of each figure) and the absolute error across both (bottom of each figure). The spatial domain is given in toroidal geometry characterised by $R$ in the $x$-axis and $Z$ in the $y$-axis.}
    \label{fig: iso_mhd}
\end{figure}

\subsubsection{Single Blob with Non-uniform Temperature}~\\

Moving up in complexity, in the case of a single density and temperature blob, the FNO is adapted to model density, electric potential, and temperature, three highly correlated variables as described in \cref{simulation Data}. Even considering the additional physics, the multi-variable FNO demonstrates the ability to learn the global features of blob diffusion. It characterises the movements of the blob, its interaction with the walls, and its eventual dissipation. We notice that with the added field variables, the FNO performs well in predicting the evolution within the initial timesteps, but loses rigour of structure as we roll further out in time. Despite the challenge of preserving minor features further into time, the multi-variable FNO achieves an MSE (in the normalised domain) of $2.56 \times 10^{-5}$. The model trains in 1 hour 37 minutes on a single Nvidia A100 chip. The multi-variable FNO is capable of (surrogate) modelling the evolution of the plasma up to the $50th$ time step in 0.03 seconds. Further plots showing the contrast of the FNO output with that of JOREK by way of discrepancy plots can be found in \ref{appendix_errorplots}. 

\begin{figure}[h!]
    \centering
    \subfloat[Density]{\includegraphics[width=7.5cm]{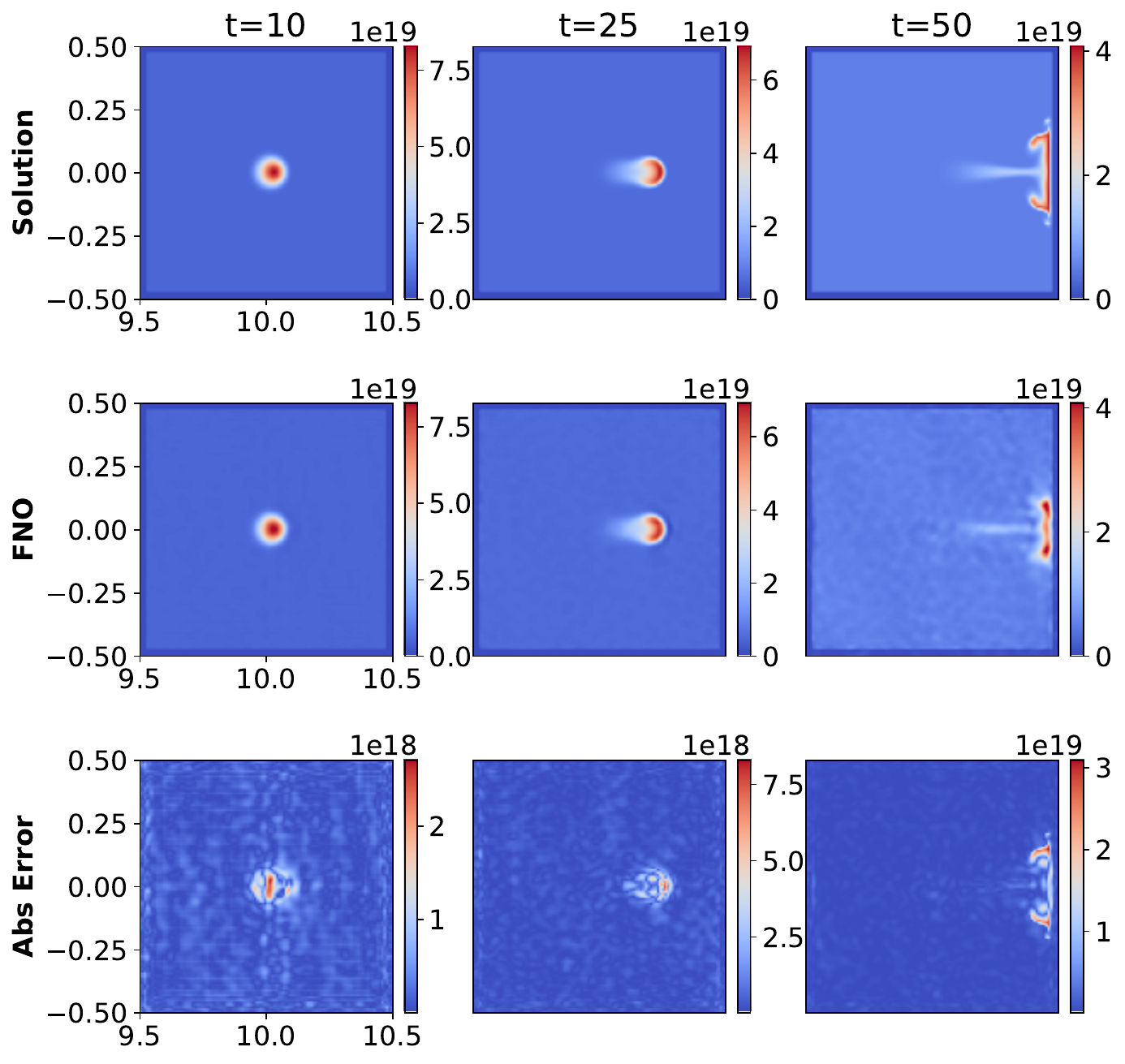}}
    \label{fig: single_rho}
    \centering
    \subfloat[Electric Potential]{\includegraphics[width=7.7cm]{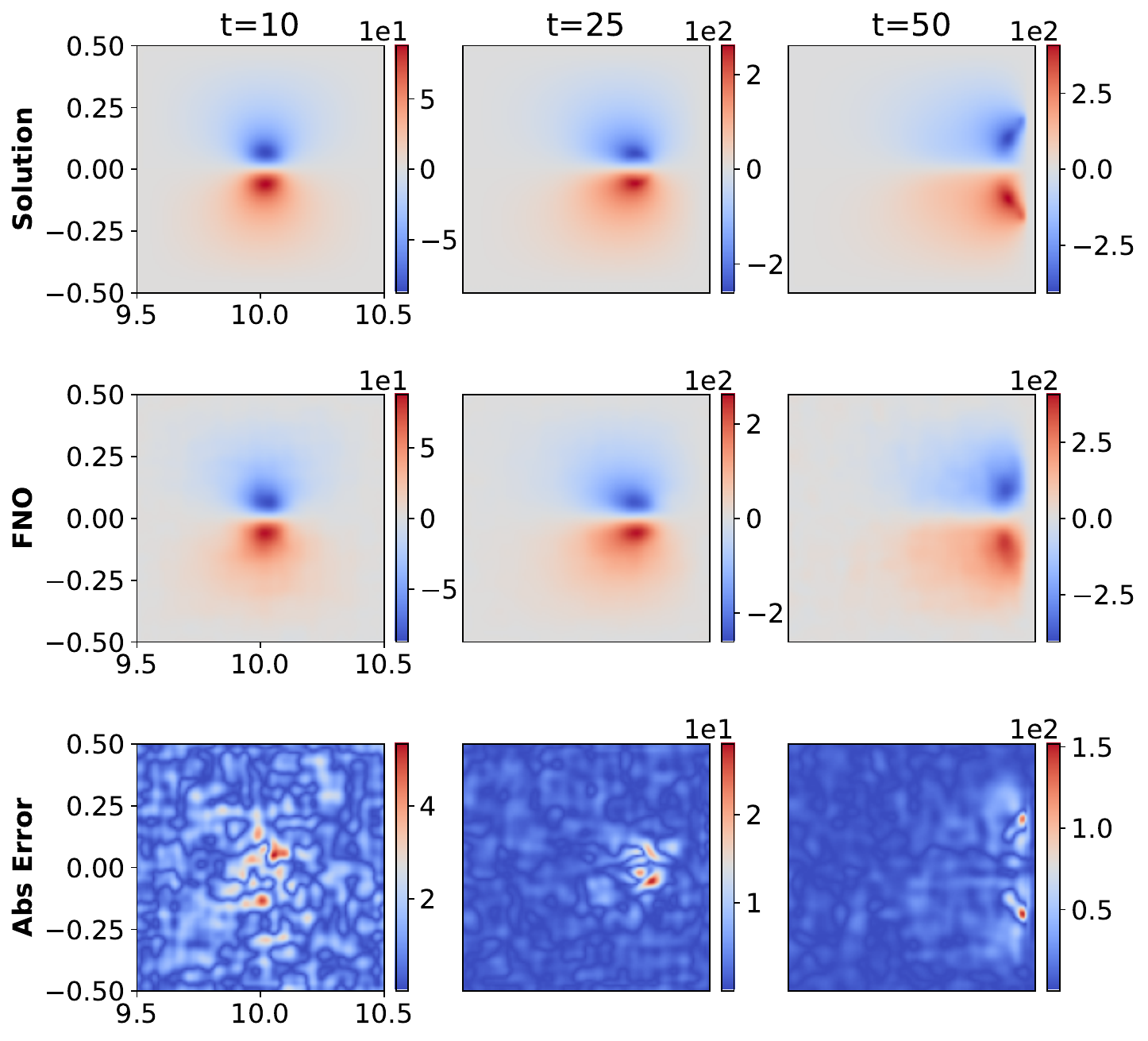}}
    \label{fig: single_phi}
    \caption{Single Blob : Temporal evolution of (a) density and (b) electric potential and (c) temperature of the plasma evolution as obtained using the JOREK code (top of each image), that of the trained multi-variable FNO (middle of each figure) and the absolute error across both (bottom of each figure). The spatial domain is given in toroidal geometry characterised by $R$ in the $x$-axis and $Z$ in the $y$-axis.}
    \label{fig: single_mhd}
\end{figure}

\subsubsection{Multi-Blobs with Non-uniform Temperature}~\\
\label{multi_blobs}

\begin{figure}[h!]
    \centering
    \subfloat[Density]{\includegraphics[width=7.5cm]{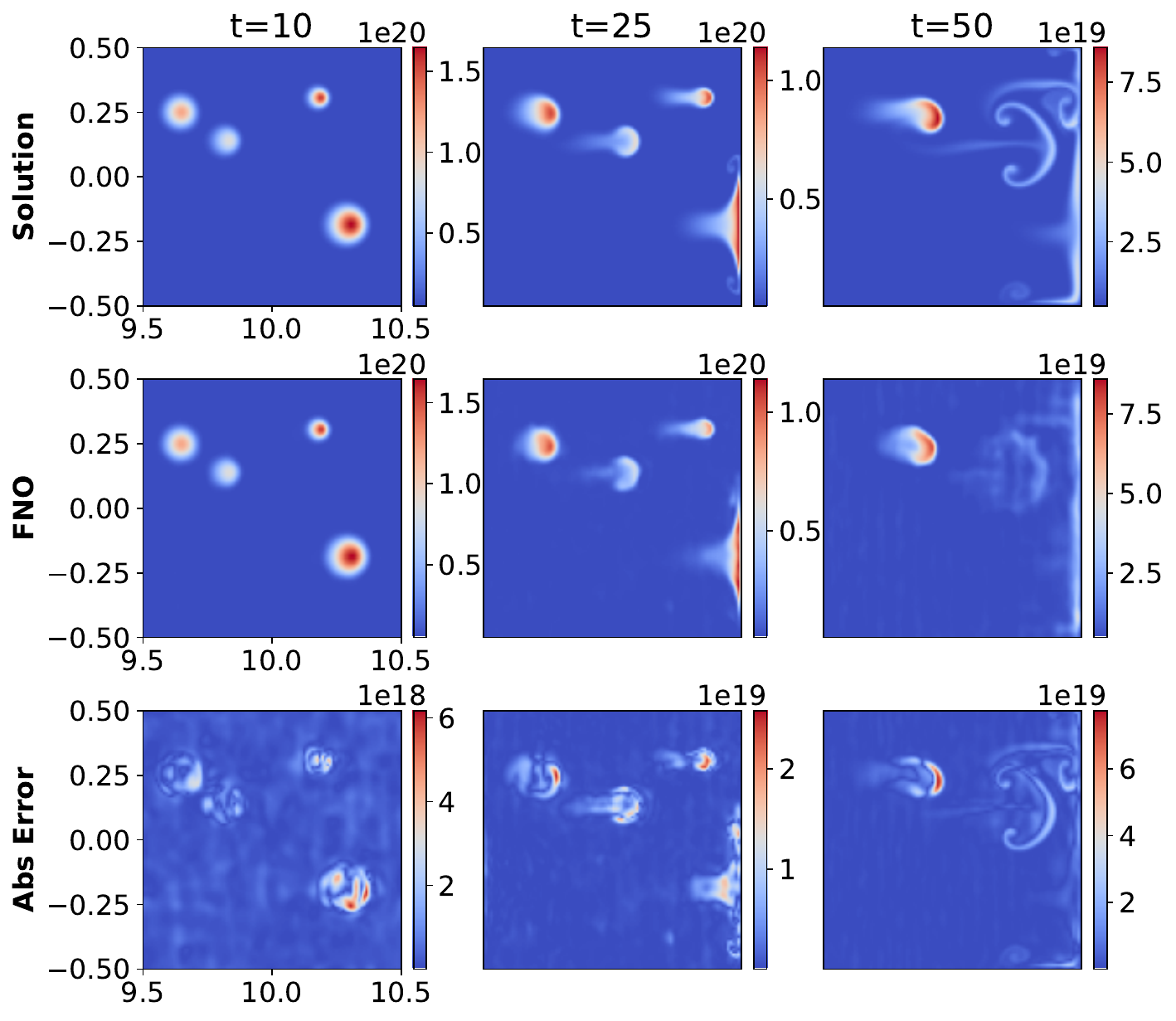}}
    \label{fig: multi_rho}
    \centering
    \subfloat[Potential]{\includegraphics[width=7.5cm]{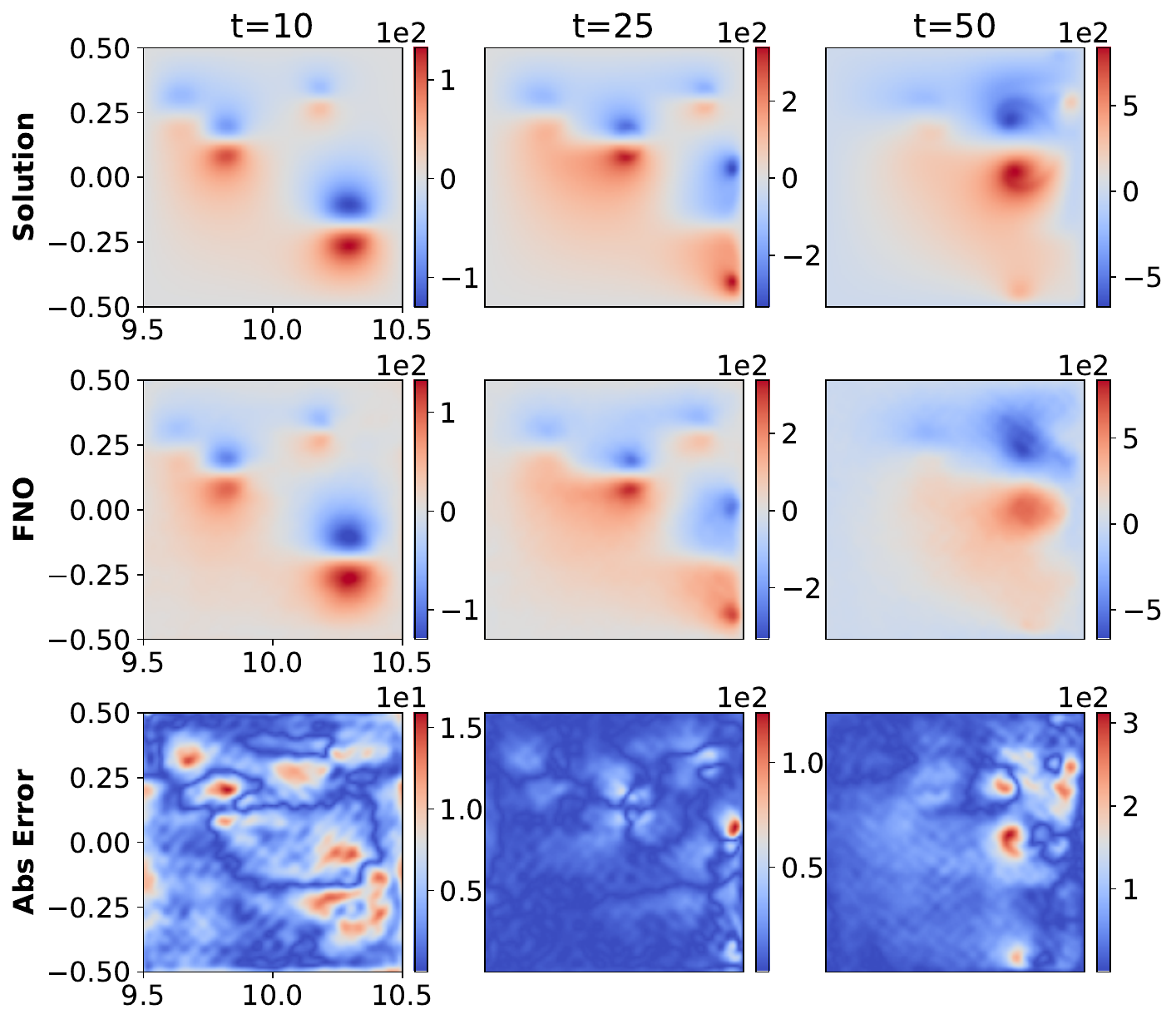}}
    \label{fig: multi_phi}
    \centering
    \subfloat[Temperature]{\includegraphics[width=7.5cm]{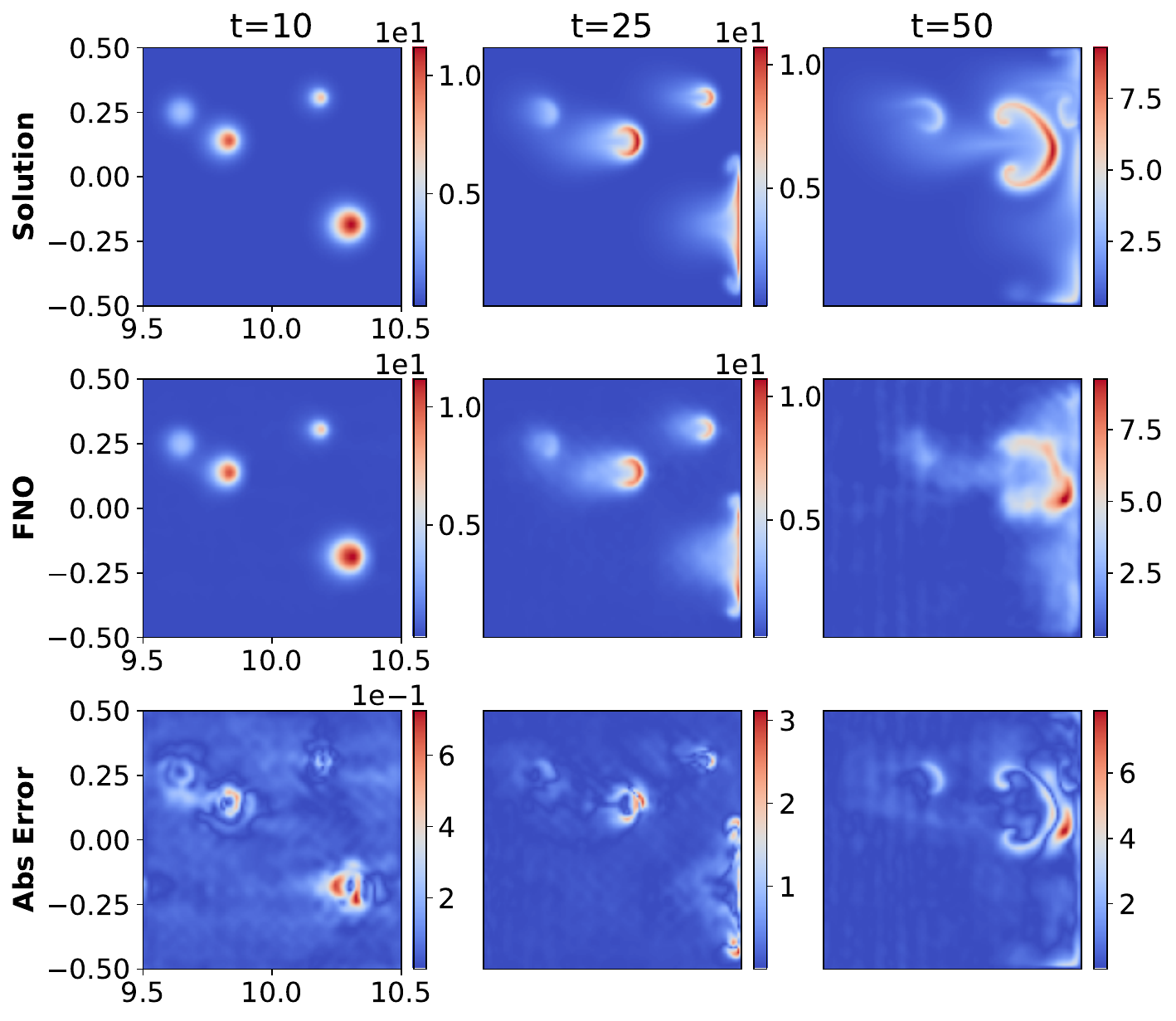}}
    \label{fig: multi_t}
    \caption{Multiple Blobs with non-uniform temperature: Temporal evolution of (a) density (b) electric potential and (c) temperature of the plasma evolution as obtained using the JOREK code (top of each image), that of the trained multi-variable FNO (middle of each figure) and the absolute error across both (bottom of each figure). The spatial domain is given in toroidal geometry characterised by $R$ in the $x$-axis and $Z$ in the $y$-axis.}
    \label{fig: multi_mhd}
\end{figure}

\cref{fig: multi_mhd} shows the capability of our newly designed multi-variable FNO in emulating the evolution of density, electric potential, and temperature for the case with multiple density and temperature blobs. This presents a more challenging case to model as the distribution of the training data is wide due to it being generated from a much larger parameter space as described in \cref{table: data_generation}.  With a training dataset of 1500 simulations, the model trains in 11 hours and 22 minutes on a single Nvidia A100 chip and achieves MSE(in the normalised domain) $\sim$ $0.0001$, capturing the larger, more global, features associated with the evolution of plasma characteristics. However, due to the truncated nature of the implementation of the linear operator over the Fourier Transform, the modelling gets limited to the more dominant modes, smoothing out the less dominant, minor, finer features. The multi-variable FNO is capable of (surrogate) modelling the evolution of the plasma up to the $50th$ time step in 0.03 seconds. For further plots showcasing the inference capabilities of the multi-variable FNO modelling MHD at different initial conditions as well as the discrepancy to the actual solution, refer to \ref{appendix_mb_errorplots}.

\begin{figure}
    \centering
    \includegraphics[width=\textwidth]{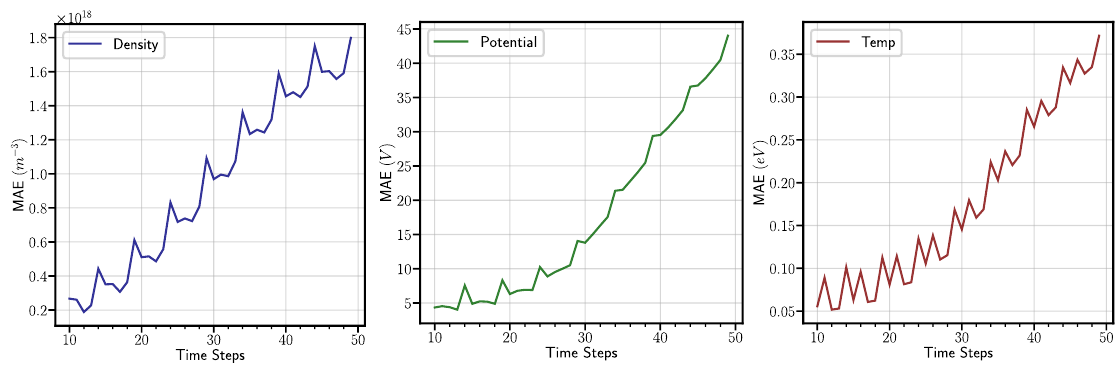}
    \caption{Error growth across time roll-out of the FNO in the physical space. The error in emulating each variable of the reduced MHD model increases as we roll further out in time.}
    \label{fig: mb_error_growth}
\end{figure}

The FNO deployed in the autoregressive manner is well equipped to model the evolution of the plasma for short time roll-outs, but as we extend the prediction further in the temporal domain, it moves further away from the ground truth. This arises mainly due to the training regime which we deploy the FNO. Being deployed in an autoregressive structure, the prediction errors accumulate over longer time roll-outs. This is illustrated in \cref{fig: mb_error_growth}. Though we demonstrate the results obtained with 1500 simulation data points, for the rest of the ablation studies (except for \cref{data_impact} and \ref{zero shot super resolution}) demonstrated in this work, we restrict ourselves to the initial 250 as it allows us to compares against the prior 2 cases where we operate in a relatively data-scarce setting. 



\subsubsection{Individual FNO vs Multi-variable FNO}~\\
\label{ind_vs_multi}
Magnetohydrodynamics governing the evolution of a plasma state is characterised by the amalgamation of Navier-Stokes equation of fluid dynamics and Maxwell's equations of electromagnetism \cite{bellan2006fundamentals}. The evolution of MHD fields involves modelling multi-physics environments with coupled variables that are integrated together. Considering this interplay of different field variables, it was imperative that we devise the multi-variable FNO, where all the interested and dependent variables can be modelled together. 

\begin{figure}[h!]
    \centering
        \includegraphics[width=\textwidth]{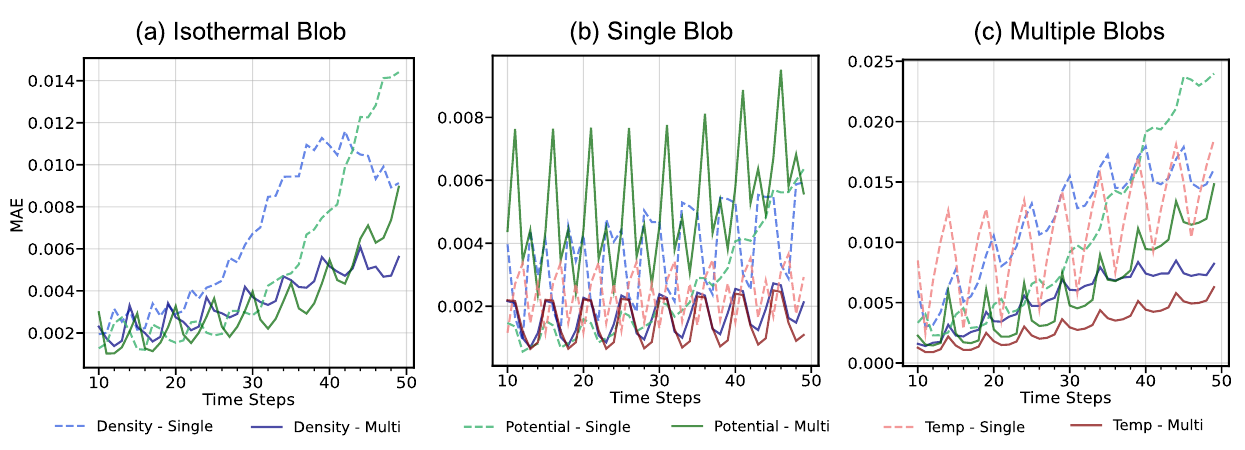}
    \caption{Comparing the error growth (mean absolute error in the normalised domain) of different variables across time roll-out of the Individual FNO (Single) and the Multi-variable FNO (Multi) for the various MHD cases under study. (a) Single Isothermal Blob (b) Single Blob with non-uniform temperature (c) Multi-blobs with non-uniform temperature. The multi-variable FNO with the same parameters as the individual FNOs demonstrates much better performance over time. The $y$-axis represents the mean absolute error for each case, and the $x$-axis represents the time iteration of the autoregressive rollout.}
    \label{fig: error_growth}
\end{figure}

Our multi-variable FNO, as described in \cref{mv_fno}, achieves the required multi-physics modelling, with the same computational costs as training individual FNOs for each variable. The multi-variable FNO has approximately the same number of parameters as the different individual FNOs combined, and thus takes the same time as training two or three different FNO models. As shown in \cref{fig: error_growth}, the multi-variable FNO achieves better performance with much less error than the individual FNOs. \cref{fig: error_growth} demonstrates that by integrating the different field variables within the model, we can reduce the error associated with the autoregressive time rollout, dampening the error growth. It must be noted that since each variable has a different normalisation applied to it (the same scheme, but different normalisation), the comparison within the normalised domain is consistent within the same variable. 

Thus, we conclude that for highly correlated multi-physics systems, such as those described by the MHD equations, multi-variable FNO, which captures the interdependence of the variables within a single model, yields a much more physically consistent result than the variables modelled individually by separate FNOs with no methods of exploiting the relationship across them. \\

\subsubsection{Impact of Step Size}~\\
\label{step_size}

The FNO, devised to be utilised in an autoregressive configuration, is only exposed to the first \textit{T\_in} time steps as inputs to the model. The rest of the inputs for emulating the time evolution are obtained by feeding back the output into the model in a recursive manner. The model follows the layout of a Markov decision process \cite{mdp} as it relies on the previously generated time steps to provide all the information pertaining to the state of the system, without having a memory component attached to it. As the model trains, the autoregressive output generated from the input is compared against that of the simulation data to learn the mapping in a supervising manner. When deployed at inference, the FNO relies purely on the initial time steps to perform the time roll-outs of the plasma evolution. 

\begin{figure}[h!]  
    \centering
    \subfloat[Isothermal Blob]{\includegraphics[width=5.0cm]{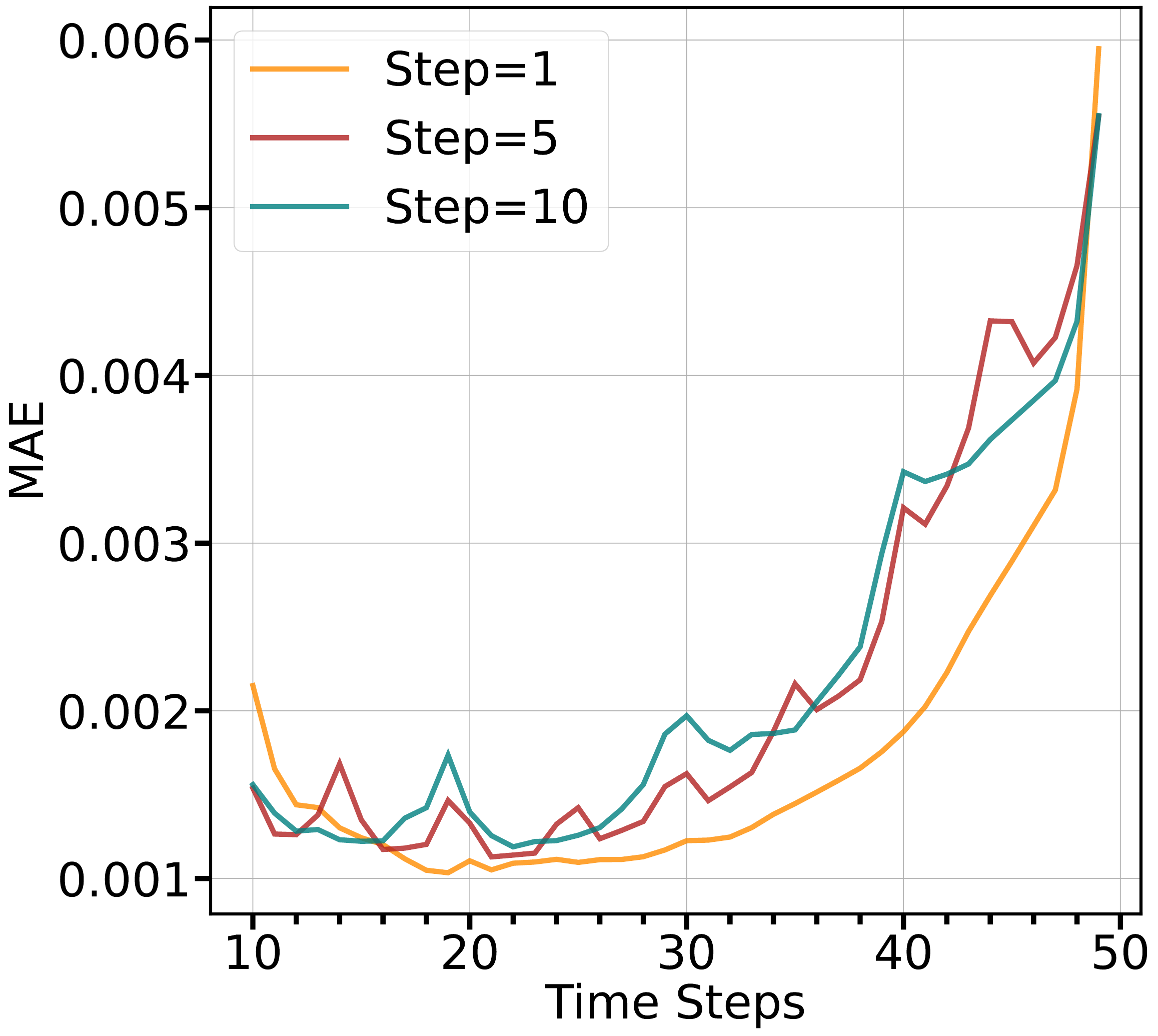}
    \label{fig: step_isothmeral}}
    \centering
    \subfloat[Single Blob]{\includegraphics[width=5.15cm]{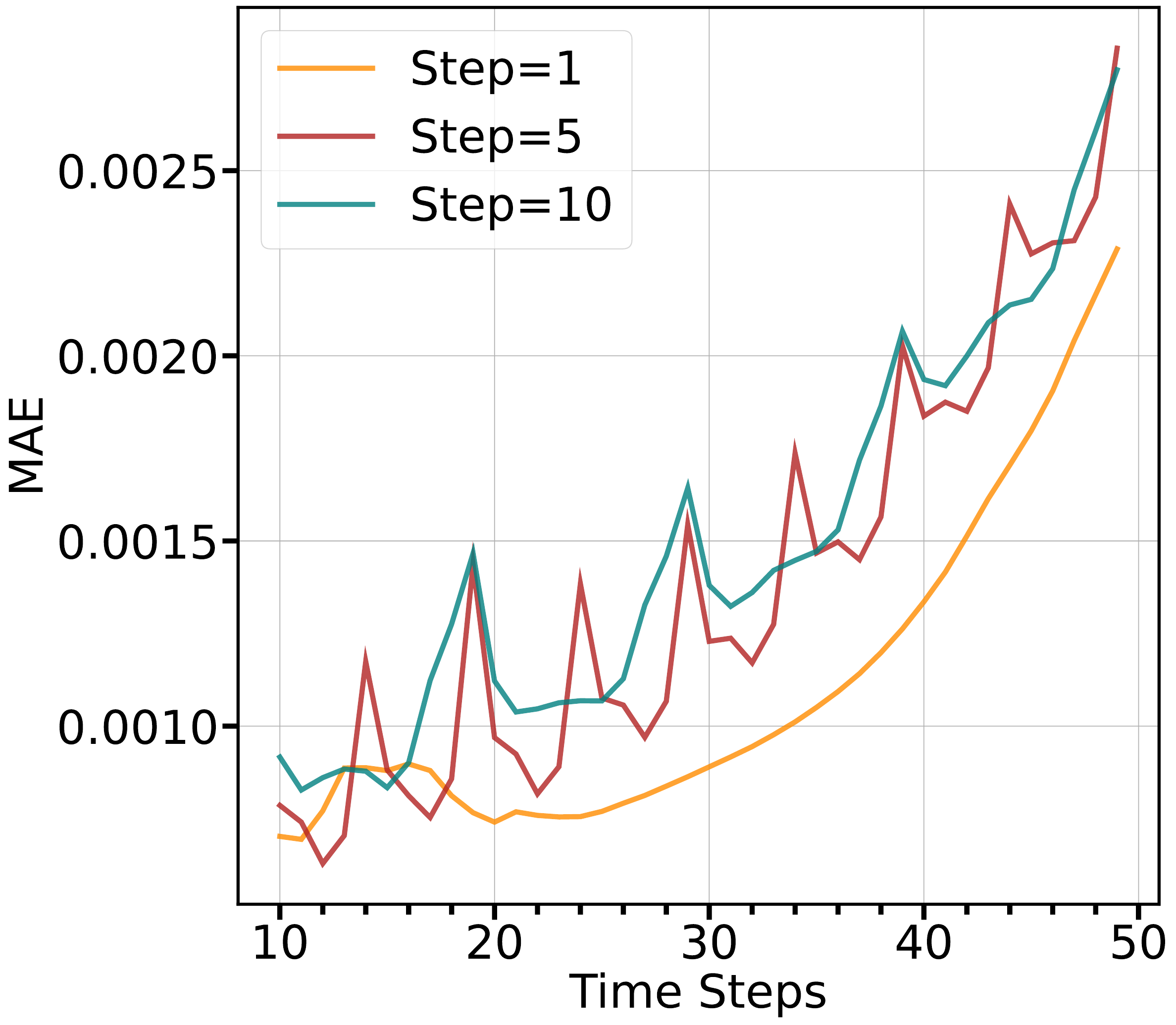}
    \label{fig: step_single}}
    \centering
    \subfloat[Multi-blobs]{\includegraphics[width=5.0cm]{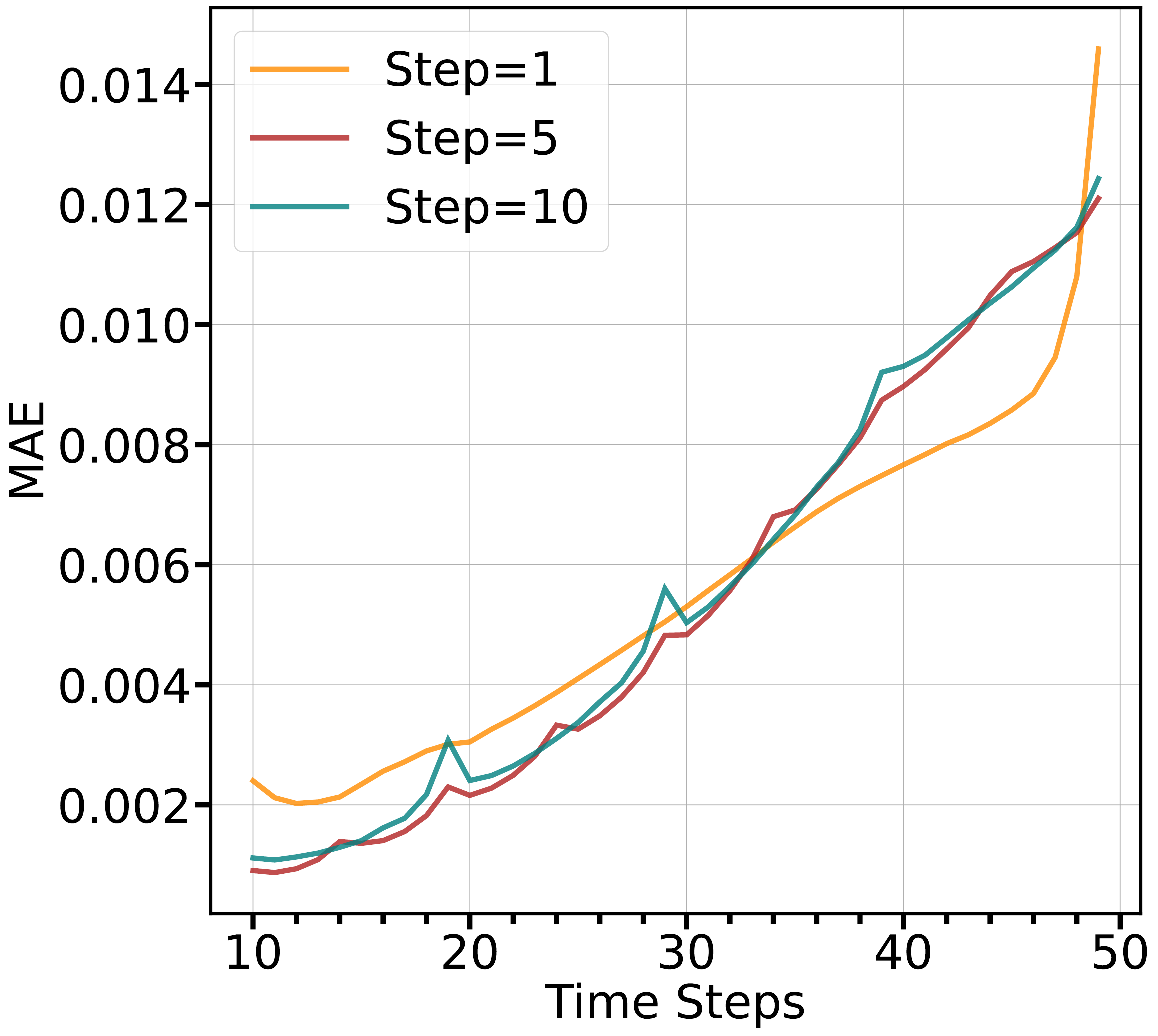}
    \label{fig: step_multi}}
    \caption{Comparing the cumulative error growth of different variables across time roll-out for multi-variable FNOs with different step sizes. The autoregressive nature of the FNO brings about a spike in error, characterising a disconnection at the time step that is a multiple of the output size. (a) Isothermal blob (b) Single blob with non-uniform temperature (c) Multi-blobs with non-uniform temperature. The $y$-axis represents the mean absolute error for each case, and the $x$-axis represents the time iteration of the autoregressive rollout.}
    \label{fig: step_growth}
\end{figure}

Thus, the choice of the output size (step size), i.e, how much further in time the FNO predicts the time evolution with each forward pass, essentially provides disconnects across time for the model, which is evident by looking at \cref{fig: step_growth}. At each time step that is a multiple of the step size, we notice a spike in the cumulative error of the model. This spike denotes a disconnection that occurs as the model performs the autoregressive time evolution. In \cref{fig: step_growth}, the orange line indicates the error growth for a step size of 1, where there is no spike characterising the disconnection. For the red line with a step size of 5, we notice that there is a spike at every $5th$ time step starting form 10 and similarly for the green line indicative of models with a step size of 10, the spike is observed at every $10th$ time step. Throughout the scope of this study for surrogate modelling plasma simulations, we restrict ourselves to step size = 5 as we find it offers the best training efficiency, optimum performance with less computational time. 

Regardless of this error spike, the FNO performs well in (surrogate) modelling the various MHD cases under study. We notice that the multi-variable FNO with a step size of 1 performs the best with a smooth and dampened error growth. This is expected as the network parameters are being used to perform mapping for one time step rather than several steps, effectively providing more parameters per step. We notice that the error decreases initially before it increases and believe it is because of the autoregressive structure of the training regime. In forward propagation, the model uses the time step data first from the data and then from the model itself. This results in a distribution shift problem because the inputs are no longer solely from ground truth data, the distribution learned during training will always be an approximation of the true data distribution \cite{lee2023autoregressiverenaissancein}. The pushforward trick demonstrated in \cite{brandstetter2023message}, offers an alternative training regime that allows us to account for this behaviour.

\subsubsection{Mode Ablations}~\\
\label{mode_ablations}

\begin{figure}[h!]
    \centering
    \includegraphics[width=6.0cm]{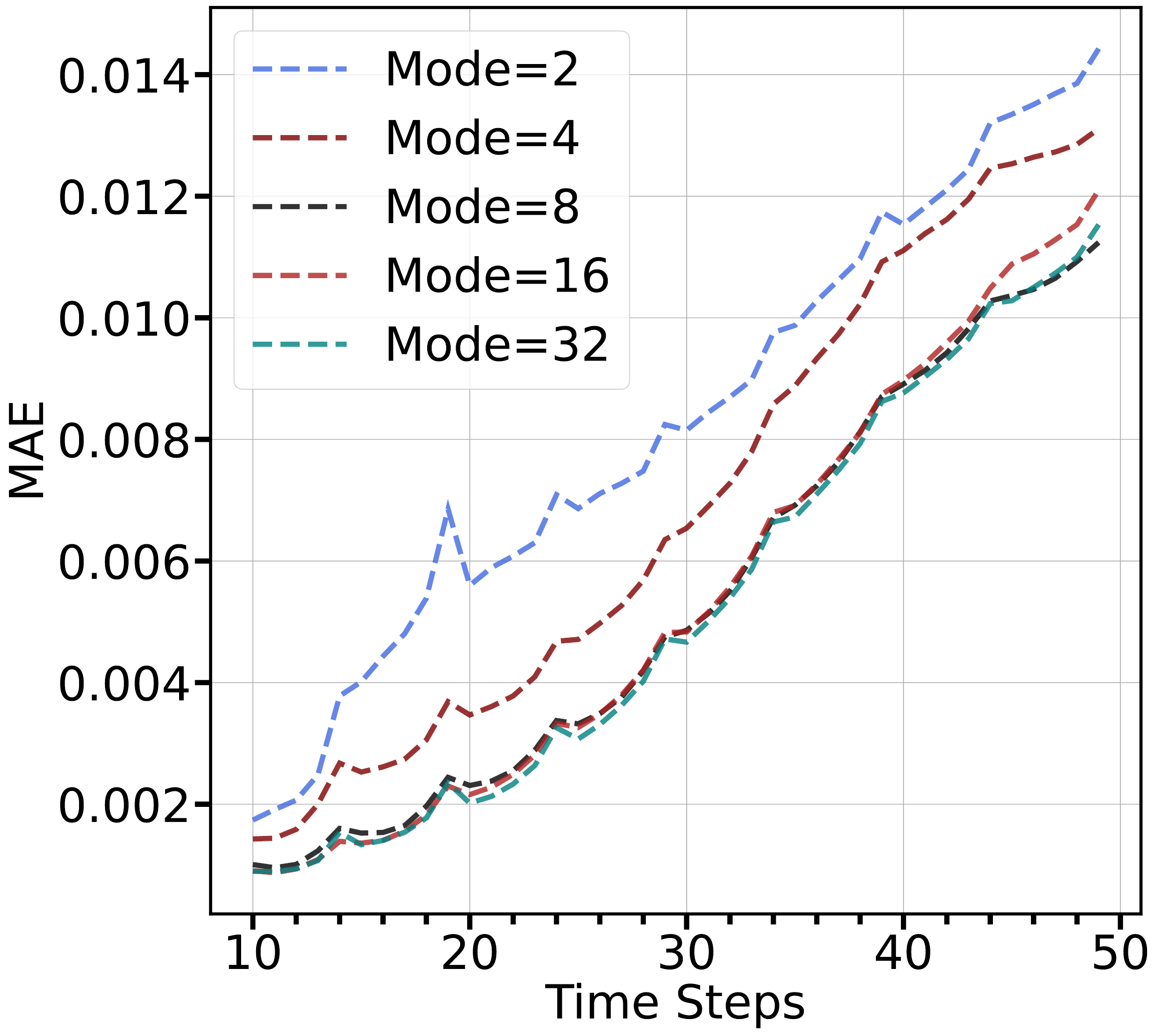}
    \caption{Ablation study of multi-variable FNOs with different truncated mode numbers. Error growth of FNOs with different mode numbers as they are autoregressively evolved in time for the Multiblob case. Errors decrease with a higher number of modes, and performance nearly plateaus at higher modes, indicating the existence of dominant features for the problem up to a certain number of modes. For each model, the width of the temporal channel was fixed at 32, and that of the variable channel was fixed at 3.}
    \label{fig: mode-ablation}
\end{figure}

The Fourier integral operator within the Fourier layer of the FNO truncates the Fourier series up to a maximal number of modes $k_{max}$ \cite{li2021fourier}. This presents an additional hyperparameter to each Fourier layer of the FNO that determines the extent to which Fourier modes are required to be learned to map the spatio-temporal evolution. Within our experiments, we ensure that for each multi-variable FNO, the modes at each Fourier layer are fixed. \cref{fig: mode-ablation} shows the impact the number of modes has on the FNO in emulating the MHD case with multiple density and temperature blobs (Case 3). For a smaller number of modes, the FNO demonstrates a higher error value at each time step, whereas for higher mode numbers, we notice that the error is relatively lower. We also notice that the performance of the FNO plateaus at 8 modes and does not increase further as the mode numbers are increased. This arises due to the fact that the dominant features that are characteristic of the system we are learning saturates at the $8^{th}$ mode, and no more dominant features are found at higher modes for the multi-blob case. This is consistent with the mode selection that one has to perform for each problem that a neural operator is applied to. The effect of this is studied in detail \cite{zhao2023incremental}. 

\begin{figure}[h!]
    \centering
        \subfloat[Solution and Modes = 2]{\includegraphics[width=7.5cm]{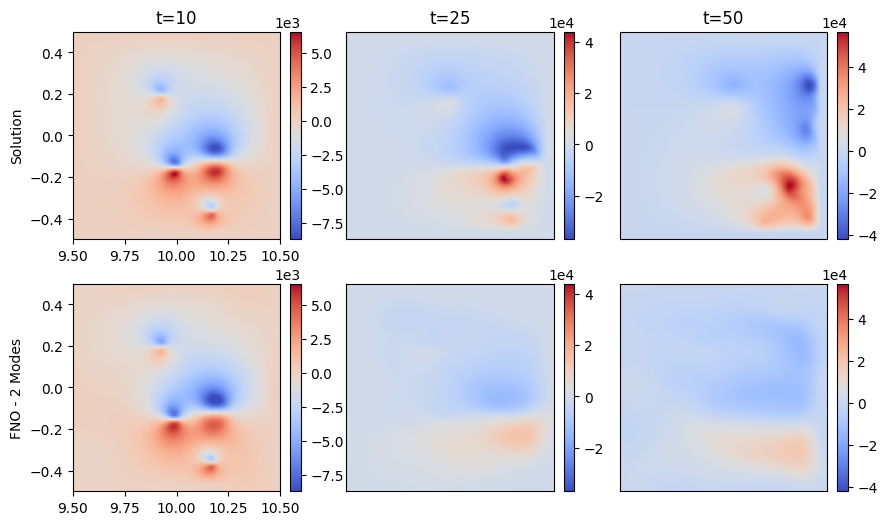}}
    \label{fig: multi_modes_1}
    \centering
    \subfloat[Modes = 8 and Modes = 32]{\includegraphics[width=7.5cm]{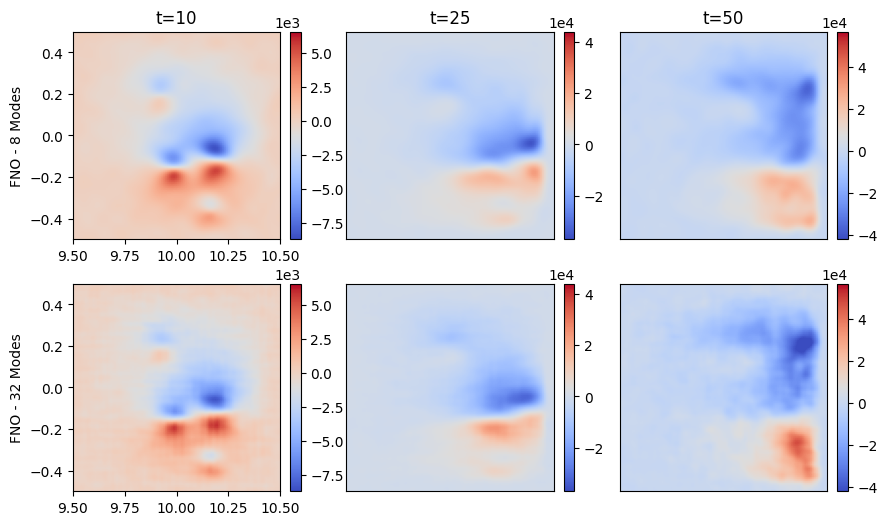}} 
    \label{fig: multi_modes_2}
    \caption{Evolution of the electric potential (Multiblobs) in time as modelled by multi-variable FNOs at different modes. Top row of (a) shows the actual solution modelled using JOREK. The bottom row of (a),  plots the time evolution of the FNO with 2 modes, top row of (b) is that of a model with 8 modes, and the bottom row of (b) is the FNO with 32 modes. FNO with lower modes produce less noise, but fail to predict the longer evolution structures, while those with higher modes are inherently noisier and capture the long-term predictions better.}
    \label{fig: modes_phi}
\end{figure}

Within the mode ablation study, we notice that at lower modes, though the error of the model is higher, it produces smoother field profiles than those at higher modes. In \cref{fig: modes_phi}, we see that the FNO with 2 modes, generates fields that are smoother and less noisy, especially at the initial time steps. However, they soon lose the general structure as they progress further in time. In the FNO with higher modes ($k_{max}$ = 8, 32), the FNO output is much noisier than those at lower modes, but preserves the general structure of the temporal evolution better. 

\subsubsection{Impact of Training Dataset Size}~\\
\label{data_impact}
\begin{figure}[h!]
    \centering
    \includegraphics[width=7.5cm]{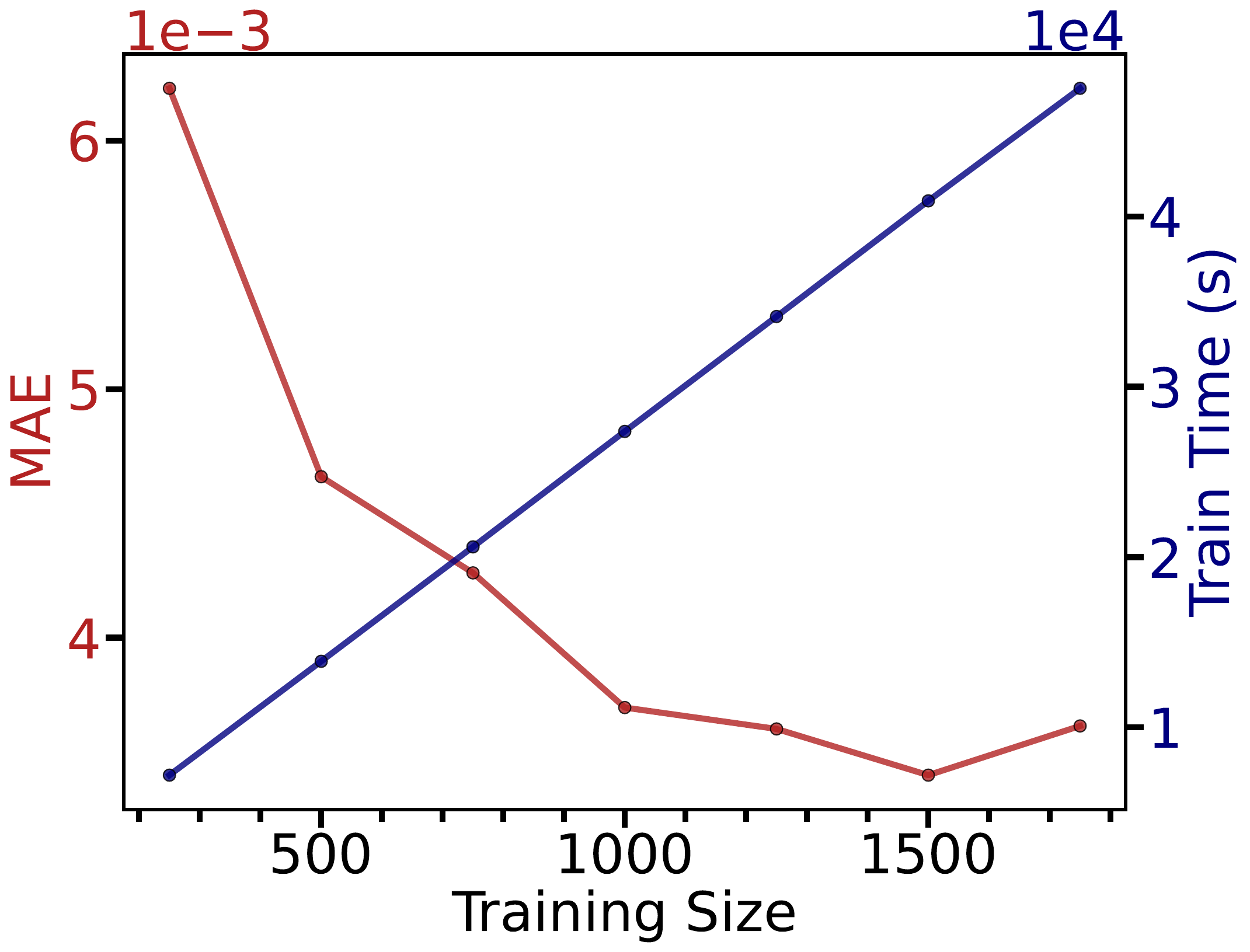}
    \caption{Impact of the size of the training dataset on the performance of the multi-variable FNO for simulating the multiple plasma blobs in a non-uniform temperature field. As we increase the availability of the training data for the multi-variable FNO, the model performance increases across the same test data. However, the gains obtained by increasing the training data are marginal compared to the additional compute required. The number of simulations used for training is given on the $x$-axis while the mean squared error is given on the $y$-axis. All training is performed on a single Nvidia A100.}
    \label{fig: data_impact}
\end{figure}


\cref{fig: data_impact} visualises the impact the size of the training dataset has on the performance of the multi-variable FNO. For this purpose, we generate a larger dataset using the same scheme as given in \cref{table: data_generation}. We generate a total of 1835 JOREK simulation data points, out of which we set aside 85 as the test dataset. We train multiple multi-variable FNOs with training sizes ranging from 250 to 1750 simulations and compare their performance on the test dataset. All the models are constructed and trained under identical conditions, with the exception of the training data. We find that the performance of the FNO increases as we increase the training size, with model error decreasing by nearly half. However, we notice the gains made by added simulation data do not scale well at all. As shown in \cref{fig: data_impact}, improvement in performance is only marginal in comparison to the added computational costs associated with the training as well as the simulation run times.

\begin{figure}[h!]
    \centering
    \includegraphics[width=7.5cm]{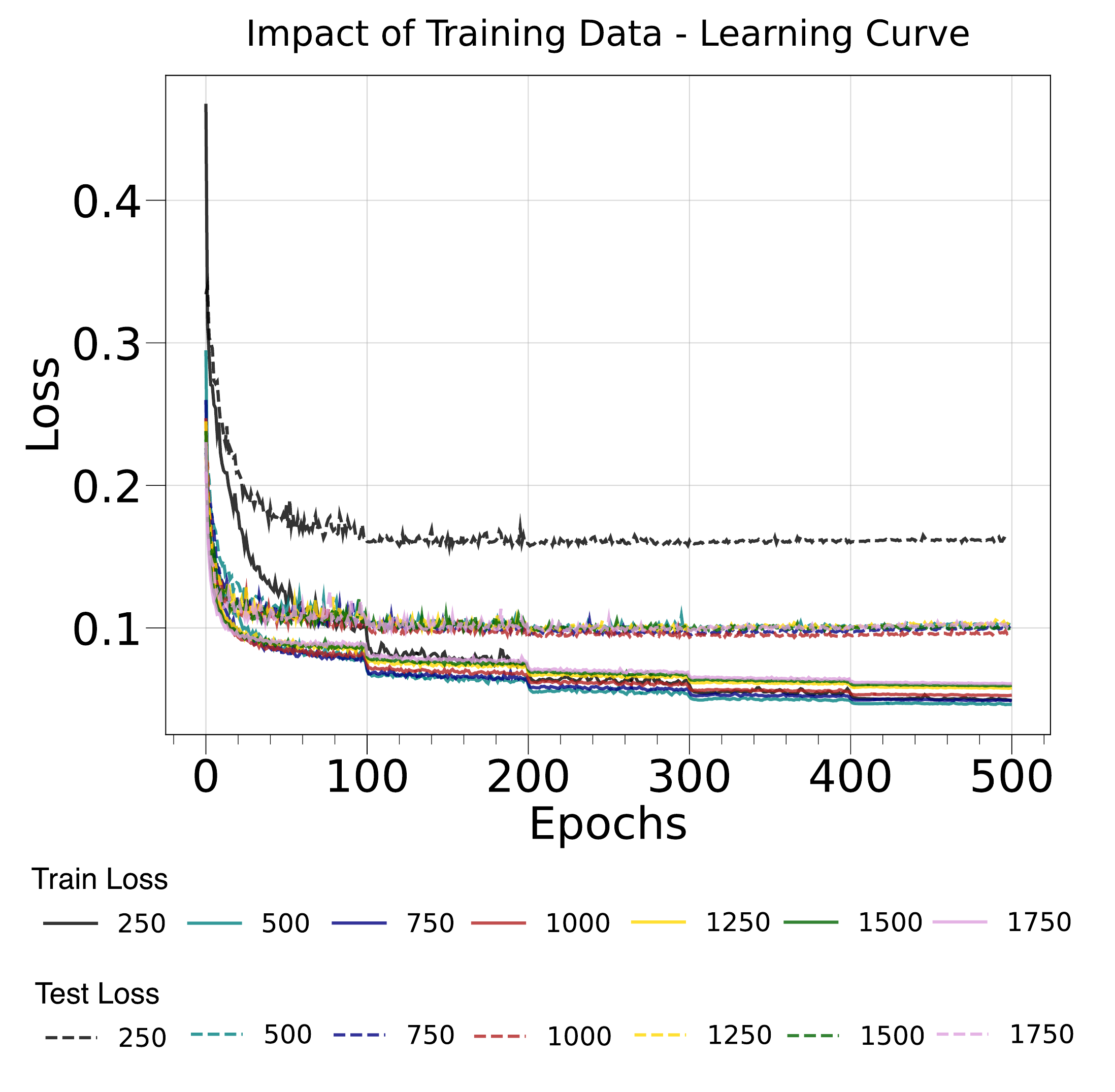}
    \caption{Learning curves associated with the models trained with data of different sizes. Training loss is represented by full lines, while the test loss is represented by dashed lines. With just 500 simulation instances, the model learns the physics of the model really well, which is only marginally improved by more training data. We also notice the quick over-fitting tendency of the FNO as observed in \cite{lanthaler2023nonlocal}}
    \label{fig: data_impact_learning_curve}
\end{figure}

\subsubsection{Zero-Shot Super-Resolution} ~\\
\label{zero shot super resolution}

As described in \cref{fno}, the NO framework allows learning within the function space as opposed to the Euclidean space as with traditional neural networks \cite{kovachki2021neural}. By learning the mapping between infinite-dimensional function spaces, the framework performs operator learning, allowing for characterising discretisation within our model. Within the FNO, discretisation invariance is built in by learning within the Fourier space rather than the Euclidean space on which the solutions exist, as demonstrated in \cref{fno_eqn}. Being discretisation-convergent, the FNO learns information up to a certain refinement of the mesh resolution, higher than what we had provided in the initial training dataset \cite{bartolucci2023neural}. The truncated nature of the FNO characterises the model as a band-limited function. Coupled with the discrete modal representation in the Fourier space, the FNO converges to a fixed performance at higher resolution \cite{bartolucci2023neural}.

\begin{figure}[h!]  
    \centering
    \subfloat[Density]{\includegraphics[width=7.5cm]{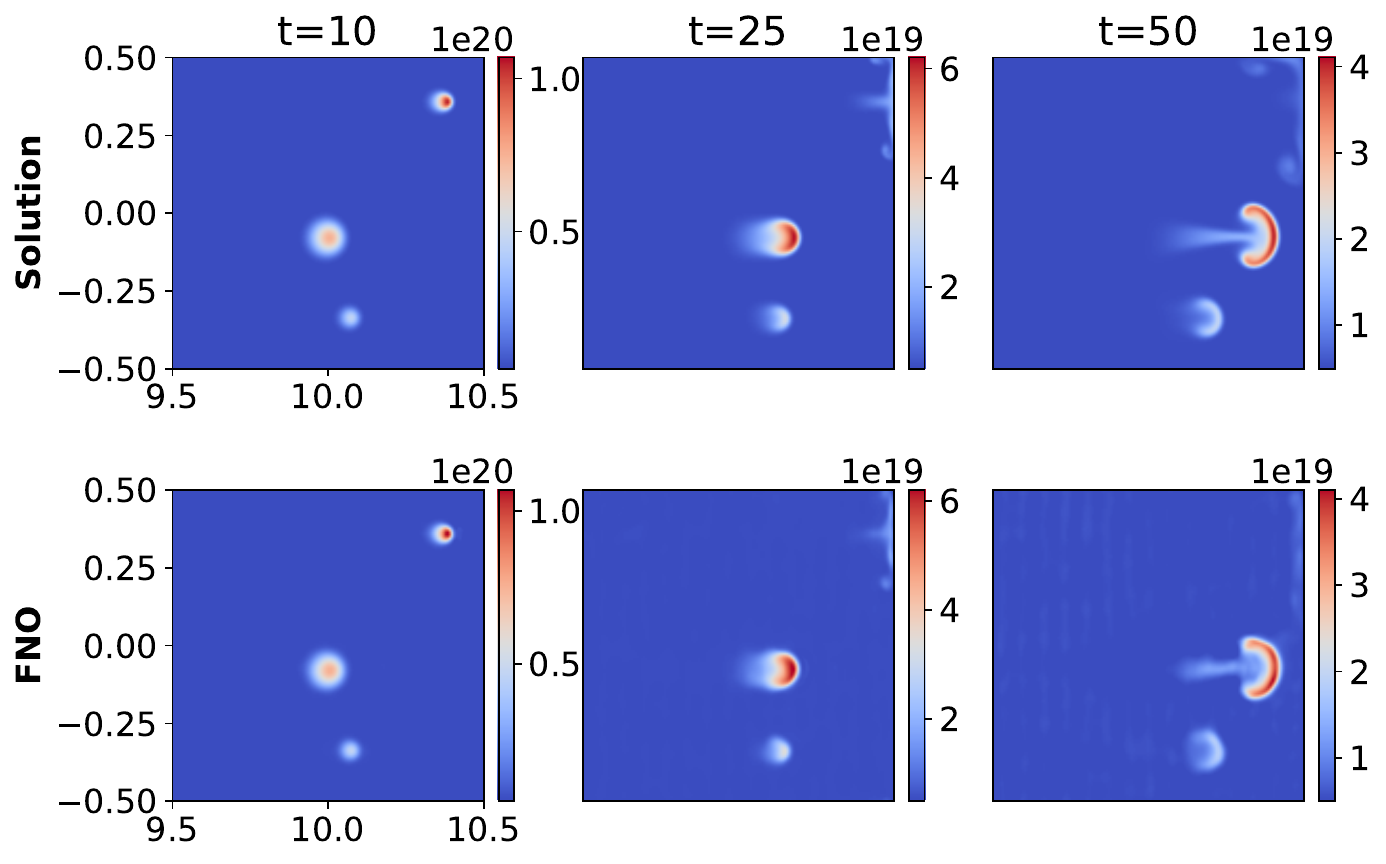}}
    \label{fig: density_zssr}
    \centering
    \subfloat[Potential]{\includegraphics[width=7.5cm]{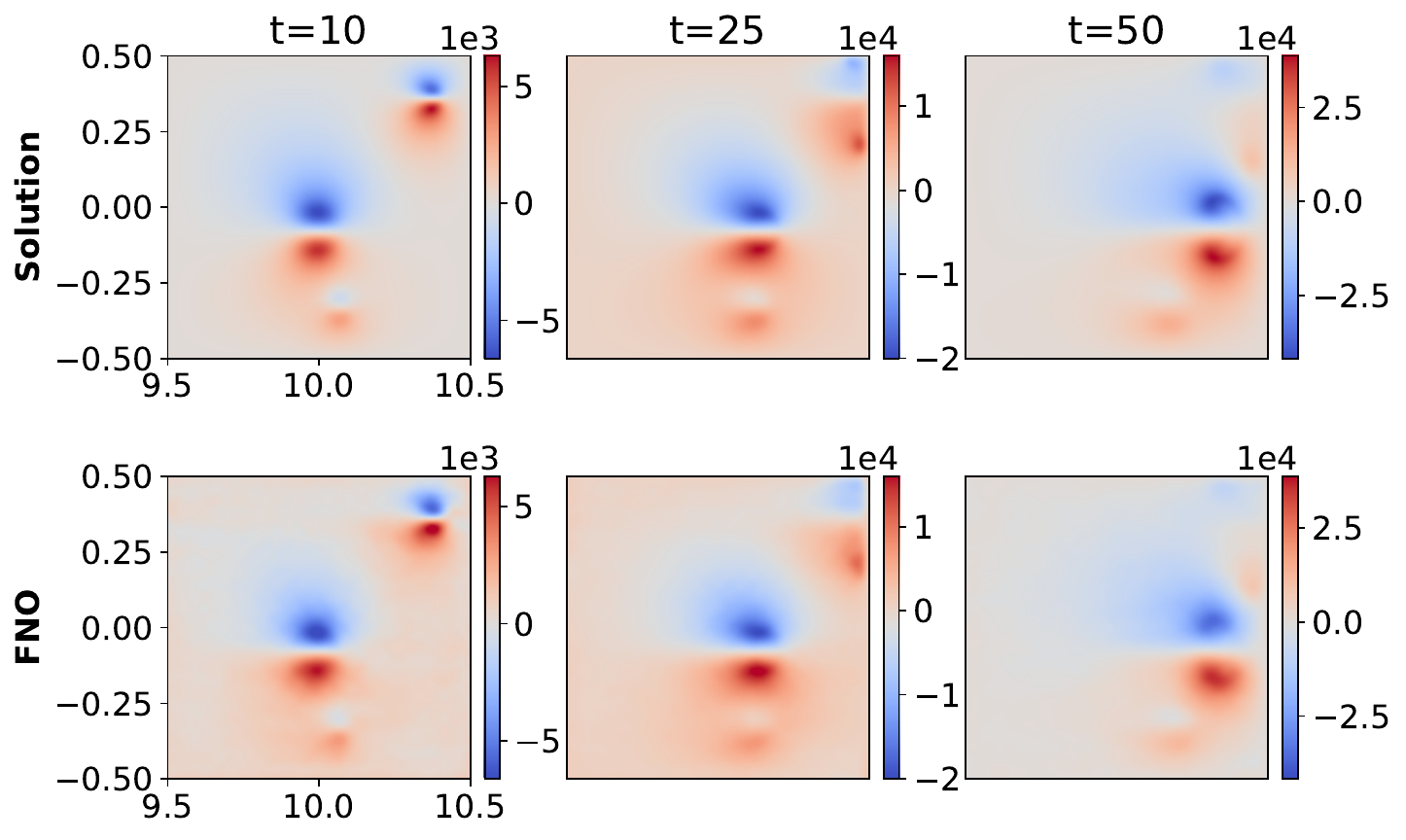}}
    \label{fig: potential_zssr}
    \centering
    \subfloat[Temperature]{\includegraphics[width=7.5cm]{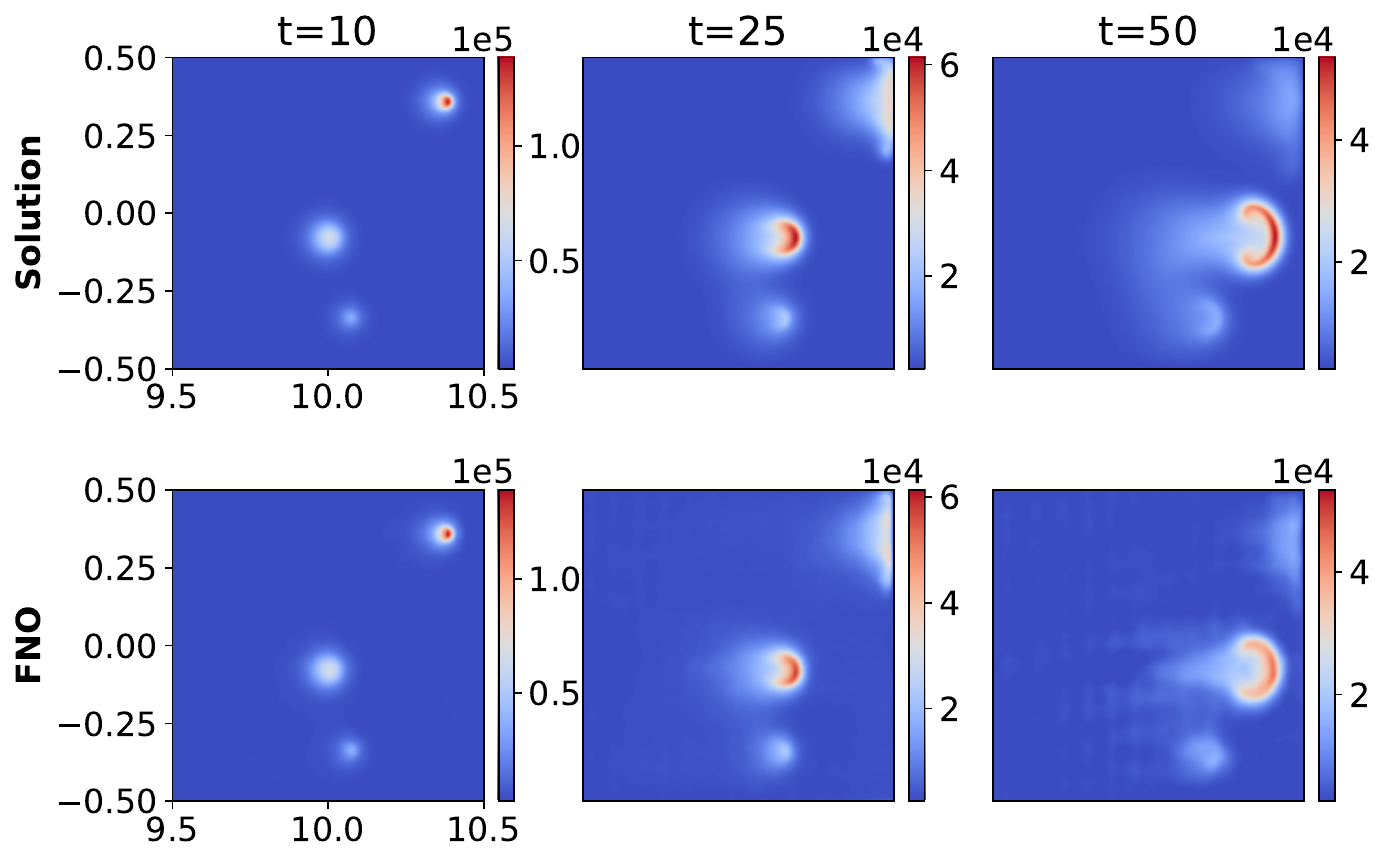}}
    \label{fig: temp_zssr}
    \caption{Demonstrating Zero-Shot Super-Resolution on the multi-variable FNO. The FNO being trained on a grid resolution of 100x100 is further deployed to model the evolution the the multiple blobs at a resolution of 500x500. The model is capable of performing zero-shot super-resolution, achieving an MSE(in the normalised domain) of $7.90\times 10^{-5}$ at a grid resolution $5 \times$ more resolved than the one it was being trained on without the requirement of any model engineering or fine-tuning. The plots demonstrate the evolution of density, electric potential, and temperature as emulated by the FNO. For plots (a), (b), and (c), the top series of plots demonstrate the time evolution of the ground truth while the bottom series of plots demonstrate the time evolution as emulated by the multi-variable FNO.}
    \label{fig: zero-shot-super-resultion}
\end{figure}

This discretisation-convergence allows us to perform zero-shot super-resolution, trained on a lower resolution, and directly evaluated on a higher resolution \cite{li2021fourier}. In \cref{fig: zero-shot-super-resultion}, we demonstrate this capability for zero-shot super-resolution, where we take the model trained on a spatial discretisation of $100\times 100$ used for the prior experiments and then deploy it to emulate the evolution of multiple blobs at a spatial discretisation of $500\times 500$. Even for grids at 5 times larger resolution, the FNO is capable of emulating the plasma with considerable accuracy, deviating from the true solution by an MSE (in the normalised domain)of only  $7.90\times 10^{-5}$ in the normalised domain. They share the same model parameters among different discretisations of the underlying function space that we are attempting to learn. Plots showing the contrast of the super-resolved FNO output with that of JOREK by way of discrepancy plots can be found in \cref{fig: zero-shot-super-resultion_error} in \ref{appendix_errorplots}.

\subsubsection{Extrapolating outside Training Domain} ~\\
\label{out_domain}
Neural Networks trained for modelling a particular task have been found to suffer from a general lack of extrapolation capabilities \cite{Courtois2023}. The FNO relying on a continuous gradient-descent based optimisation of the deep learning parameters exhibit the same behaviour. The models perform well within the domain of the training data, as can be explained by the associated Neural Tangent Kernels (NTK) \cite{jacot2020neural} and fail to perform well outside the training regime. 

\begin{figure}[h!]
    \centering
    \subfloat[Density]{\includegraphics[width=7.5cm]{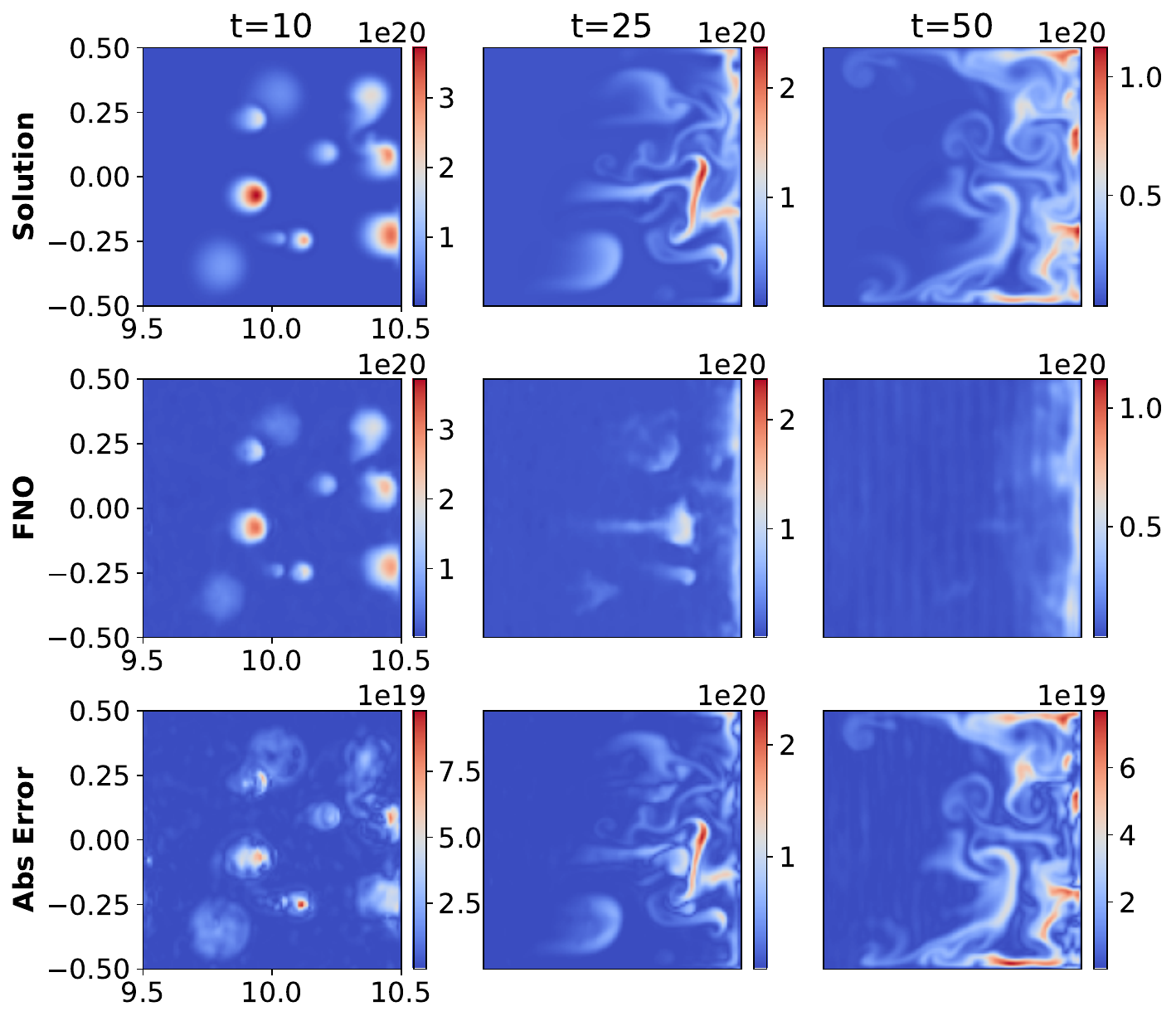}}
    \label{fig: multi_rho_out}
    \centering
    \subfloat[Potential]{\includegraphics[width=7.5cm]{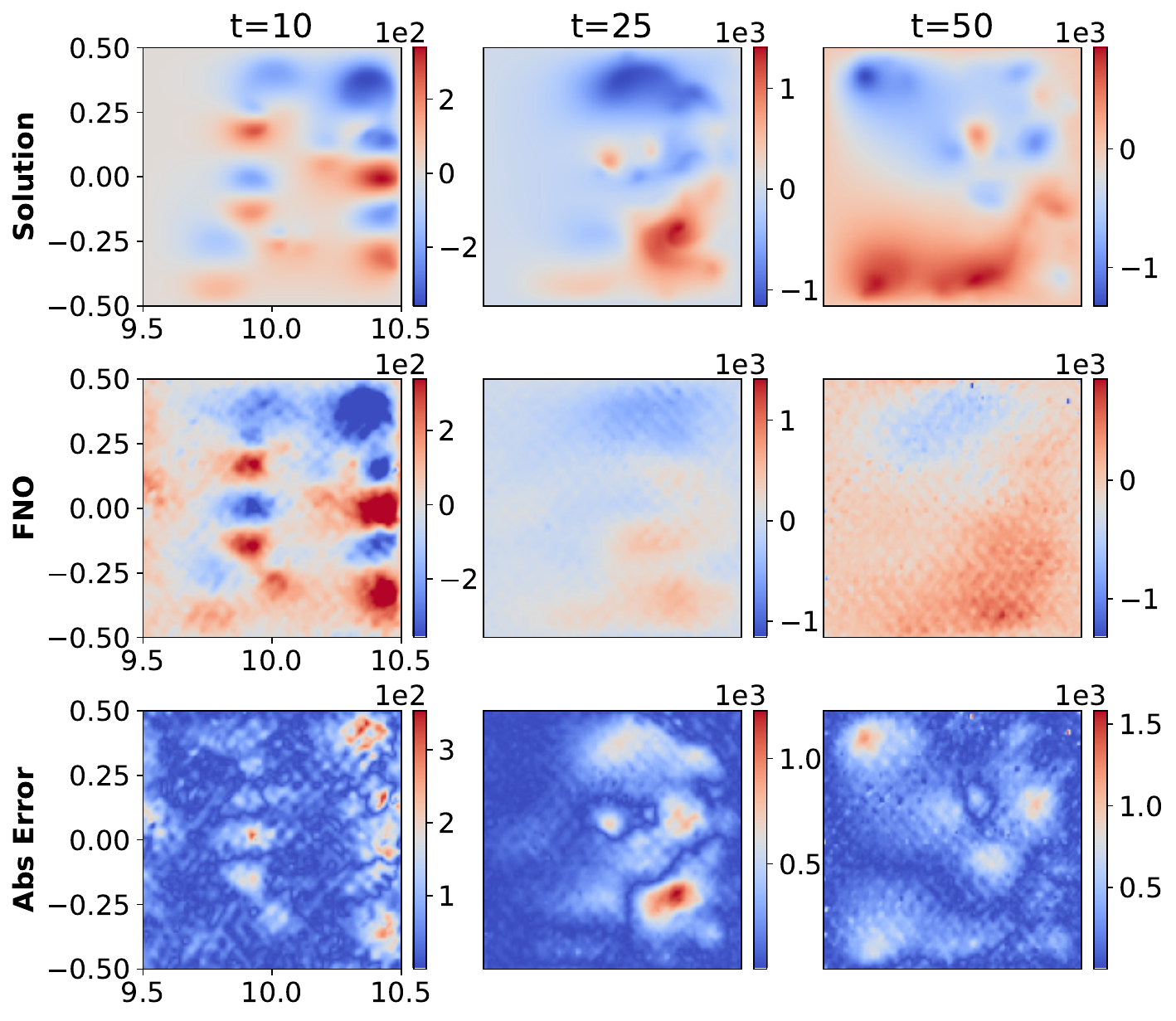}}
    \label{fig: multi_phi_out}
    \centering
    \subfloat[Temperature]{\includegraphics[width=7.5cm]{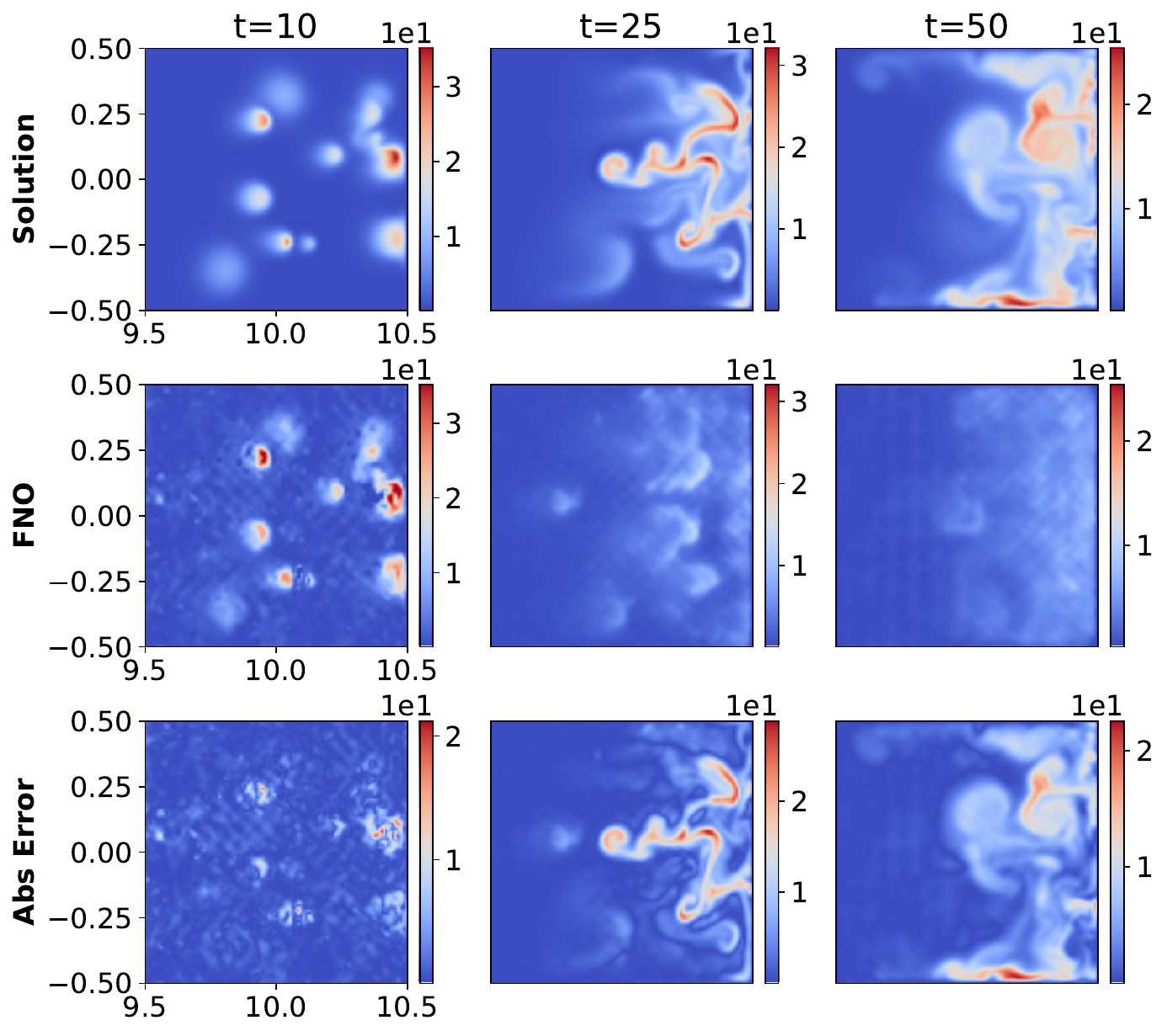}}
    \label{fig: multi_t_out}
    \caption{Outside the training domain: Inferring on a case with 13 blobs with amplitude and density twice more and the blob width 1.5 times more than the maximum used within the training data (see \cref{table: data_generation}). Temporal evolution of (a) density (b) electric potential and (c) temperature of the plasma evolution as obtained using the JOREK code (top of each image), that of the trained multi-variable FNO (middle of each figure) and the absolute error across both (bottom of each figure). The spatial domain is given in toroidal geometry characterised by $R$ in the $x$-axis and $Z$ in the $y$-axis.}
    \label{fig: multi_mhd_out}
\end{figure}

For our multi-variable FNO that models the evolution of multiple blobs in reduced MHD, we can empirically validate the above-mentioned statements. When inferred on a case where there are 13 blobs with amplitude and density twice more and width 1.5 times more than the maximum used within the training data, we observe that the FNO can model the plasma evolution at the earliest prediction time instance and fails to capture the evolution of the plasma behaviour further in time as evidenced in the \cref{fig: multi_mhd_out}.

\subsubsection{Scaling to 3D} ~\\
\label{challenges_scale}

There exist several challenges in the extending the FNO to model the evolution of multi-variable PDEs to 3D settings. The primary concern is of the memory constraints needed for a 3D multi-variable FNO when being evaluated autoregressively in time, leading to GPU bottlenecks. Recent work in performing Tensor factorisation over the FNO weights, allows representing the network in a lower rank approximation \cite{kossaifi2023multigrid}, offers potential to those GPU training bottlenecks. 

Another challenge associated with the scaling is that of the computational costs associated with data generation. In 3D, the numerical solvers that generate the training datasets are exponentially more intensive than in the 2D settings, limiting the amount of training data that could be generated. Deploying Active Learning (AL) methods, which allow for more informed sampling of the parameter domain space offers a solution to this. Within AL, the model is initially trained on a coarsely sampled, limited dataset and then evaluated across the interested domain. Performance (or uncertainty) of the model within the interested domain informs the next data generation points, which are added to the training data, improving the model accuracy. AL allows the model to maximise performance, minimise uncertainty within data scarce scenarios \cite{charu_AL}. Within the scope of plasma physics, AL has been found to reduce the data requirement of surrogate models by a factor of 20 in modelling Gyrokinetic Turbulence \cite{Zanisi_2024}. The issue of data-scarcity arising from the profound computational intensiveness of 3D simulations could also be combat by using physics-informed models, where the information regarding the PDE is hard-baked into the model architecture or the training regime. Recent works on Physics Informed Neural Operators (PINOs) has shown promise in fine-tuning coarsely trained models with PDE constraints within MHD scenarios \cite{Rosofsky_2023}. 

While scaling to 3D settings it might be the case that the choice of operator within the Neural Operator might be limiting and not representative of the true solution, thus there might be a requirement of a combination of different operators implemented over different families of basis functions within the model to account for the varied physics that is found within the PDEs. One last challenge to consider is that proposed by the autoregressive layout of the FNO. The addition of another dimension will contribute further to the error growth in time as noted in \cref{fig: mb_error_growth}. Though this remains a standing problem within the domain of Neural-PDEs, recent works have shown promise in solving this challenge by providing further conditioning of the longer time roll-outs \cite{kohl2023turbulent}.

\subsection{FNO over MAST Camera} \label{fno_over_camera}

As described in \cref{experimental_data}, we conduct exploratory research in looking at using the FNO as an emulator for performing predictive inference in real-time of the plasma evolution within the MAST fusion device. For this purpose, we deploy FNOs with the same architecture as that used for emulating the Individual FNOs (given in \ref{appendix_fno_individual_arch}). However, considering the nature of the camera dataset, we implement a different training and inference regime from that of the autoregressive structure used to solve PDEs. 

In the case of the MHD simulations, the evolution of plasma is a Markovian process, with no element of noise found within the dataset. However, within the case of the experimental data, the images are still time-series data, but they are non-markovian, noisy, and represent only partial information about the state of the system. In addition, the dataset does not offer any information about the causal contributors, such as the various control inputs that influence the state of the system, and is assumed to be implicitly defined within the time evolution of the data. Taking this into account, implementing an autoregressive structure as demonstrated in \cite{gopakumar2023fourier}, provided limited capabilities as the error growth across the time roll-out would be too large, not allowing us to perform longer time roll-outs. Within the scale of the experimental data, it might be the case that there might be new information fed into the system midway through it. By utilising an autoregressive rollout that relies solely on the initial time steps, the model will not be able to account for this new information added later to the system. Unlike in the case of modelling simulations, there is no actual need within the experimental space to rely just on the initial conditions and state of the system to perform the prediction. Data characterising the temporal evolution of the plasma is generated and captured by the diagnostic as the experiment is run and does not need any additional intervention to obtain it. Our training and inference setup takes advantage of this continuously generated data to build our predictive model.

Instead of performing an autoregressive time rollout, we perform a fixed time window mapping (as outlined in figure \cref{fig:ar_tw}), where the FNO takes in the camera information for fixed time steps (\textit{T\_in}) and outputs the next set of frames characterising the future plasma state (\textit{step}). Being trained in this sequential pipeline, the model initially takes in the first 10 time instances (1--10 frames) to output the next 10 time instances (11--20 frames). With the next forward pass, the model takes in the other 10 time instances, but this time, forward by one frame (2--11 frames) to predict the next 10 frames to that input (12--21 frames). This input, characterised by a 10-frame window slid by one each forward pass, is run until we output the last required frame (\textit{T\_out}).  Our trained FNO models the plasma at ramp-up, entering the flat-top regime, and in most cases, across the L-Mode and H-Mode operational regimes, covering the full (available) plasma duration. Since we are designing this to be run in real-time with the experiment, future time windows, as they become available, will be fed into the model to predict the future plasma states. The time resolution of the fast camera diagnostic is 1.2\,ms, where the FNO only requires 6\,ms to predict the next 12\,ms of the plasma evolution, allowing us to run the model efficiently in real time. 

Even though we keep our focus on a limited number of shots curated according to the criteria given in \cref{camera_data_selection}, as we structure them into input-output pairs by way of the sliding windows as described above, we essentially turn each shot characterising of approximately 200 frames into 180 input-output pairs. The FNO is exposed to dynamics of the plasma evolution across the interested time window of MAST operation, learning to predict from any intermediary states of the system to the next states of the system. The FNO is trained to learn how to predict the evolution of the plasma as witnessed by the camera for the full (available) range of operation of the Tokamak pulse.  

For each camera FNO, we chose the input feed to consist of 10 time steps of the field information (\textit{T\_in = 10}), with an output size of 10 (step=10). Each Fourier layer within the FNO has 8 modes and a temporal channel width of 16. The grid discretisation associated with the field data (obtained from the reactor CAD model) is added to the training data within each model's forward call as an addition to the temporal channel, providing further field information to the model. The data is normalised by following a linear range scaling, allowing the field values to lie between $-1$ and $1$. Each model is trained for 500 epochs using the Adam optimiser \cite{adam} with a step-decaying learning rate. The learning rate is initially set to 0.005 and scheduled to decrease by half after every 100 epochs. The model was trained using a Relative L2 loss function over the reconstruction error alone (see \cref{eqn: loss_function}). 

\subsubsection{Data Selection} 
\begin{figure}[h!]  
    \centering
    \subfloat[Plasma Current]{\includegraphics[width=7.5cm]{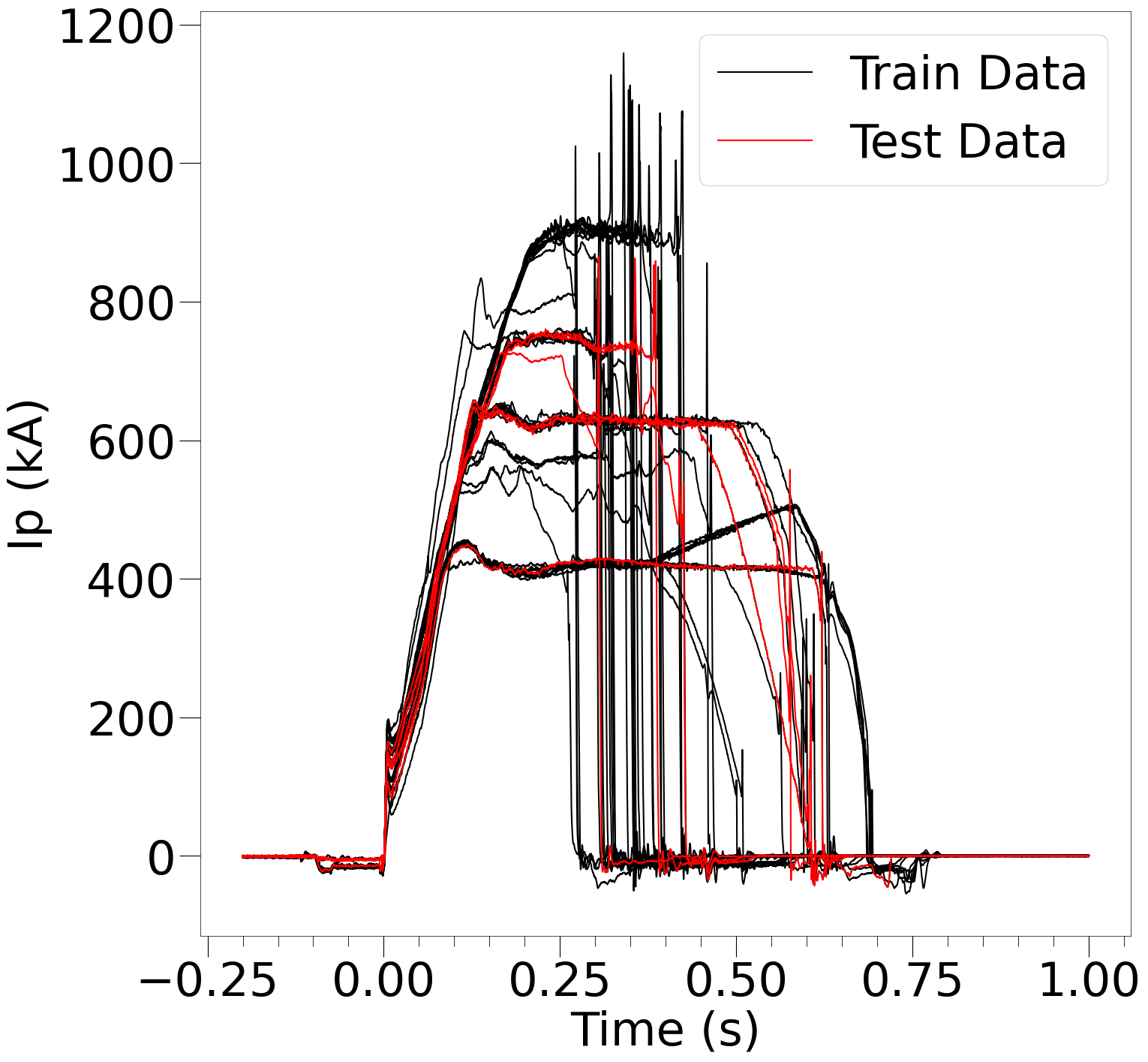}}
    \label{fig: ip_rbb}
    \centering
    \subfloat[Total Heating Power]{\includegraphics[width=6.9cm]{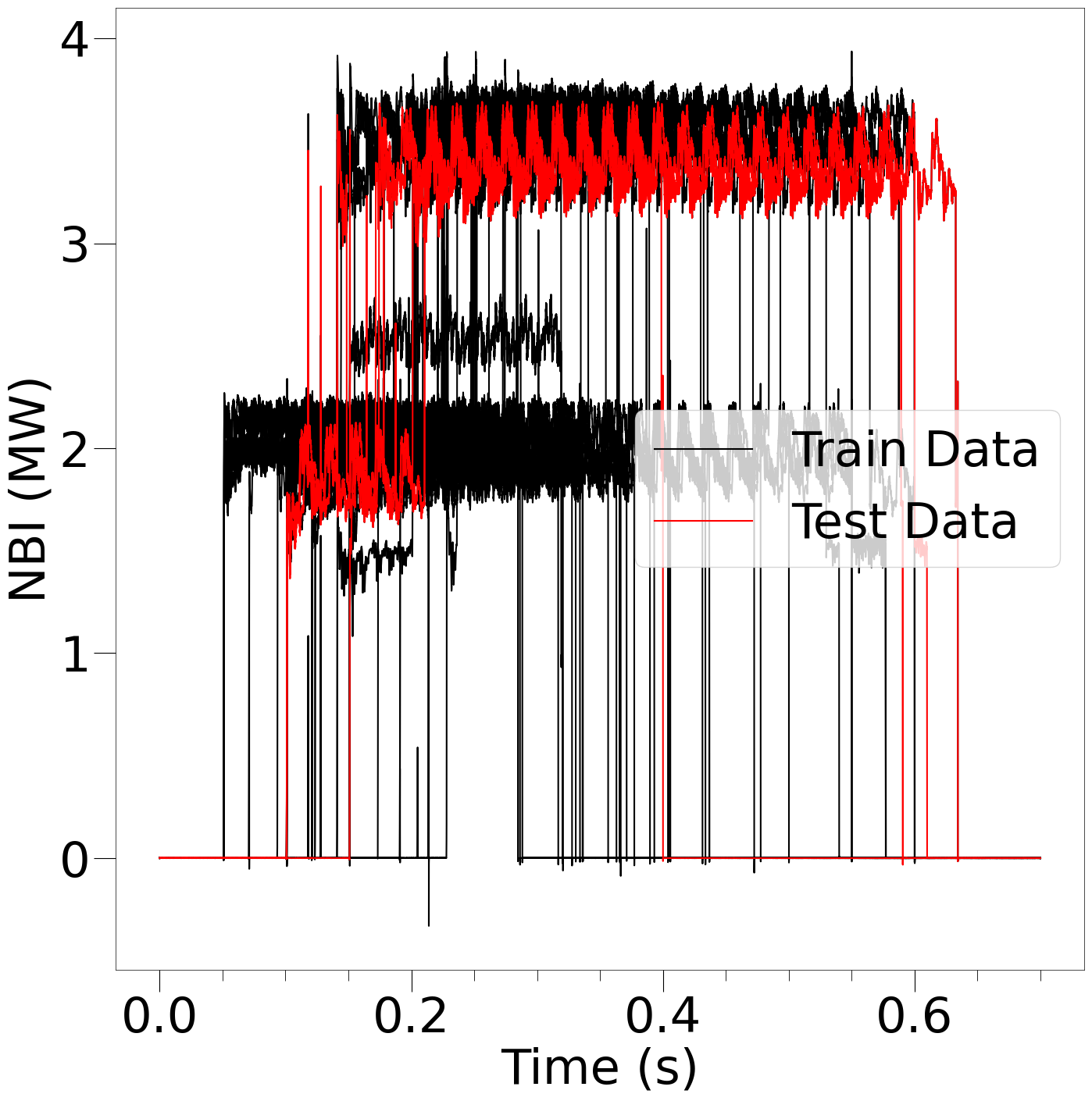}}
    \label{fig: nbi_rbb}
    \caption{Temporal plots of (a) Plasma Current and (b) Total Heating Power of the shots selected for training and testing the FNO for modelling the camera data. The shots depicting the training data is given in black, whereas the test data is given in red. The plots help convey the diversity of plasma scenarios the FNO is trained to model over. The training data consists of multiple different scenarios, allowing for building a general model. However, the test data is selected to be largely representative of the scenarios found within the training data.}
    \label{fig: ip_nbi_rbb}
\end{figure}

~\\
\label{camera_data_selection}
Throughout this study, we work on a selected number of plasma shots for each cameras, restricting ourselves to the M9 campaign on MAST. For the \textit{rbb camera}, we restrict ourselves to 55 shots lying within the range: 30250--30431 from the MAST experimental database. For the \textit{rba camera} we look at 48 shots lying within the same range 30250--30431. The shots were carefully curated from the above shot range by considering several factors. The first criterion was to ensure that the selected shot actually contained a confined plasma and was not a dummy shot like commissioning pulses. The second criterion was whether the shot had the desired duration of more than 100 time steps - looking at forecasting longer-term plasma evolution. The third one was selecting shots within the above-mentioned ranges that had the same camera calibration. Considering the FNO relied on a grid discretisation to leverage the Fourier modes, it was crucial to ensure that the pixels from the camera belonged to the same line of sight vectors. The camera calibrated for specific experimental runs captures the information of the plasma onto a uniform pixel grid, allowing us to perform Fourier transformations over them. By using \textit{calcam}, a calibration tool developed by the European Atomic Energy Community \cite{scott_silburn_2022_6891504} combined with 3D CAD models of MAST, we determine the domain range covered by each camera calibration. The upper and lower limits of the domain range of these images are then fixed onto uniform grids in both the $R$ and $Z$ axis, allowing us to create uniform grid discretisations underlying each camera image. The last criterion was to ensure that the camera images obtained from the diagnostic had the same spatial resolution across the dataset that could be fed into the FNO. For the \textit{rbb} camera, we chose the shots with a resolution of (448x640), and for the \textit{rbb} camera, we chose those with the resolution (400x512). However, the FNO offers discretisation convergence, the training setup benefits greatly from ensuring coherence along the spatial resolution. \cref{fig: ip_nbi_rbb} shows the plasma signals allows to demonstrate the diversity of the scenarios used to train and test the FNO modelling the cameras. The training data consists of multiple different scenarios, allowing for building a general model. However, the test data is selected to be largely representative of the scenarios found within the training data. For a full list of the selected shots, refer \ref{appendix_camera_shots}.

\subsubsection{Camera Viewing the Central Solenoid (\textit{rbb})}~\\

\begin{figure}[h!]
    \centering
    \includegraphics[width=16.0cm]{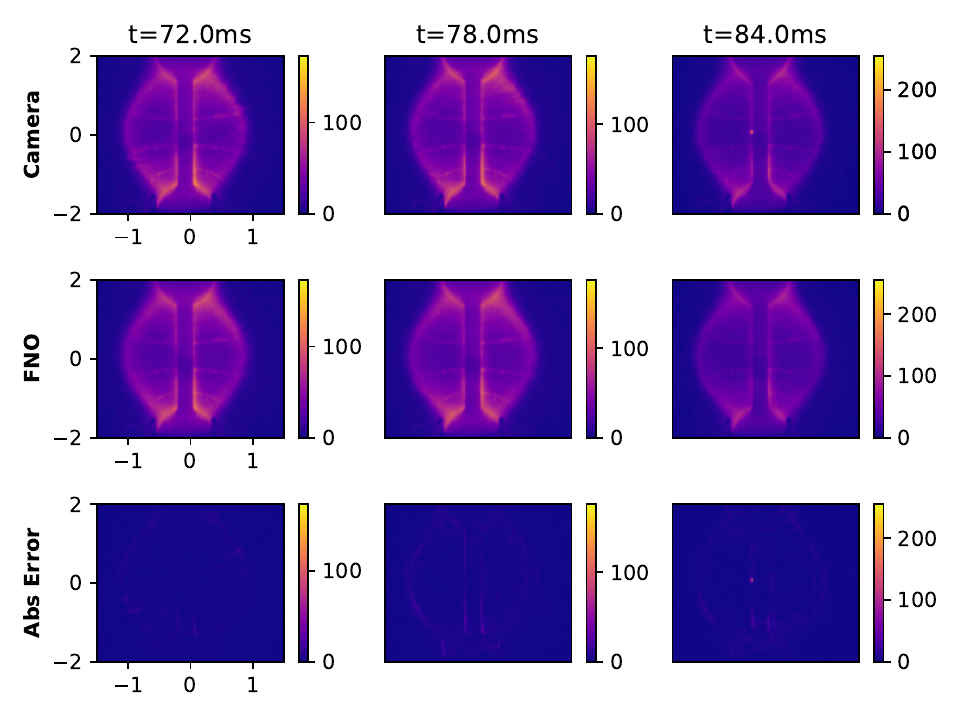}
    \caption{Predicting in L-mode: Temporal evolution of the plasma around the central solenoid within MAST as perceived by the main camera with the plasma in L-mode. The image compares the plasma evolution across one feed-forward operation of the FNO spanning 10 frames $\sim$ 12 ms. The top layer shows the actual data obtained from the experiment. The middle layer displays the emulation performed by the FNO. The bottom layer showcases the absolute error across the prediction and the camera.}
    \label{fig: rbb_ff_Lmode}
\end{figure}

\begin{figure}[h!]
    \centering
    \includegraphics[width=16.0cm]{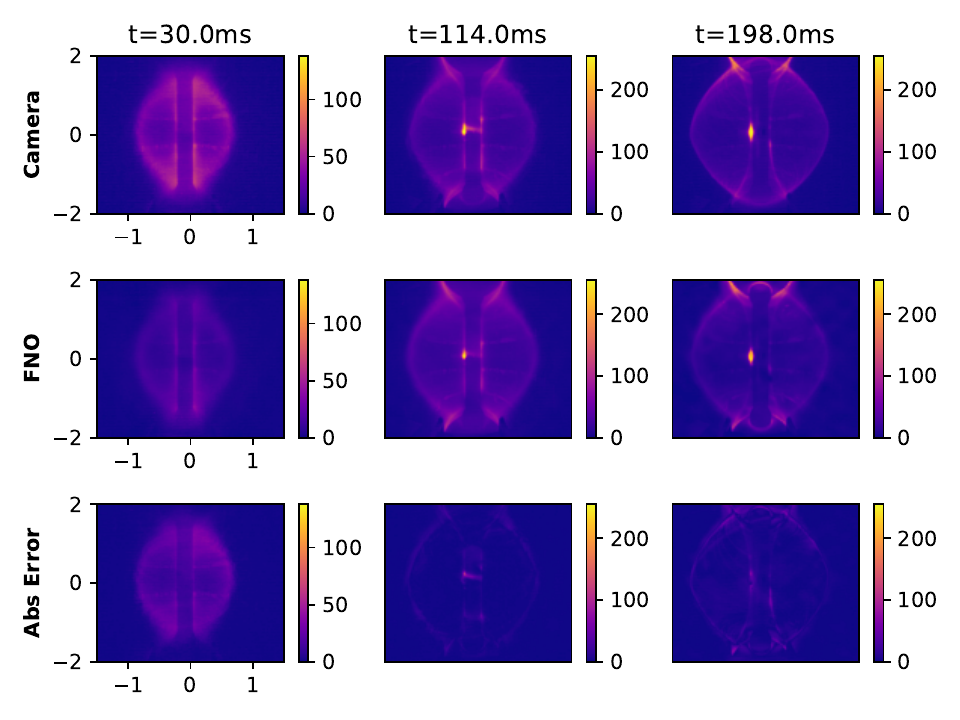}
    \caption{Long-term prediction: Temporal evolution of the plasma around the central solenoid within MAST as perceived by the main camera. The image compares the last frame of the predicted plasma evolution across multiple feed-forward instances across the duration of the pulse.  The top layer shows the actual data obtained from the experiment, the middle layer displays the emulation performed by the FNO. The bottom layer showcases the absolute error across the prediction and the camera. Though the camera captures the confined plasma shape, it fails to capture the localised filaments}
    \label{fig: rbb_last10}
\end{figure}

The evolution of the plasma within the Tokamak, as seen by the main camera, can be shown in \cref{fig: rbb_ff_Lmode}. The plot shows the comparisons of the evolution as observed by the FNO with that of the camera spanning 10 frames predicted in one feed-forward iteration. The central solenoid serves as the backbone of the MAST magnet system, generating the majority of the magnetic fluxes as well as the plasma current. In \cref{fig: rbb_last10}, we portray the predictive capability of the FNO by comparing the last frame from multiple feed-forward instances taken from data across the interested length of the pulse. This allows us to compare that even for the longest time prediction, we can get reasonable predictions. In \cref{fig: rbb_last10}, we demonstrate that the FNO is capable of predicting the plasma in L-mode as well as in H-mode, characterising the plasma evolution across the full (available) range of the plasma shot. The main camera view, as shown in \cref{fig: elms} provides a panoramic view of the reactor, showing the poloidal cross-sectional layout on both sides of the central solenoid. The main camera has a spatial resolution of 448x640 for the experiments that we sampled. We sampled a total of 55 plasma shots from the MAST database, used 50 for training and 5 for testing. Once we structure the data into windowed input-output pairs, the training size of the dataset grows to 7661, and the test points grow to 828 data points. The model takes 27 hours 45 minutes on a Nvidia A100 chip. With the above-mentioned training configuration, the model achieves an MSE (in the normalised domain) of $1.7\times 10^{-3}$. For further plots showcasing the forecasting capabilities of the FNO at the plasma around the central solenoid, refer to \ref{appendix_camera_solenoid}

\begin{figure}[h!]
    \centering
    \includegraphics[width=16.0cm]{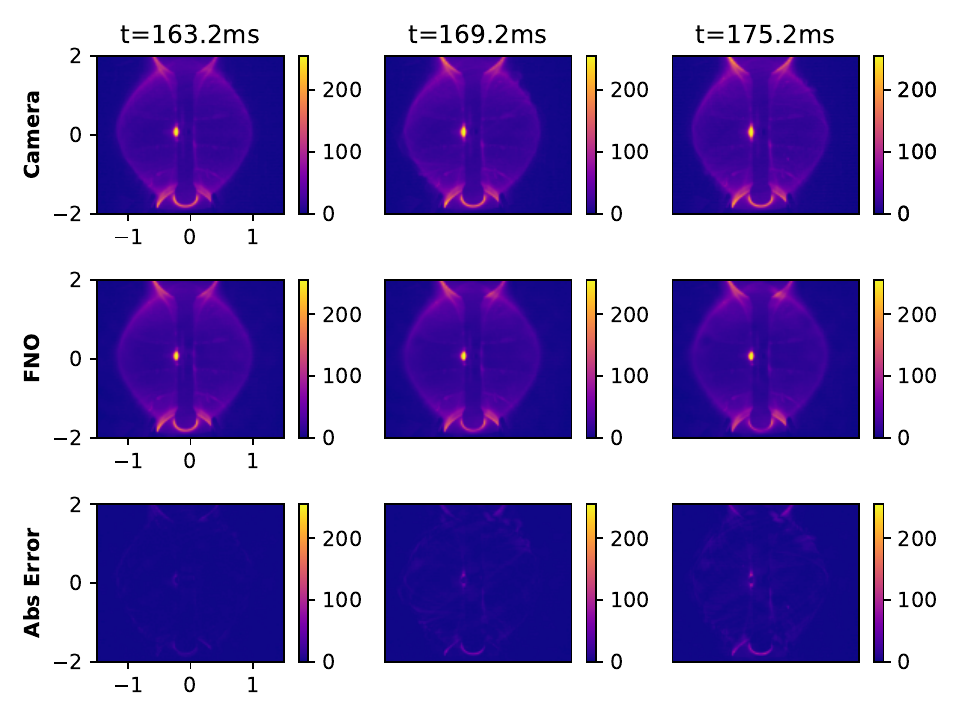}
    \caption{Predicting in H-mode: Temporal evolution of the plasma around the central solenoid within MAST as perceived by the main camera wiht the plasma in H-mod. The image compares the plasma evolution across one feed-forward operation of the FNO spanning 10 frames $\sim$ 12 ms. The domain is selected to showcase the performance of the FNO in mapping the plasma evolution in H-mode. The top layer shows the actual data obtained from the experiment. The middle layer displays the emulation performed by the FNO. The bottom layer showcases the absolute error across the prediction and the camera.}
    \label{fig: rbb_ff_Hmode}
\end{figure}

The FNO deployed with the time-window pipeline (see \cref{fig:ar_tw}), is able to capture the long-term dependencies that arise within the plasma confinement. The model is able to predict the evolution of the plasma from L-mode to H-mode as seen in \cref{fig: rbb_last10}. The prediction around the transition regime can be seen in  Though the camera captures the confined plasma shape, it fails to capture the localised filaments. In \cref{fig: rbb_ff_Lmode}, the trained FNO is deployed to model the evolution of plasma within the L-Mode of operation, whereas in \cref{fig: rbb_ff_Hmode} the FNO is modelling the region of H-mode within the same plasma shot. 

\begin{figure}[h!]
    \centering
    \includegraphics[width=16.0cm]{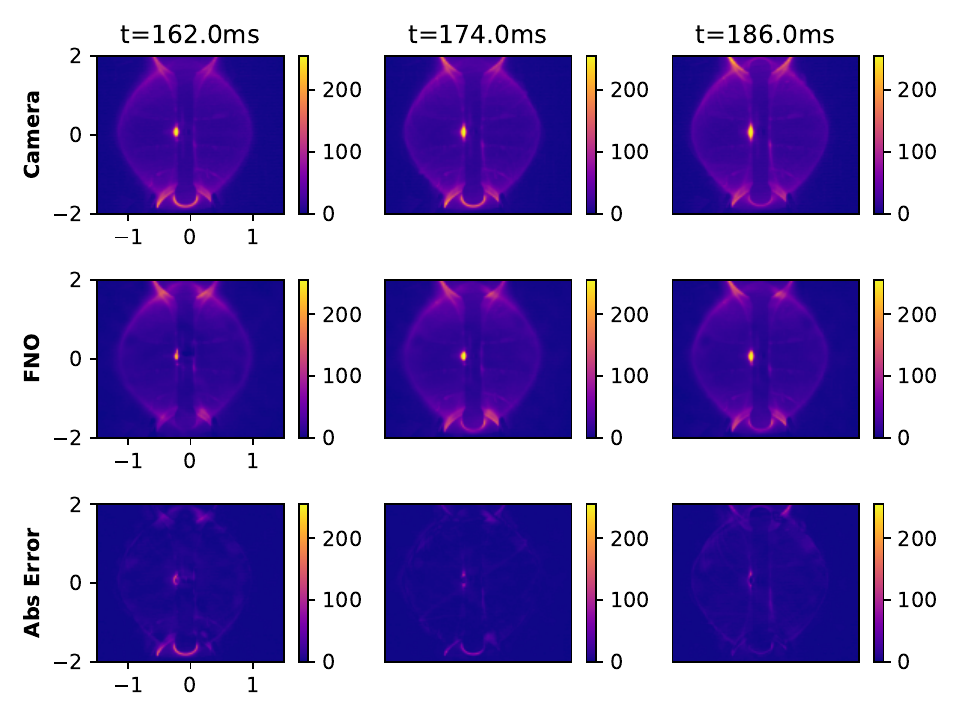}
    \caption{Predicting around the L-H transition: Temporal evolution of the plasma around the central solenoid within MAST as perceived by the main camera. The image compares the last frame of the predicted plasma evolution across multiple feed-forward instances across the duration of the pulse.  The top layer shows the actual data obtained from the experiment, the middle layer displays the emulation performed by the FNO. The bottom layer showcases the absolute error across the prediction and the camera. Though the camera captures the confined plasma shape, it fails to capture the localised filaments}
    \label{fig: lh_transition}
\end{figure}

The capability of the FNO in predicting across the L-H transition is further emphasised through \cref{fig: lh_transition}, where we show the prediction around the transition region for shot number 30428. We provide the camera image, FNO prediction and the absolute error at L-mode (162 ms), around the transition (172 ms) and with the plasma in H-mode (186 ms). The FNO is capable of modelling the plasma in both L-mode, H-mode and across the transition, but as the FNO is a regression model and not a classification model, without further analysis over the predicted solution, the model is unable to explicitly mention that it is capable of predicting the exact moment of transition. This will be a key area of focus for the next research avenue.

\subsubsection{Camera Viewing the Divertor (\textit{rba})}~\\

\begin{figure}[h!]  
    \centering
    \includegraphics[width=14.0cm]{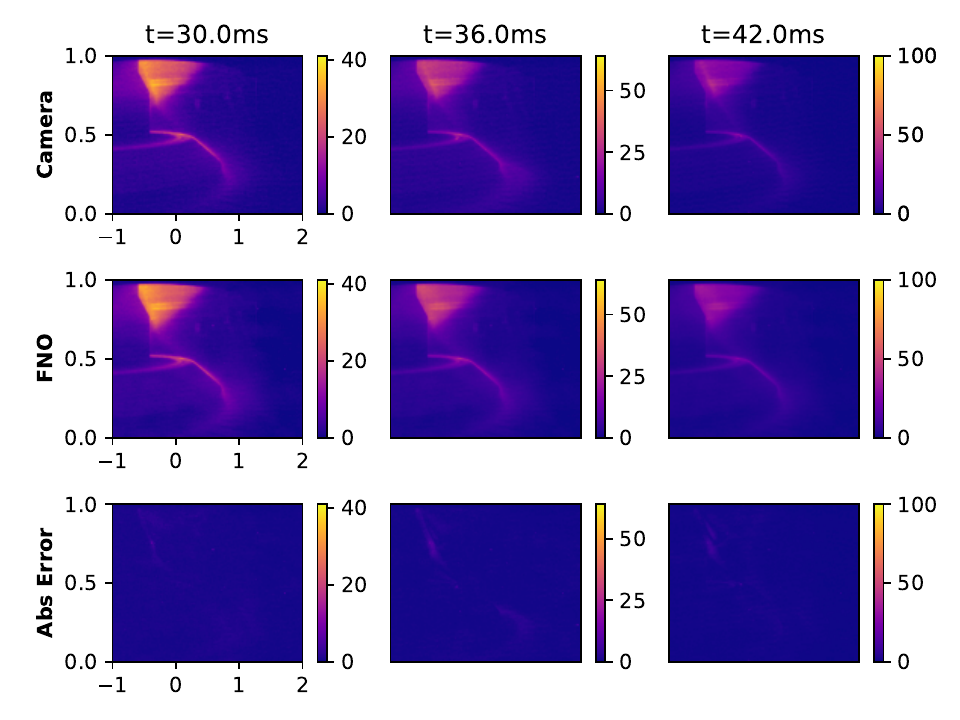}
    \caption{Single forward pass: Temporal evolution of the plasma at the divertor within MAST as perceived by the divertor camera. The image compares the plasma evolution across one feed-forward operation of the FNO spanning 10 frames $\sim$ 12 ms. The top layer shows the actual data obtained from the experiment. The middle layer displays the emulation performed by the FNO. The bottom layer showcases the absolute error across the prediction and the camera.}
    \label{fig: rba_ff}
\end{figure}

\begin{figure}[h!]
    \centering
    \includegraphics[width=14.0cm]{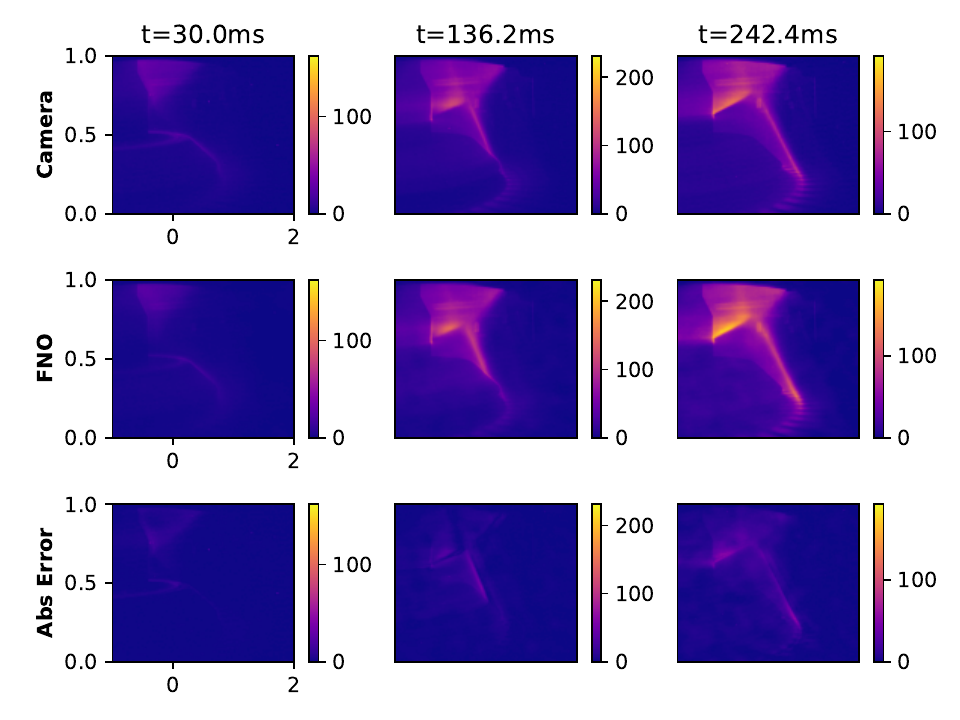}
    \caption{Long-term prediction: Temporal evolution of the plasma at the divertor within MAST as perceived by the divertor camera. The image compares the last frame of the predicted plasma evolution across multiple feed-forward instances across the duration of the pulse. The top layer shows the actual data obtained from the experiment. The middle layer displays the emulation performed by the FNO. The bottom layer showcases the absolute error across the prediction and the camera.}
    \label{fig: rba_last10}
\end{figure}

The evolution of the plasma within the Tokamak as seen by the divertor camera, along with predicted images from the FNO can be seen in \cref{fig: rba_ff}. The plot shows the comparisons of the evolution as observed by the FNO with that of the camera spanning 10 frames predicted in one feed-forward iteration. The divertor of a Tokamak serves as the exhaust of the device, where the topology of the magnetic flux surfaces is designed to isolate the confined plasma from the first-wall of the device. In \cref{fig: rba_last10}, we portray the predictive capability of the FNO by comparing the last frame from multiple feed-forward instances taken from data across the interested length of the pulse, demonstrating the predictive power of the FNO. The divertor camera looks at the lower divertor of MAST with a spatial resolution of 400x512. We sampled a total of 48 plasma shots from the MAST database, used 45 for training and 3 for testing. Once we structure the data into windowed input-output pairs, the training size of the dataset grows to 5815, and the test points grow to 701 data points. The model takes 9 hours 10 minutes on a Nvidia A100 chip. With the above-mentioned training configuration, the model achieves an MSE (in the normalised domain) of $1.70\times 10^{-4}$. For further plots showcasing the forecasting capabilities of the FNO at the divertor plasma, refer to \ref{appendix_camera_divertor}.


The FNO deployed for emulating the camera data accurately preserves the global features of the plasma evolution. It is able to clearly characterise the locations and intensities of the heat fluxes onto the solenoid and the divertors, allowing us to anticipate the heat flux in advance and initiate the appropriate actuator mechanism to minimise its impact on the various plasma-facing components. The FNO also captures the reflections generated within the device due to the interactions across the plasma and the plasma-facing components. The FNO, in its current form, smooths out the filaments but captures the general confined structure of the plasma as it evolves in time. As expected, the FNO learns characteristics extremely well, while failing to capture the localised phenomena with matching pixel-pixel accuracy. 

\subsubsection{Limitations}~\\
 We decided to look at a narrow shot corridor, since the data variability will be low, allowing us to demonstrate a functioning proof-of-concept.   Our predictive model only takes in the camera diagnostic as the input and does not take into account any of the control signals, such as coil currents and heating parameters that influence the plasma evolution. However, we argue that within the past window of plasma observed from the camera data that is taken as the input, these control parameters and their influence are implicitly defined (with a latency), and thus, the model provides a suitable predictive tool for short-term predictions. The inclusion of various control signals, time matched with the camera inputs, will potentially help improve the performance of the model. Our future research trajectory will be heading in this direction.

\section{Conclusion and Discussion}
\label{section: conclusion}

Through this work, we demonstrate the utility of using the Fourier neural operator as both a surrogate model for a plasma simulator and as a predictive emulator for a plasma experiment. We design a modified version of the FNO, that performs integrated emulation of multiple variables associated with a family of PDEs. Our analysis across a range of reduced MHD cases demonstrates that even in data-scarce scenarios, the FNO can be trained to act as a surrogate. Through this paper, we highlight the various features that make the FNO a natural choice for (surrogate) modelling MHD cases, all the while performing ablation studies on the various hyperparameters and the training regimes. We find that the FNO fits the requirements of a quick, cheap surrogate, but due to its auto-regressive nature, fails to capture the physics when deployed for long time roll-outs. Being utilised for emulating the experiment, the FNO is capable of forecasting the evolution of the plasma as seen through the cameras on a fusion device. The FNO camera models are capable of identifying the regions of plasma interaction with the central solenoid and the divertor, while effectively predicting the plasma shape evolution. The FNO has shown versatile performance in being able to model the plasma confinement across both L-mode and H-mode of plasma operation across a diverse range of shots. The FNO for the camera can predict the plasma evolution in real-time and offers us the potential to be deployed within the planning stage for various control scenarios. Our work also demonstrates how we can perform zero-shot super-resolution across the pre-trained multi-variable FNO and obtain the plasma emulation at higher resolutions than it was trained on, without requiring any fine-tuning. 

In continuing with this line of work, we will be exploring the utility of building more versatile and robust FNO models that can be deployed not just in a particular case, but across a range of simulation cases. While we build more general MHD models, we will also be looking at methods to better physics-inform the models \cite{li2022pino}, allowing us to fine-tune the general models to specific MHD instances. We will also be looking into methods of quantifying the uncertainty associated with the surrogate models and into active learning pipelines that can facilitate the entire surrogate ecosystem. With the FNO for camera data, we are currently working on extending the utility across more plasma shots with a more diverse line of sights, across multiple campaigns while exploring the utility of state space models for longer predictions \cite{gopakumar2023fourierrnns}. The work will be looking at including various control signals, to help improve the performance of the real-time predictive emulator for the MAST Tokamak, performing longer forecasting. The code for the entire project can be found in https://github.com/Plasma-FNO. 
\clearpage
\newpage
\section*{References}

\bibliographystyle{ieeetr}
\bibliography{references}

\appendix
\newpage
\section{Error Plots}
\label{appendix_errorplots}

\subsection{Single Blob with uniform Temperature}
\begin{figure}[h!]
    \centering
    \subfloat[Density]{\includegraphics[width=7.5cm]{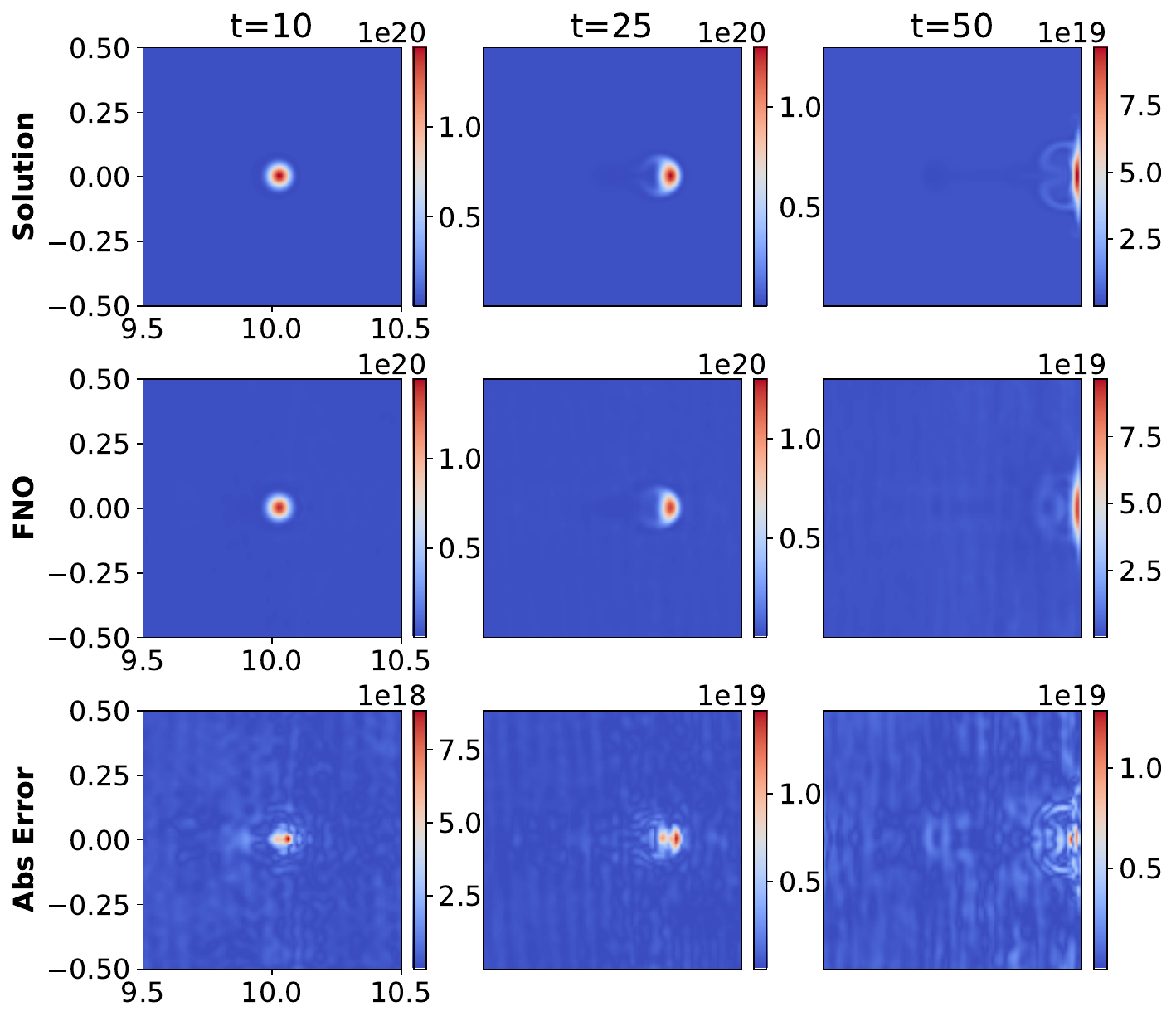}}
    \label{fig: ib_err_rho}
    \centering
    \subfloat[Potential]{\includegraphics[width=7.5cm]{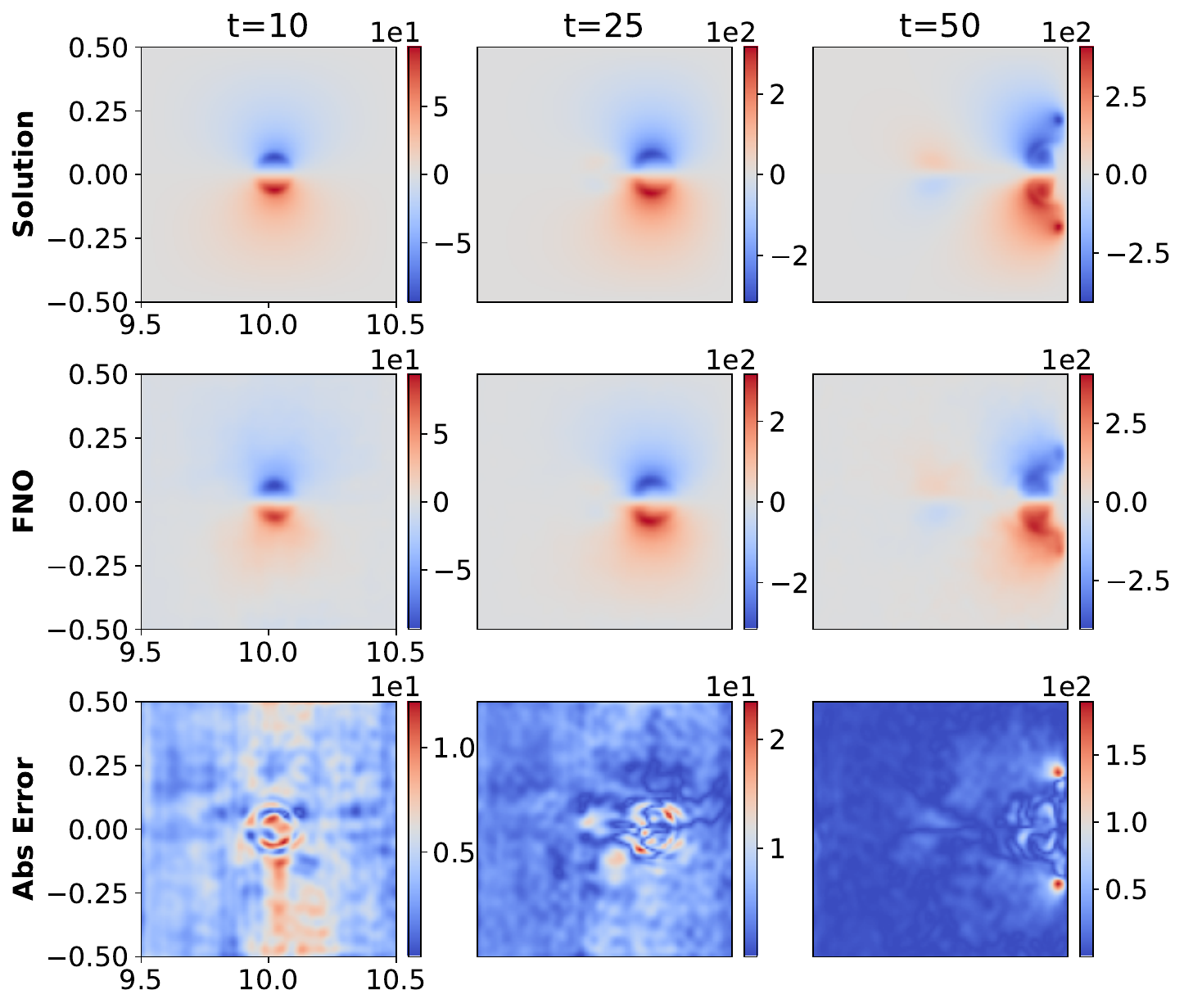}}
    \label{fig: ib_err_phi}
    \caption{Isothermal Blob : Temporal evolution of (a) the density and (b) the potential of the plasma evolution as obtained using the JOREK code (top of each image), that of the trained multi-variable FNO (middle of each figure) and the absolute error across both (bottom of each figure). The spatial domain is given in toroidal geometry characterised by $R$ in the $x$-axis and $Z$ in the $y$-axis.}
    \label{fig: ib_errors}
\end{figure}

\newpage
\subsection{Single Blob with non-uniform Temperature}
\begin{figure}[h!]
    \centering
    \subfloat[Density]{\includegraphics[width=7.5cm]{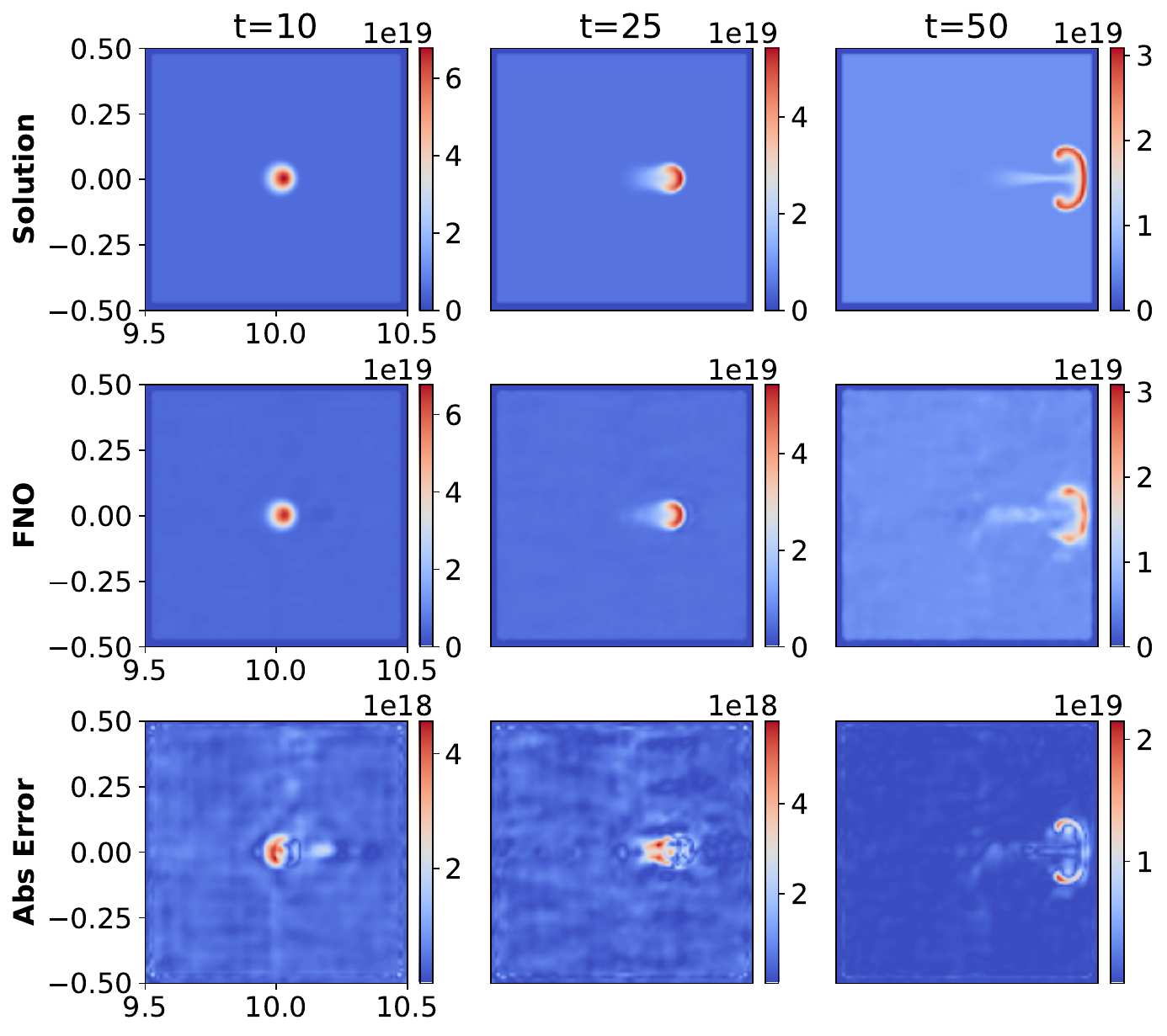}}
    \label{fig: sb_err_rho}
    \centering
    \subfloat[Potential]{\includegraphics[width=7.5cm]{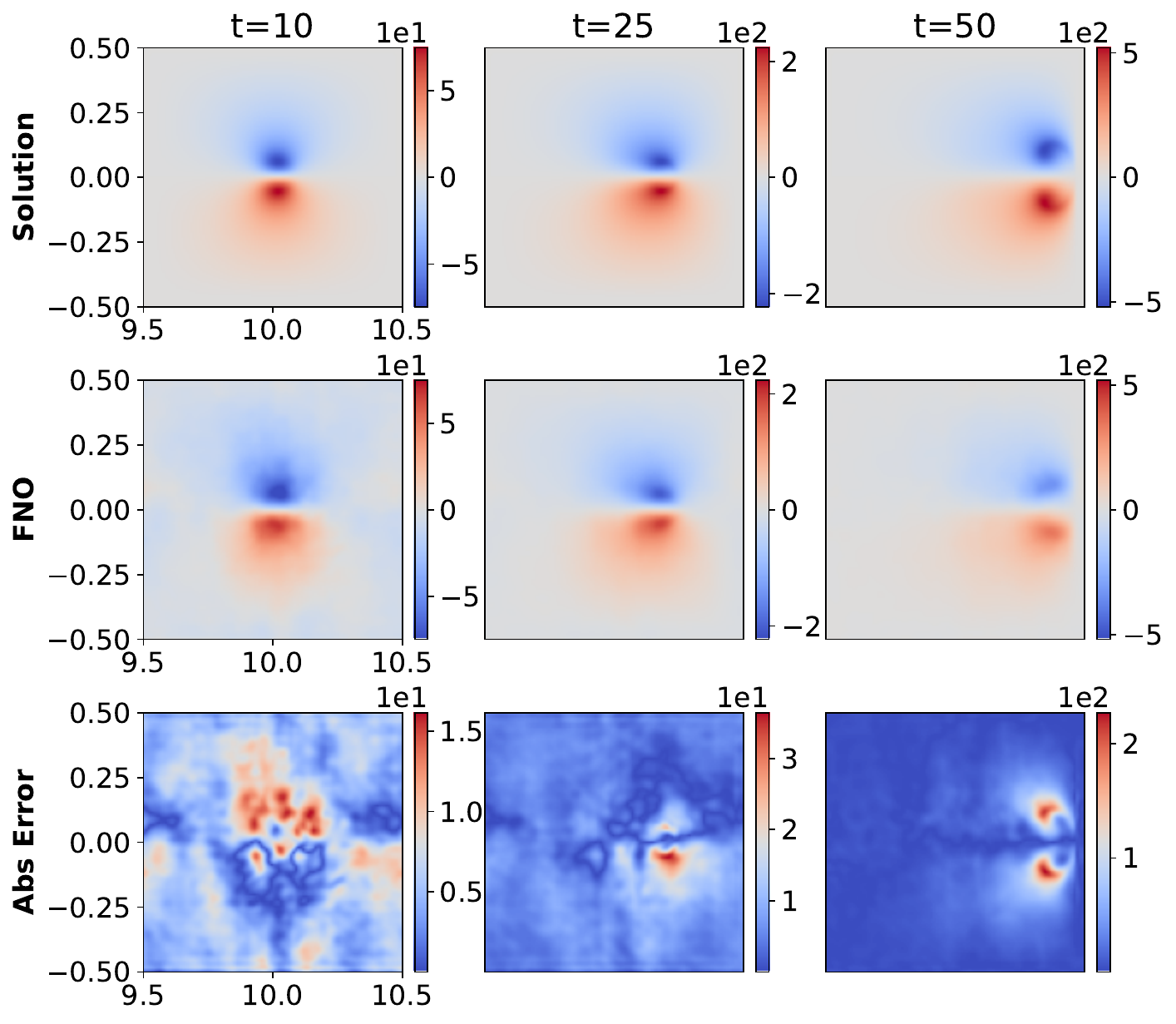}}
    \label{fig: sb_err_phi}
    \centering
    \subfloat[Temperature]{\includegraphics[width=7.5cm]{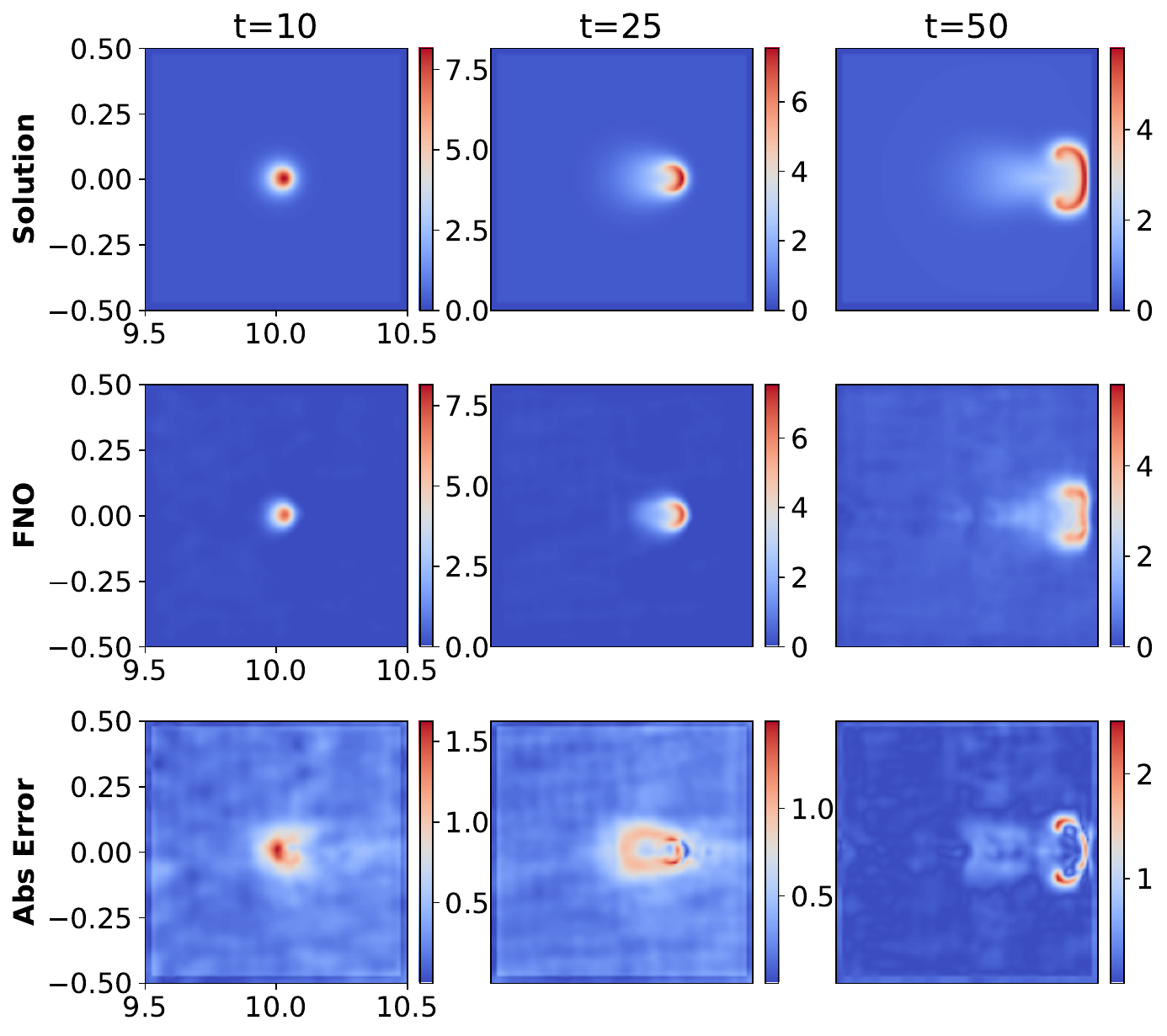}}
    \label{fig: sb_err_T}
    \caption{Single Blob with non-uniform Temperature: Temporal evolution of (a) the density and (b) the potential, (c) the Temperature of the plasma evolution as obtained using the JOREK code (top of each image), that of the trained multi-variable FNO (middle of each figure) and the absolute error across both (bottom of each figure). The spatial domain is given in toroidal geometry characterised by $R$ in the $x$-axis and $Z$ in the $y$-axis.}
    \label{fig: sb_errors}
\end{figure}

\newpage
\subsection{Multiple Blobs with non-uniform Temperature}
\label{appendix_mb_errorplots}

\begin{figure}[h!]
    \centering
    \subfloat[Density]{\includegraphics[width=7.5cm]{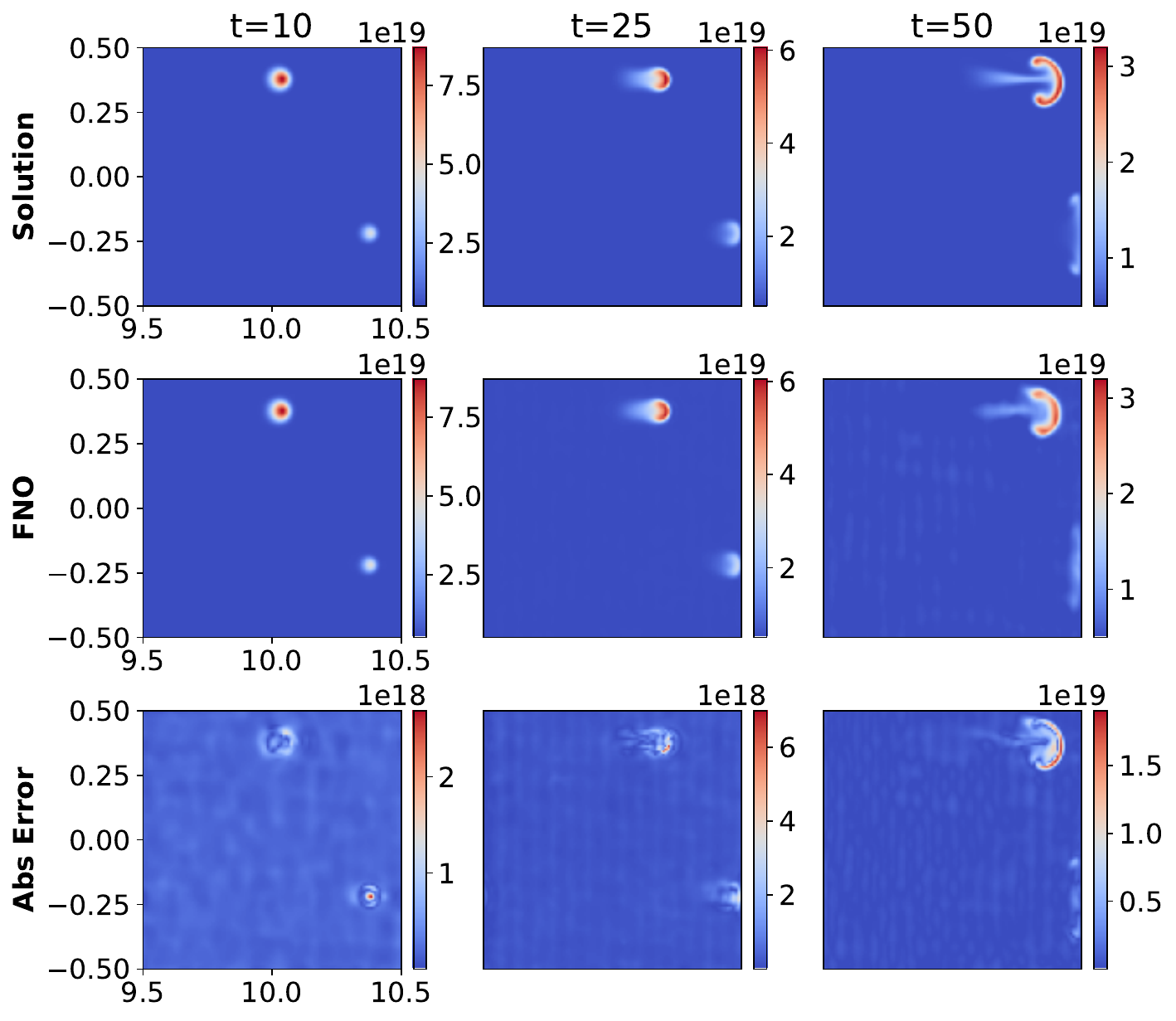}}
    \label{fig: multi_rho_84}
    \centering
    \subfloat[Potential]{\includegraphics[width=7.5cm]{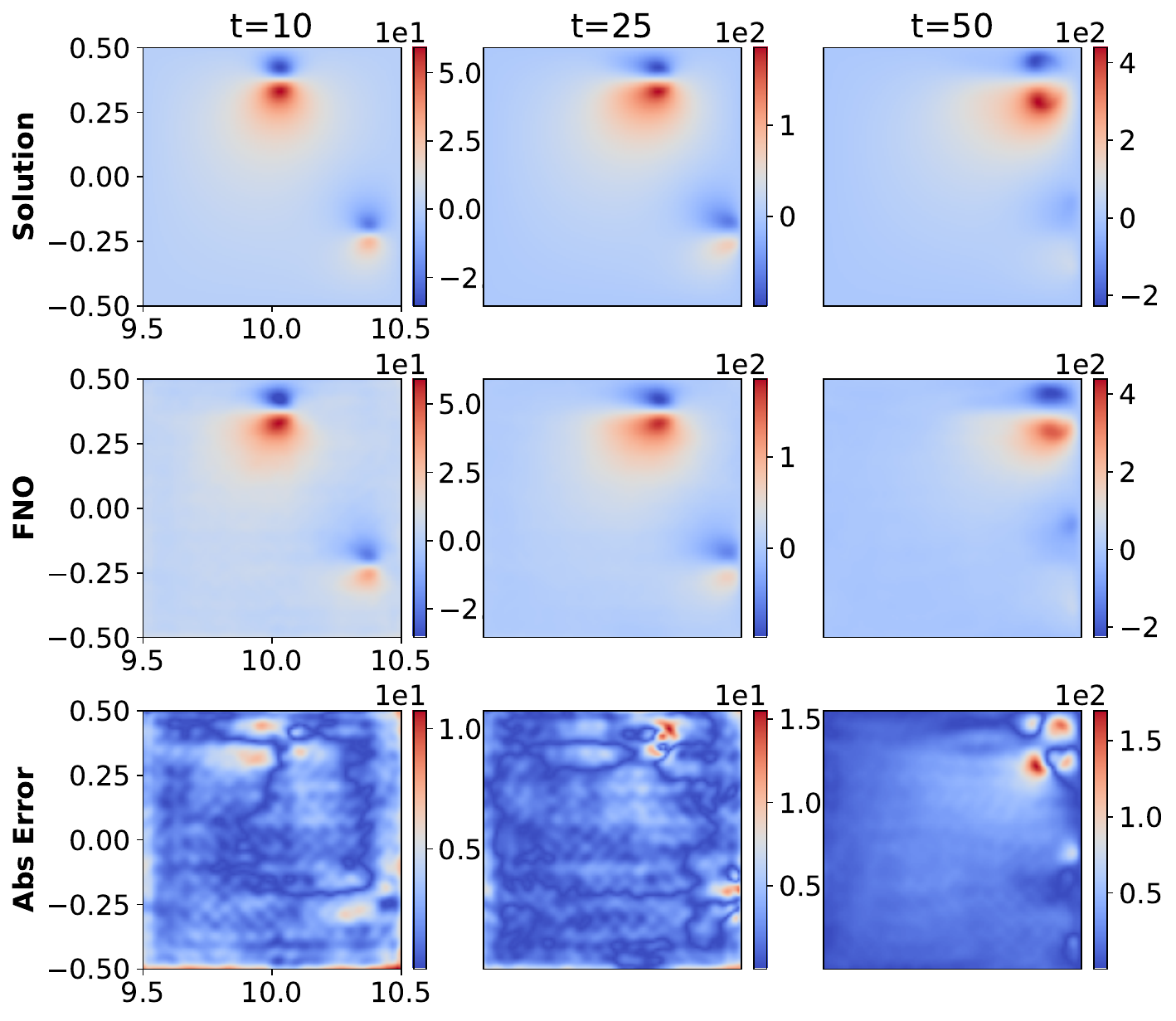}}
    \label{fig: multi_phi_84}
    \centering
    \subfloat[Temperature]{\includegraphics[width=7.5cm]{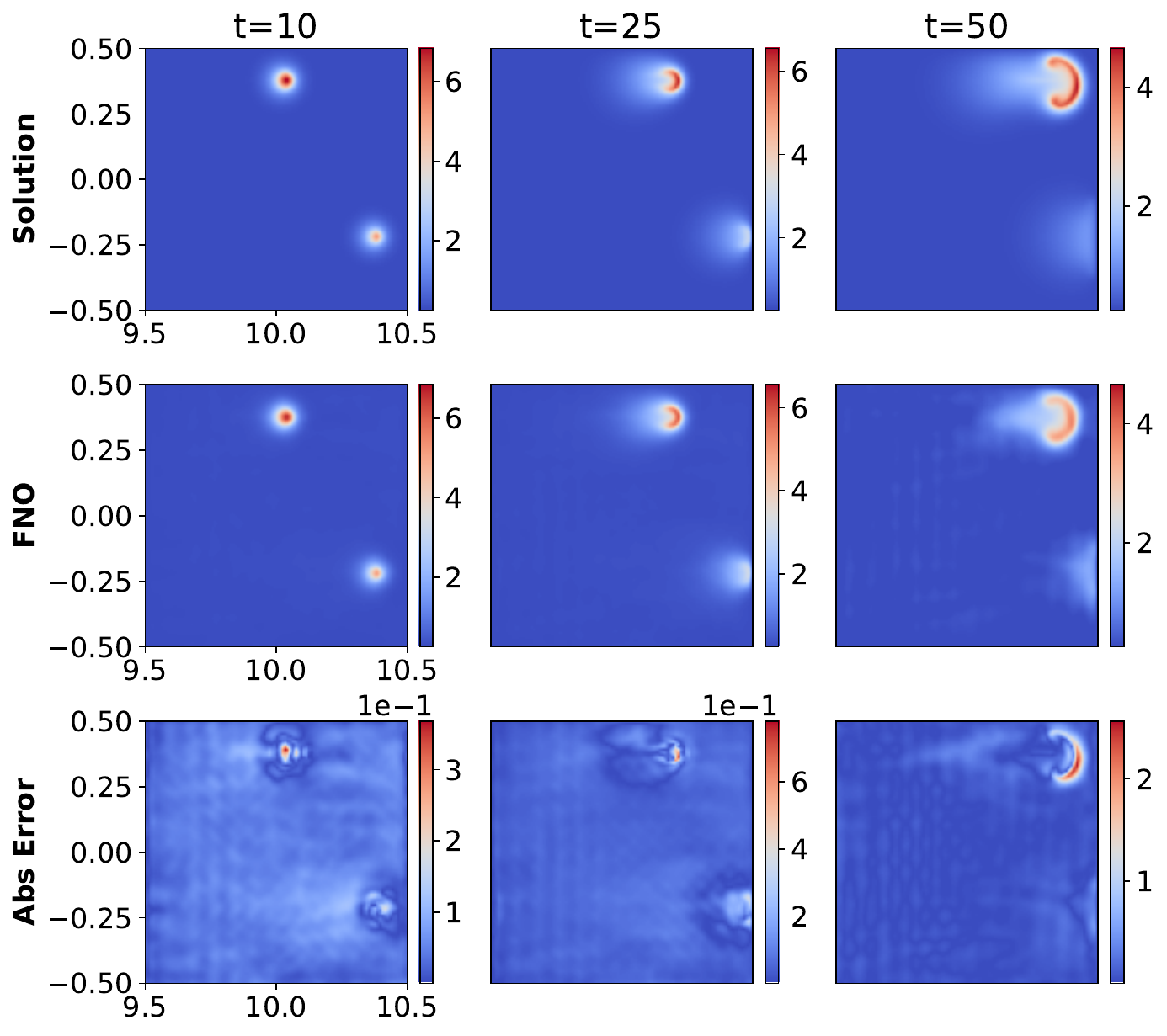}}
    \label{fig: multi_t_84}
    \caption{Multiple Blobs with non-uniform temperature: Temporal evolution of (a) the density and (b) the potential, (c) the Temperature of the plasma evolution as obtained using the JOREK code (top of each image), that of the trained multi-variable FNO (middle of each figure) and the absolute error across both (bottom of each figure). The spatial domain is given in toroidal geometry characterised by $R$ in the $x$-axis and $Z$ in the $y$-axis.}
    \label{fig: mb_84}
\end{figure}

\newpage
\begin{figure}[h!]
    \centering
    \subfloat[Density]{\includegraphics[width=7.5cm]{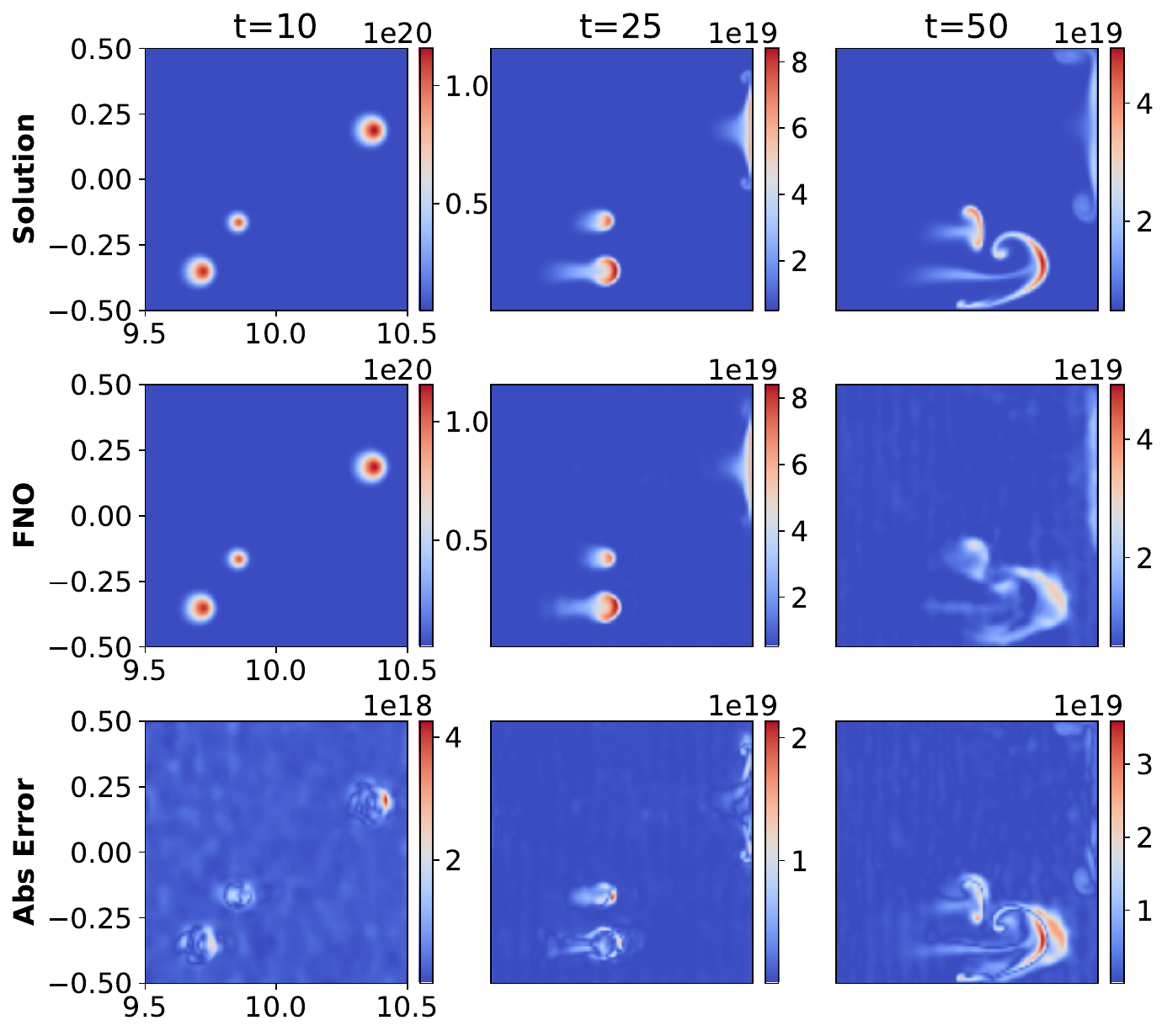}}
    \label{fig: multi_rho_64}
    \centering
    \subfloat[Potential]{\includegraphics[width=7.5cm]{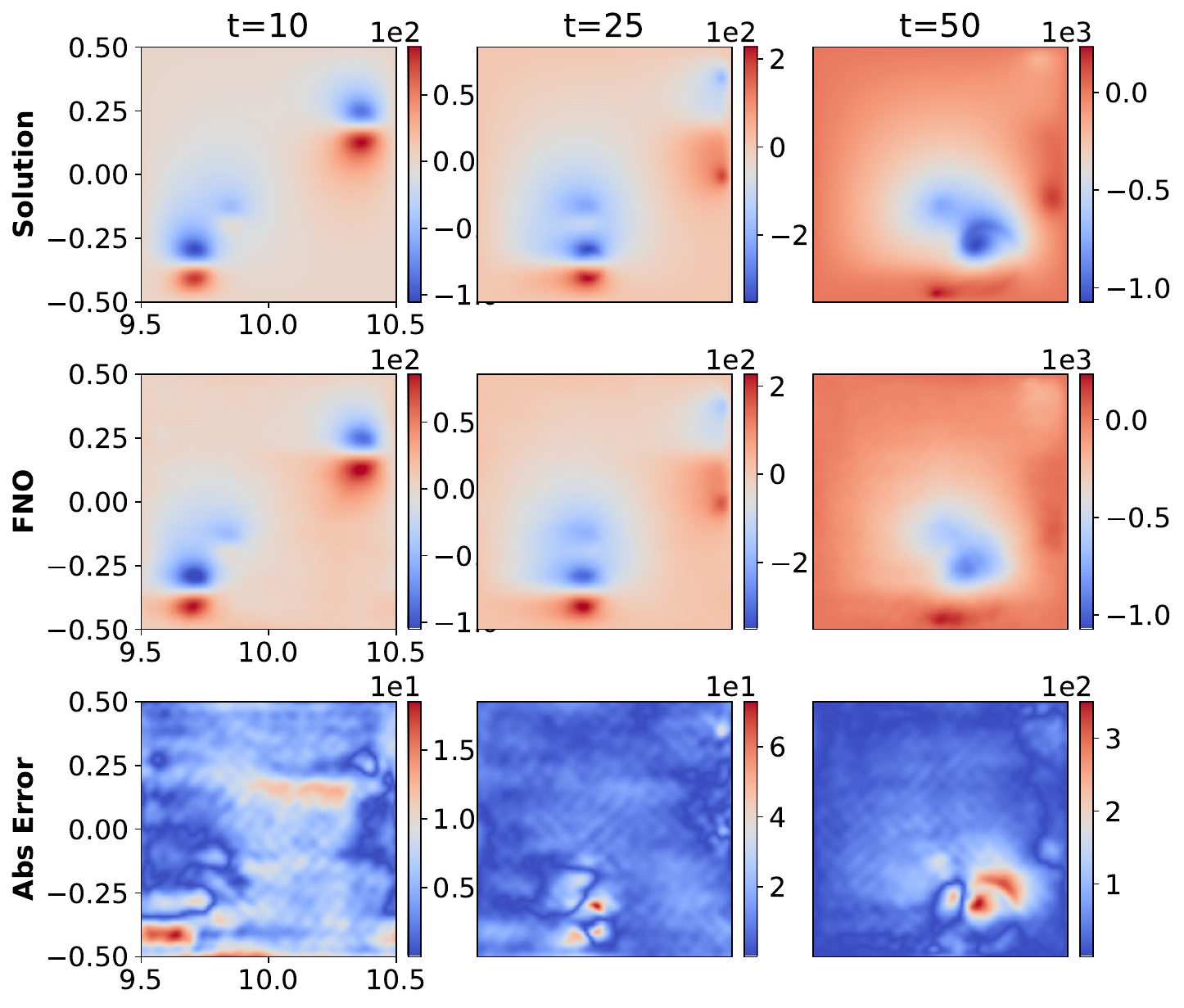}}
    \label{fig: multi_phi_64}
    \centering
    \subfloat[Temperature]{\includegraphics[width=7.5cm]{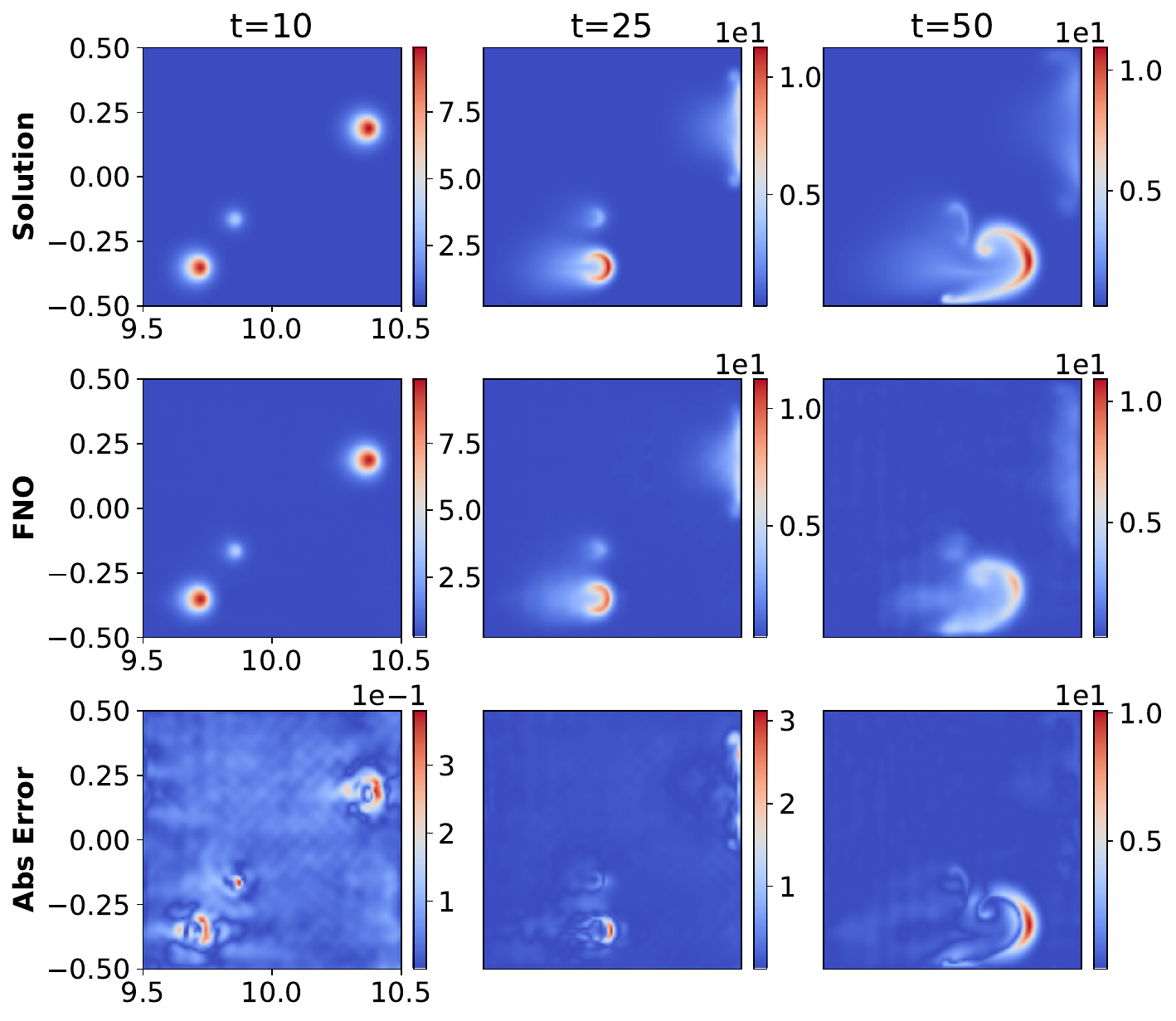}}
    \label{fig: multi_t_64}
    \caption{Multiple Blobs with non-uniform temperature: Temporal evolution of (a) the density and (b) the potential, (c) the Temperature of the plasma evolution as obtained using the JOREK code (top of each image), that of the trained multi-variable FNO (middle of each figure) and the absolute error across both (bottom of each figure). The spatial domain is given in toroidal geometry characterised by $R$ in the $x$-axis and $Z$ in the $y$-axis.}
    \label{fig: mb_64}
\end{figure}

\newpage
\begin{figure}[h!]
    \centering
    \subfloat[Density]{\includegraphics[width=7.5cm]{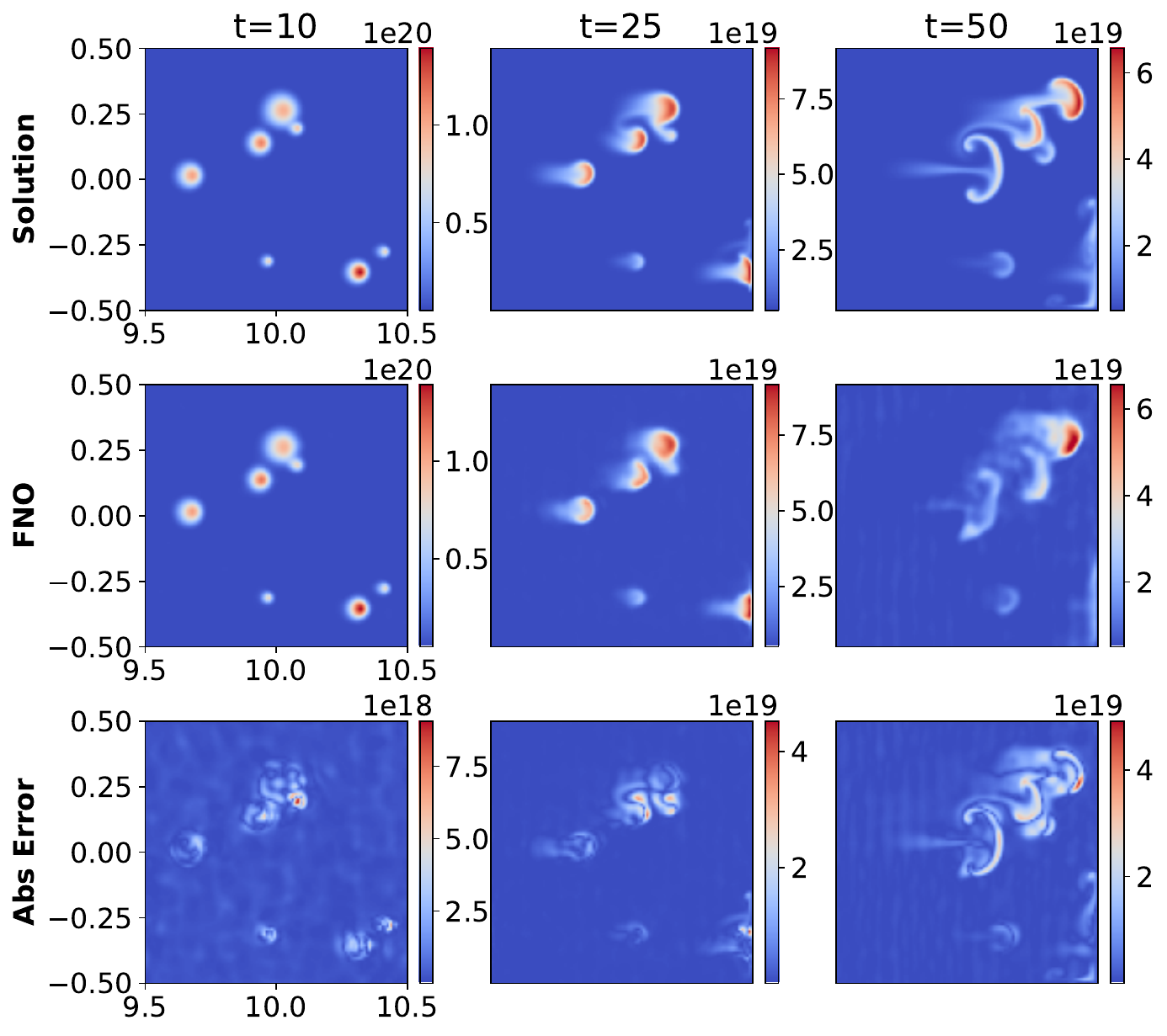}}
    \label{fig: multi_rho_35}
    \centering
    \subfloat[Potential]{\includegraphics[width=7.5cm]{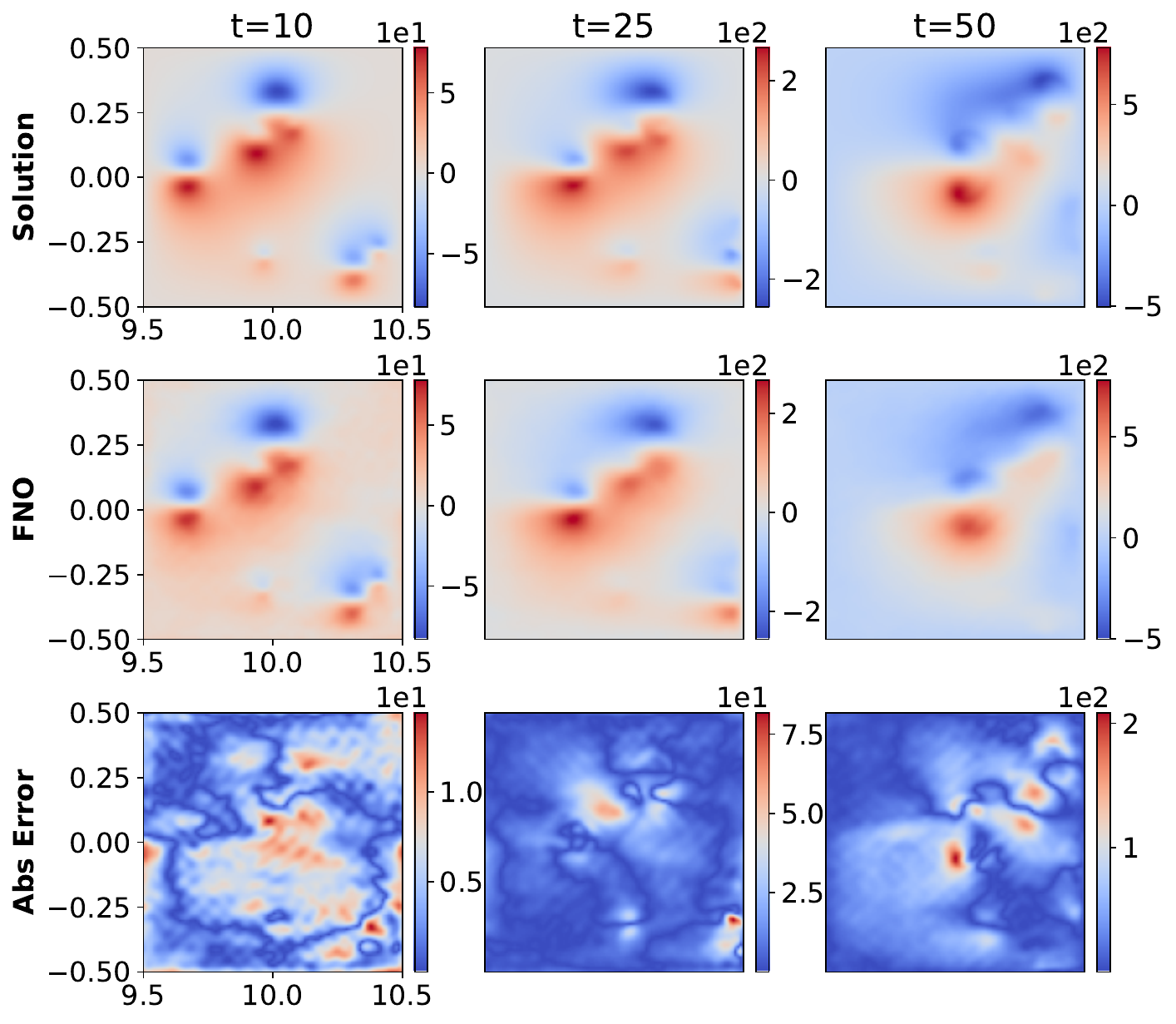}}
    \label{fig: multi_phi_35}
    \centering
    \subfloat[Temperature]{\includegraphics[width=7.5cm]{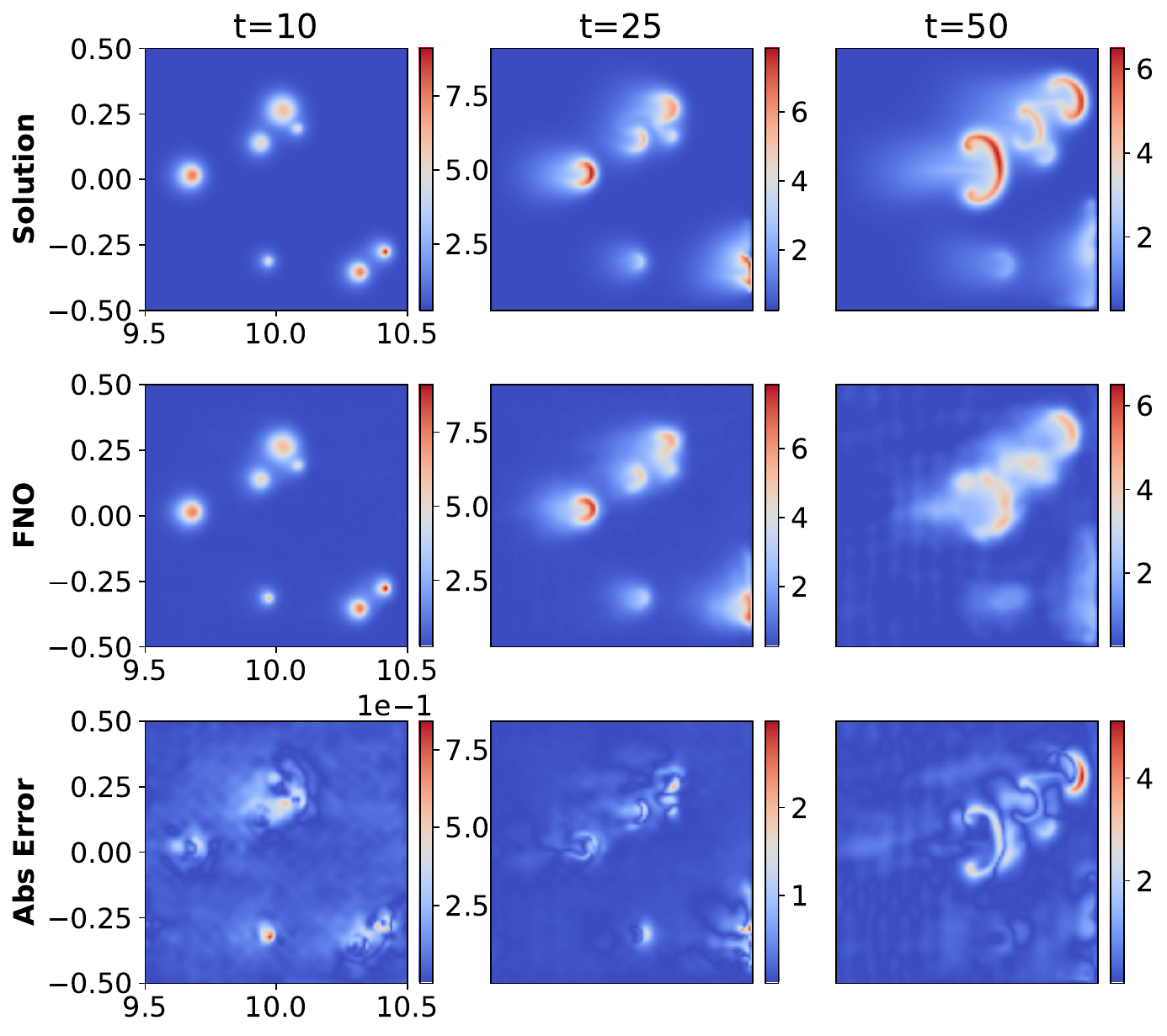}}
    \label{fig: multi_t_35}
    \caption{Multiple Blobs with non-uniform temperature: Temporal evolution of (a) the density and (b) the potential, (c) the Temperature of the plasma evolution as obtained using the JOREK code (top of each image), that of the trained multi-variable FNO (middle of each figure) and the absolute error across both (bottom of each figure). The spatial domain is given in toroidal geometry characterised by $R$ in the $x$-axis and $Z$ in the $y$-axis.}
    \label{fig: mb_35}
\end{figure}

\newpage 
\begin{figure}[h!]  
    \centering
    \subfloat[Density]{\includegraphics[width=7.5cm]{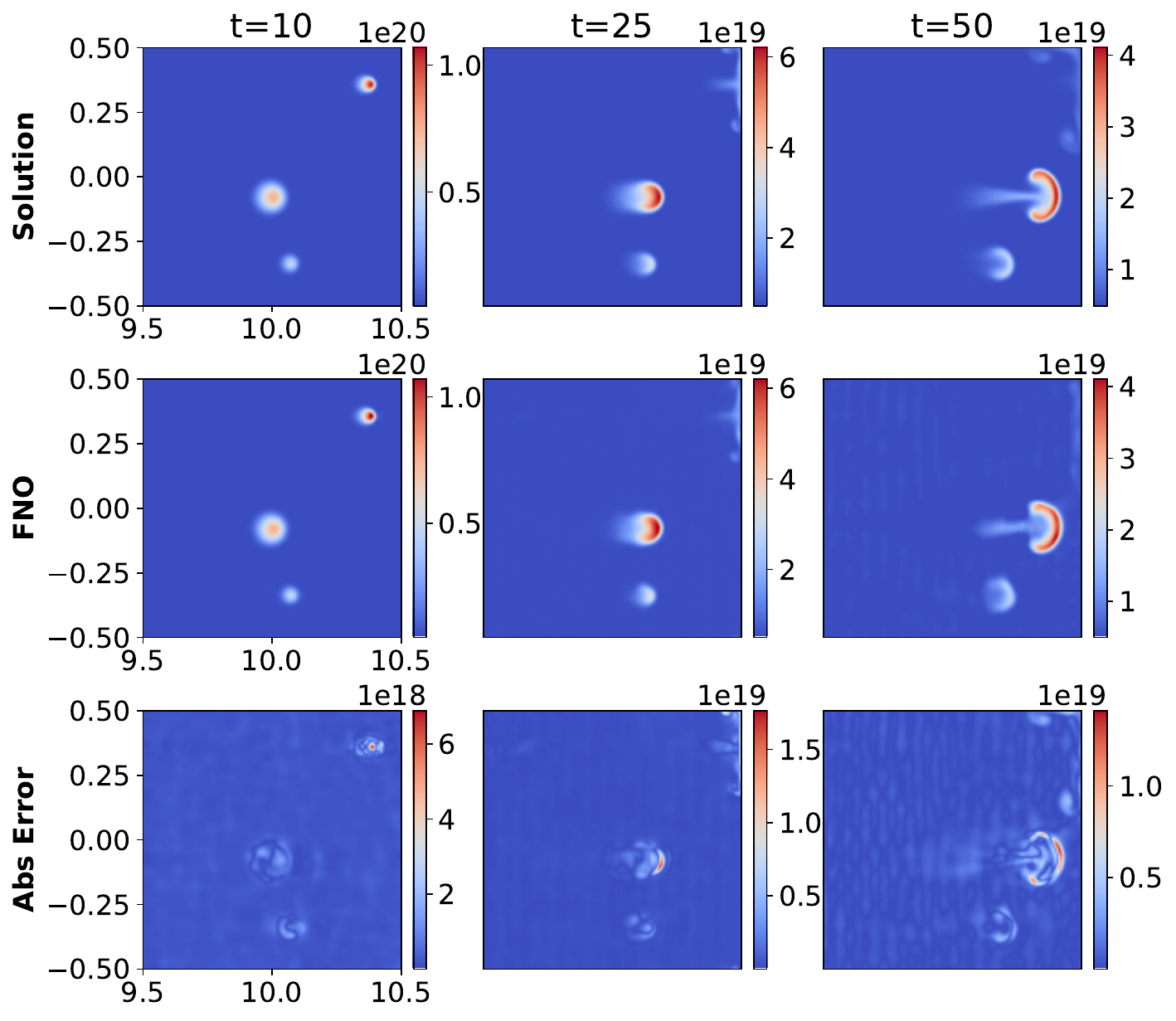}}
    \label{fig: density_zssr_error}
    \centering
    \subfloat[Potential]{\includegraphics[width=7.5cm]{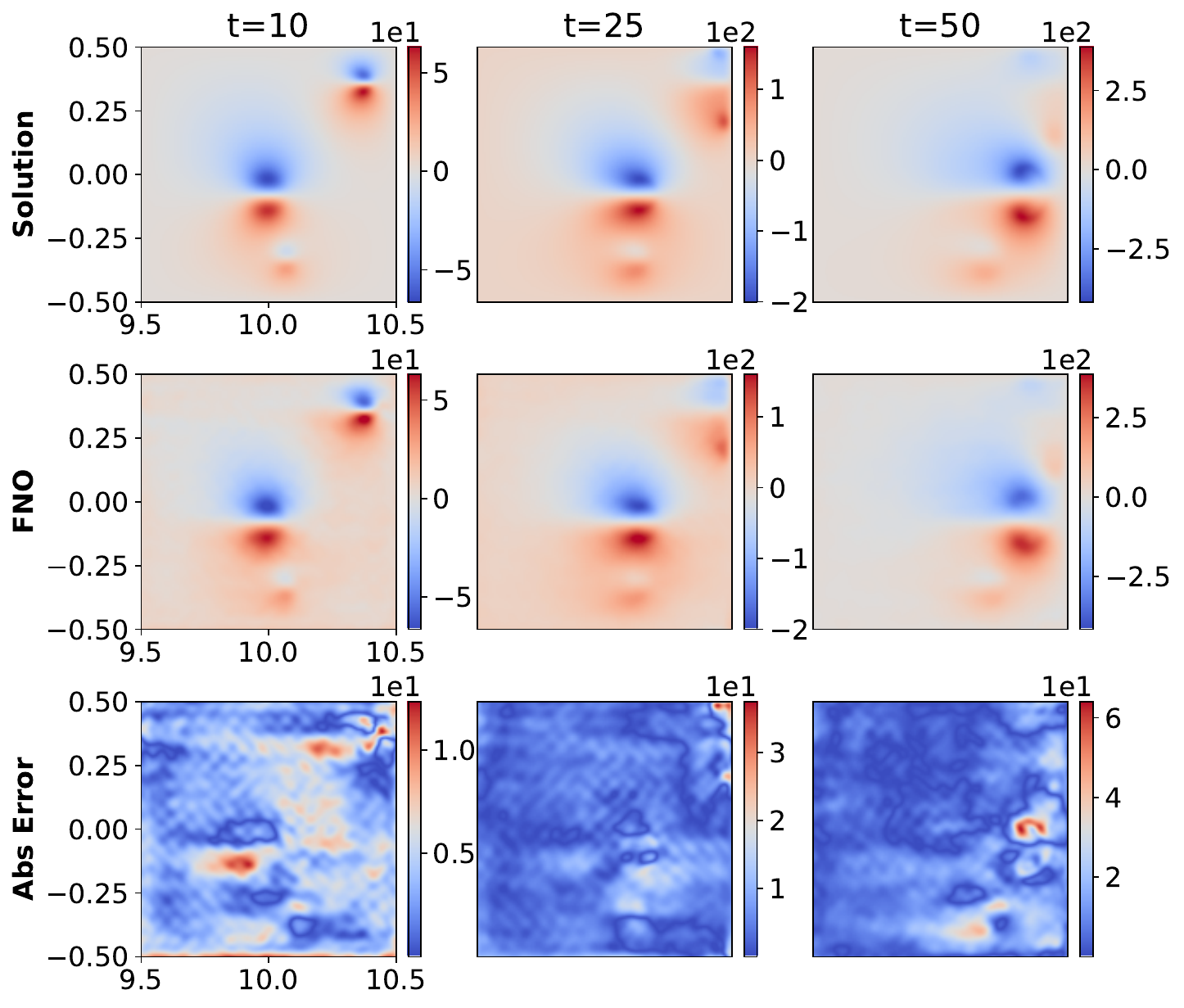}}
    \label{fig: potential_zssr_error}
    \centering
    \subfloat[Temperature]{\includegraphics[width=7.5cm]{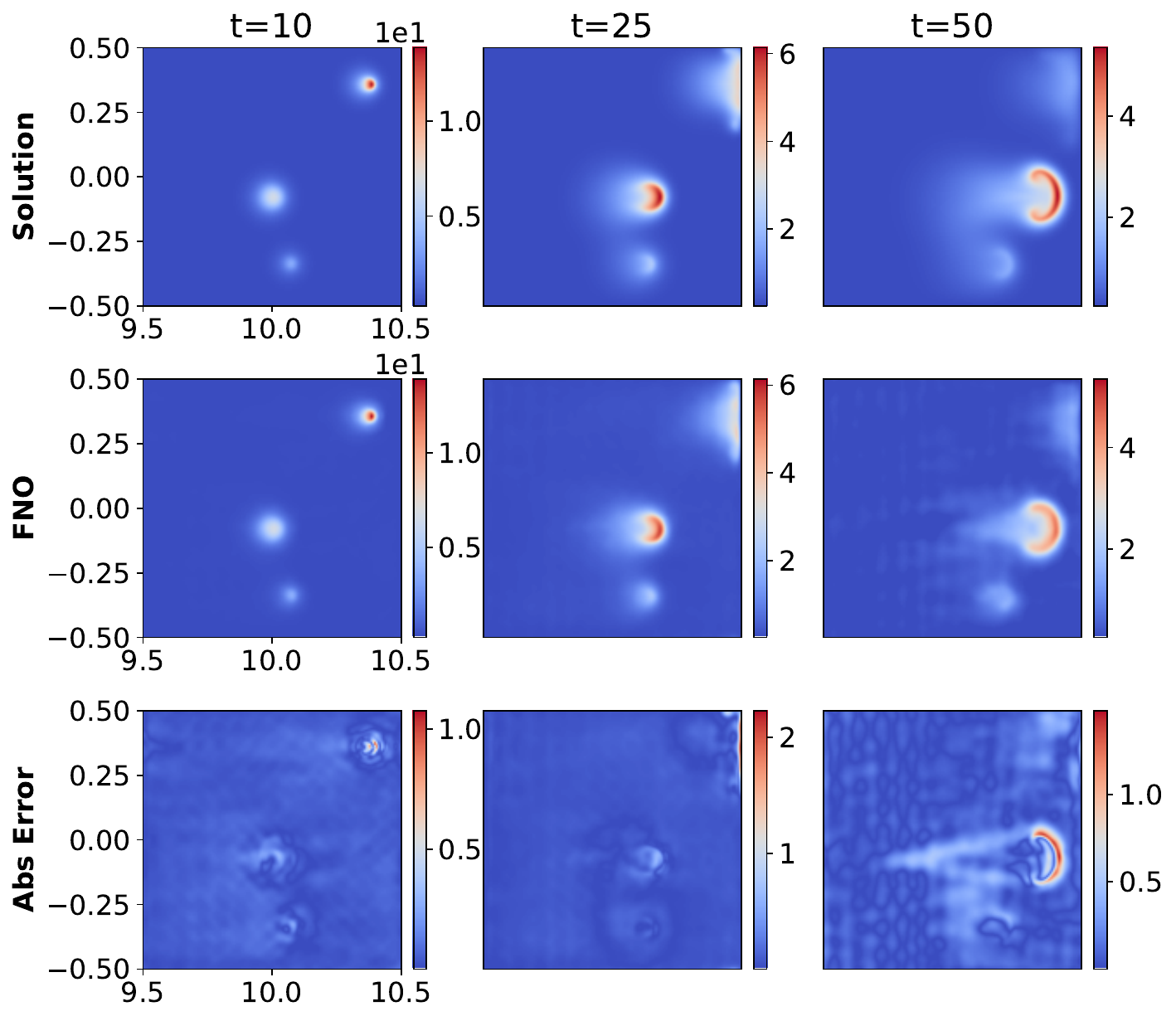}}
    \label{fig: temp_zssr_error}
    \caption{Demonstrating Zero-Shot Super-Resolution on the multi-variable FNO. The FNO being trained on a grid resolution of 100x100 is further deployed to model the evolution the the multiple blobs at a resolution of 500x500. The model is capable of performing zero-shot super-resolution, achieving an MSE (in the normalised domain) of $7.90\times 10^{-5}$ at a grid resolution $5 \times$ more resolved than the one it was being trained on without the requirement of any model engineering or fine-tuning. The plots demonstrate the evolution of density, electric potential, and temperature as emulated by the FNO. For plots (a), (b), and (c), the top series of plots demonstrate the time evolution of the ground truth while the middle series of plots demonstrate the time evolution as emulated by the multi-variable FNO and the bottom series indicates the absolute error across space.}
    \label{fig: zero-shot-super-resultion_error}
\end{figure}

\newpage
\section{Multi-variable FNO Architecture}
\label{appendix_fno_arch}

The architecture of the multi-variable FNO deployed in an auto-regressive manner for emulating the Reduced MHD equations. The model initially lifts the input data into a higher dimensional space using a linear layer. This is followed by 6 Fourier layers that learn the necessary modal mappings within the spatial domain, before being projected onto the lower dimensional space by a series of linear layers. \texttt{Fourier 2d} represents the 2D Fourier operator, features only extracted within the spatial domain, \texttt{Conv3d} represents the 3D convolutional layer across the variable channel and the spatial domain, \texttt{Add} operation adds the outputs from the Fourier layer and the convolutional layer together, \texttt{GELU} denotes the activation function: Gaussian Error Linear Unit. 

\begin{table}[h!]
\caption{\label{table: mv_fno_arch}Architecture of the Multi-variable FNO deployed for emulating Reduced MHD.} 
\begin{indented}
\lineup
\item[]\begin{tabular}{@{}*{7}{l}}
\br                              
    Part     & Layer     &  Output Shape \\
\mr
    Input & - & (10, 3, 100, 100, 12) \\    
    Lifting & \texttt{Linear} & (10, 3, 100, 100, 32) \\
    Fourier 1 & \texttt{Fourier2d/Conv3d/Add/GELU} & (10, 3, 32, 100, 100)\\
    Fourier 2 & \texttt{Fourier2d/Conv3d/Add/GELU} & (10, 3, 32, 100, 100)\\
    Fourier 3 & \texttt{Fourier2d/Conv3d/Add/GELU} & (10, 3, 32, 100, 100)\\
    Fourier 4 & \texttt{Fourier2d/Conv3d/Add/GELU} & (10, 3, 32, 100, 100)\\
    Fourier 5 & \texttt{Fourier2d/Conv3d/Add/GELU} & (10, 3, 32, 100, 100)\\
    Fourier 6 & \texttt{Fourier2d/Conv3d/Add/GELU} & (10, 3, 32, 100, 100)\\

    Projection 1 & \texttt{Linear} & (10, 3, 100, 100, 128) \\
    Projection 2 & \texttt{Linear} & (10, 3, 100, 100, 5) \\

\br
\end{tabular}
\end{indented}
\end{table}


\newpage
\section{Long time roll-outs}
\label{appendix_long_rollouts}

\begin{figure}[h!]
    \centering
    \subfloat[Density]{\includegraphics[width=7.5cm]{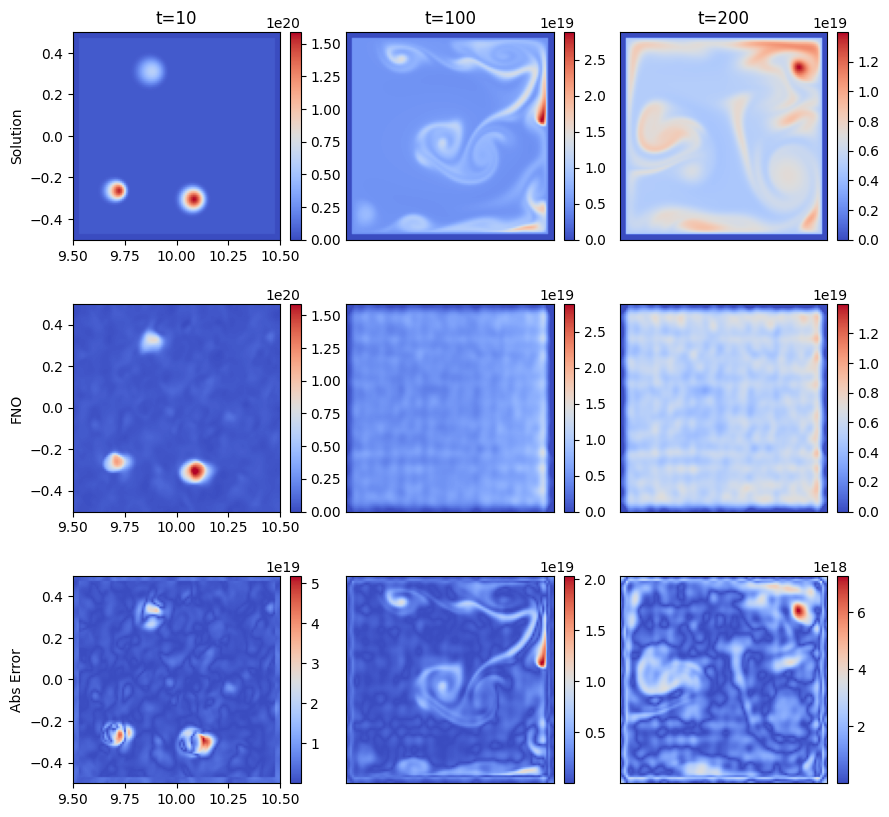}}
    \label{fig: multi_rho_long}
    \centering
    \subfloat[Potential]{\includegraphics[width=7.5cm]{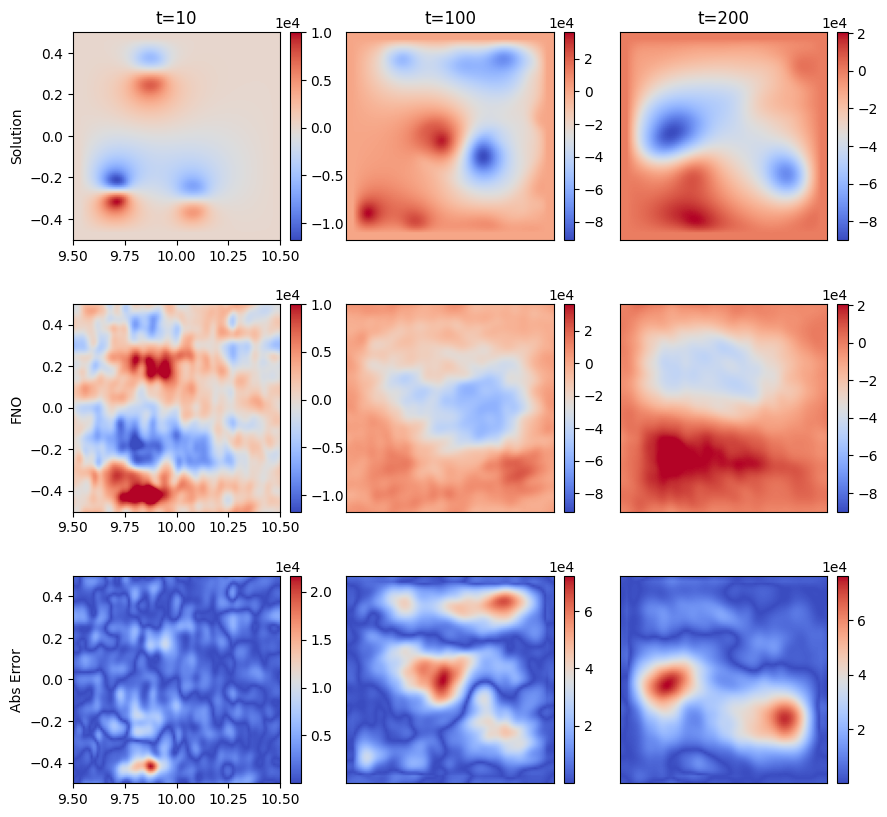}}
    \label{fig: multi_phi_long}
    \centering
    \subfloat[Temperature]{\includegraphics[width=7.5cm]{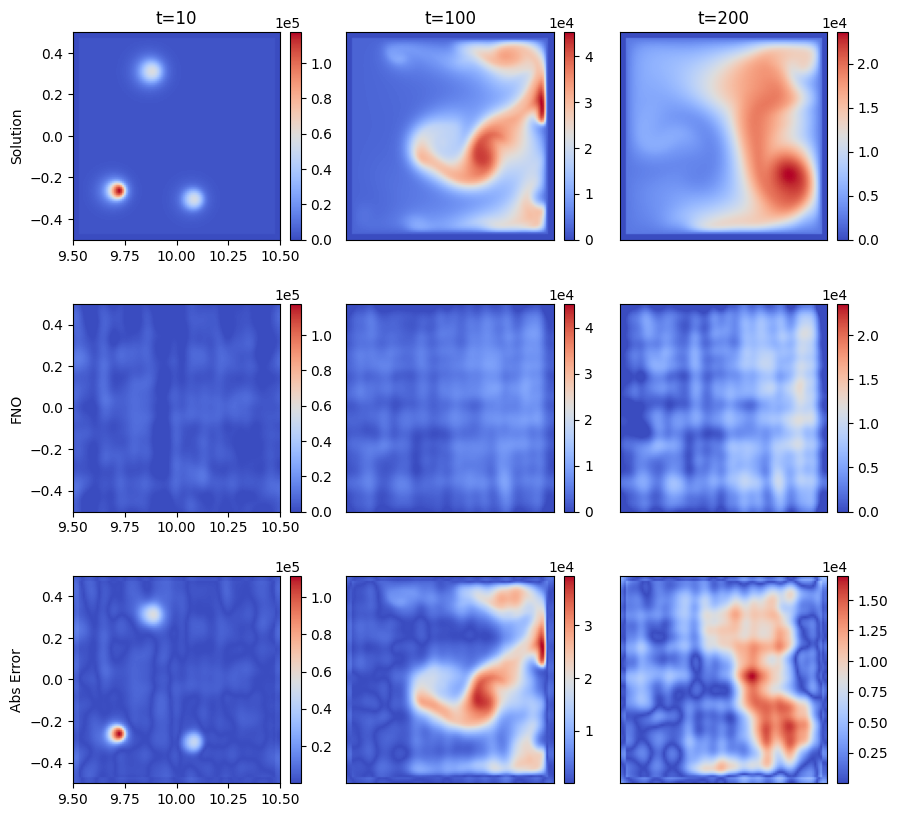}}
    \label{fig: multi_t_long}   
    \caption{Long Term roll-outs of Multiple Blobs with non-uniform temperature: Temporal evolution of (a) the density and (b) the potential, (c) the Temperature of the plasma evolution as obtained using the JOREK code (top of each image), that of the trained multi-variable FNO (middle of each figure) and the absolute error across both (bottom of each figure). The spatial domain is given in toroidal geometry characterised by $R$ in the $x$-axis and $Z$ in the $y$-axis. This model is trained to solve for up to 200 time instances, the full (available) scale of the simulation that we are interested in. Within the autoregressive roll-outs, the model quickly loses grip of the spatio-temporal dynamics as it accumulates further errors, leading to a noisy, incoherent output. }
    \label{fig: multi_mhd_long}
\end{figure}



\newpage
\section{Camera FNO Architecture}
\label{appendix_fno_individual_arch}

The architecture of the Individual FNO deployed in an auto-regressive manner for emulating the Camera data. The model initially lifts the input data into a higher dimensional space using a linear layer. This is followed by 6 Fourier layers that learn the necessary modal mappings within the spatial domain, before being projected onto the lower dimensional space by a series of linear layers. \texttt{Fourier 2d} represents the 2D Fourier operator, features only extracted within the spatial domain, \texttt{Conv2d} represents the 3D convolutional layer across the spatial domain, \texttt{Add} operation adds the outputs from the Fourier layer and the convolutional layer together, \texttt{GELU} denotes the activation function: Gaussian Error Linear Unit. 

\begin{table}[h!]
\caption{\label{table: rbb_fno_arch}Architecture of the Individual FNO deployed for emulating the \textit{rbb} Camera Data.} 
\begin{indented}
\lineup
\item[]\begin{tabular}{@{}*{7}{l}}
\br                              
    Part     & Layer     &  Output Shape \\
\mr
    Input & - & (50, 448, 640, 12) \\    
    Lifting & \texttt{Linear} & (50, 448, 640, 16) \\
    Fourier 1 & \texttt{Fourier2d/Conv2d/Add/GELU} & (50, 16, 448, 640)\\
    Fourier 2 & \texttt{Fourier2d/Conv2d/Add/GELU} & (50, 16, 448, 640)\\
    Fourier 3 & \texttt{Fourier2d/Conv2d/Add/GELU} & (50, 16  448, 640)\\
    Fourier 4 & \texttt{Fourier2d/Conv2d/Add/GELU} & (50, 16, 448, 640)\\
    Fourier 5 & \texttt{Fourier2d/Conv2d/Add/GELU} & (50, 16, 448, 640)\\
    Fourier 6 & \texttt{Fourier2d/Conv2d/Add/GELU} & (50, 16, 448, 640)\\

    Projection 1 & \texttt{Linear} & (50, 448, 648, 128) \\
    Projection 2 & \texttt{Linear} & (50, 448, 648, 10) \\

\br
\end{tabular}
\end{indented}
\end{table}

\begin{table}[h!]
\caption{\label{table: rba_fno_arch}Architecture of the Individual FNO deployed for emulating the \textit{rba} Camera Data.} 
\begin{indented}
\lineup
\item[]\begin{tabular}{@{}*{7}{l}}
\br                              
    Part     & Layer     &  Output Shape \\
\mr
    Input & - & (50, 400, 512, 12) \\    
    Lifting & \texttt{Linear} & (50, 400, 512, 16) \\
    Fourier 1 & \texttt{Fourier2d/Conv2d/Add/GELU} & (50, 16, 400, 512)\\
    Fourier 2 & \texttt{Fourier2d/Conv2d/Add/GELU} & (50, 16, 400, 512)\\
    Fourier 3 & \texttt{Fourier2d/Conv2d/Add/GELU} & (50, 16  400, 512)\\
    Fourier 4 & \texttt{Fourier2d/Conv2d/Add/GELU} & (50, 16, 400, 512)\\
    Fourier 5 & \texttt{Fourier2d/Conv2d/Add/GELU} & (50, 16, 400, 512)\\
    Fourier 6 & \texttt{Fourier2d/Conv2d/Add/GELU} & (50, 16, 400, 512)\\

    Projection 1 & \texttt{Linear} & (50, 400, 512, 128) \\
    Projection 2 & \texttt{Linear} & (50, 400, 512, 10) \\

\br
\end{tabular}
\end{indented}
\end{table}

\newpage
\section{Camera at the Central Solenoid}
\label{appendix_camera_solenoid}
\begin{figure}[h!]
    \centering
    \includegraphics[width=10.0cm]{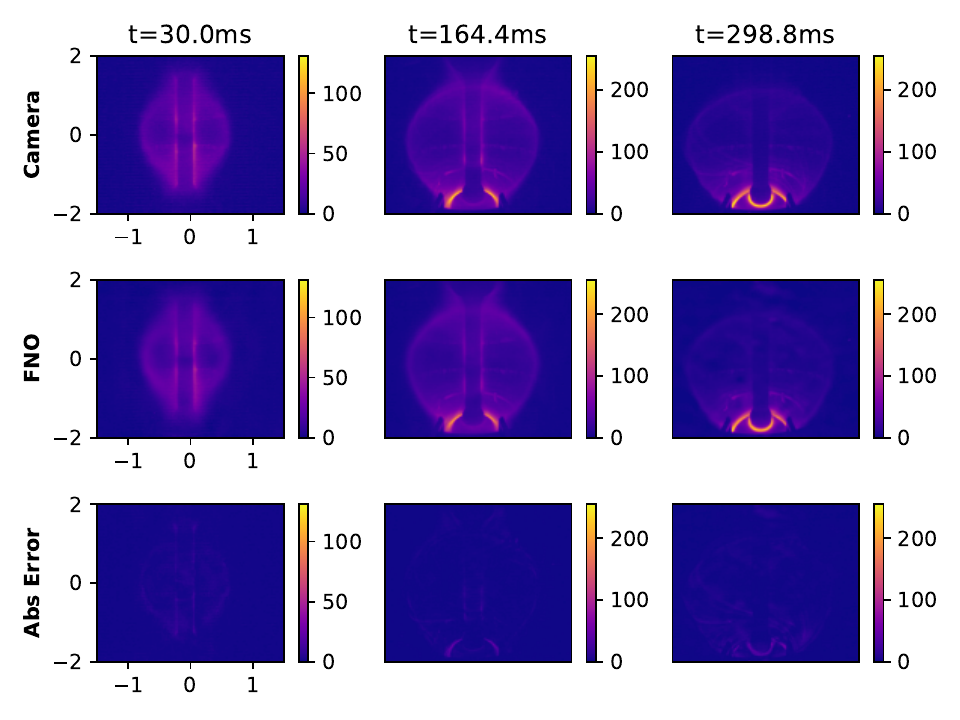}
    \caption{Temporal evolution of the plasma around the central solenoid within MAST as perceived by the main camera. The image compares the last frame of the predicted plasma evolution across multiple feed-forward instances across the duration of the pulse.}
    \label{fig: rbb_last10_0}
\end{figure}

\begin{figure}[h!]
    \centering
    \includegraphics[width=10.0cm]{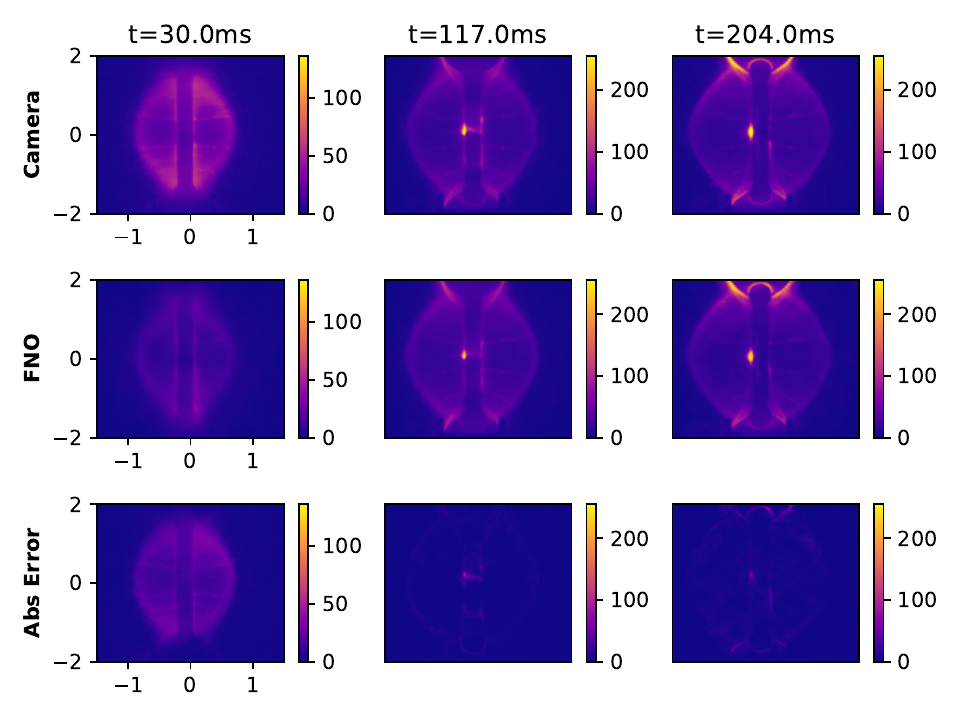}
    \caption{Temporal evolution of the plasma around the central solenoid within MAST as perceived by the main camera. The image compares the last frame of the predicted plasma evolution across multiple feed-forward instances across the duration of the pulse. }
    \label{fig: rbb_last10_3}
\end{figure}

\newpage
\section{Camera at the Divertor}
\label{appendix_camera_divertor}
\begin{figure}[h!]
    \centering
    \includegraphics[width=10.0cm]{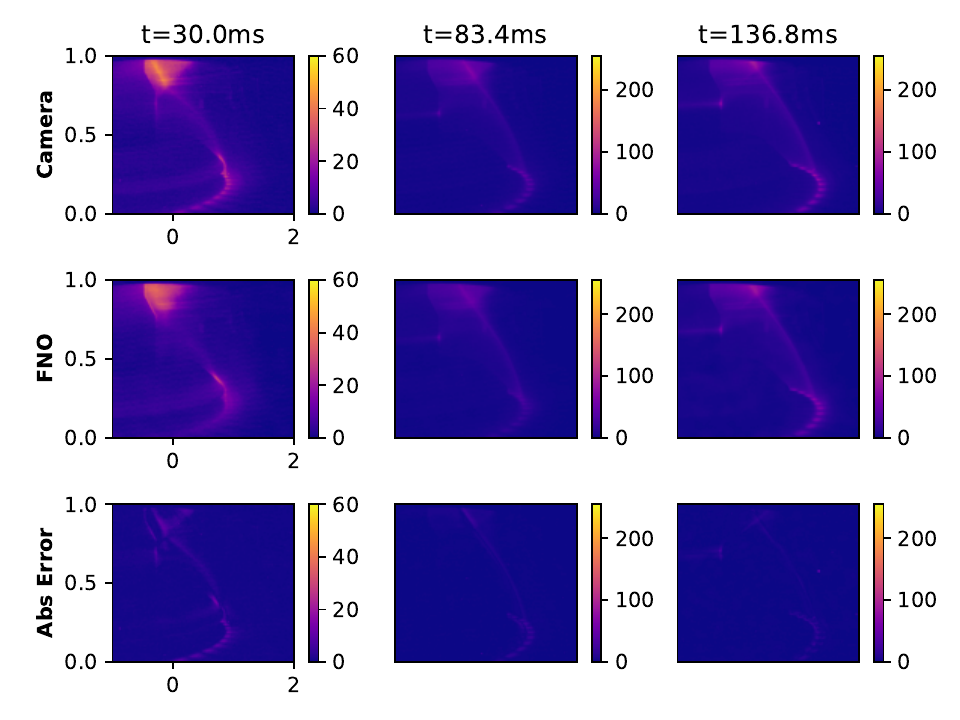}
    \caption{Temporal evolution of the plasma at the divertor within MAST as perceived by the divertor camera. The image compares the last frame of the predicted plasma evolution across multiple feed-forward instances across the duration of the pulse.}
    \label{fig: rba_last10_0}
\end{figure}

\begin{figure}[h!]
    \centering
    \includegraphics[width=10.0cm]{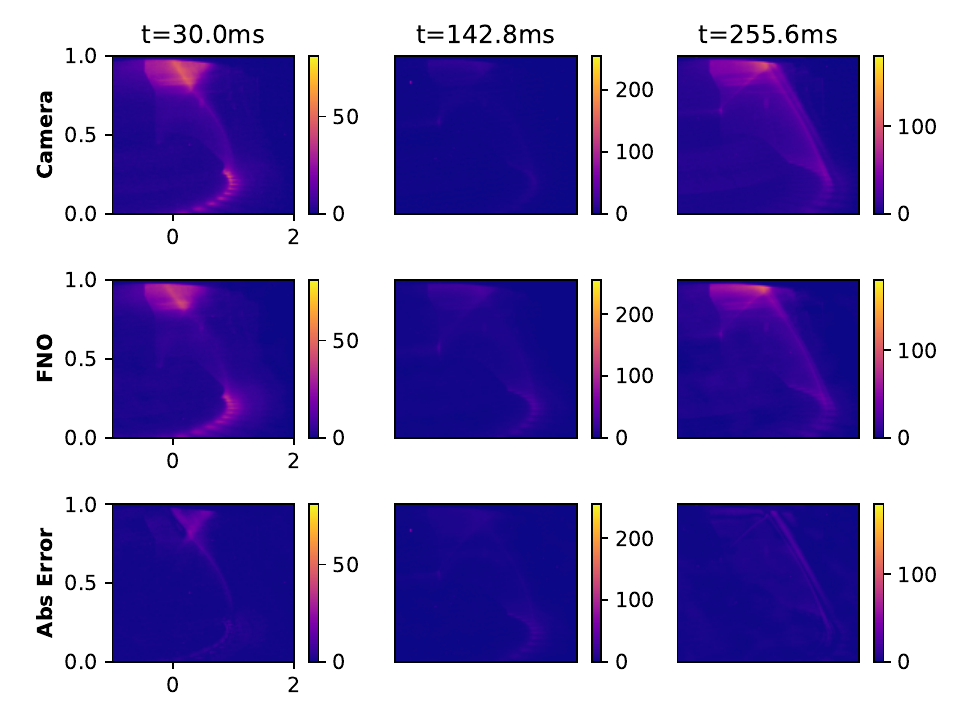}
    \caption{Temporal evolution of the plasma at the divertor within MAST as perceived by the divertor camera. The image compares the last frame of the predicted plasma evolution across multiple feed-forward instances across the duration of the pulse.}
    \label{fig: rba_last10_2}
\end{figure}


\newpage
\section{MAST Shots used for Camera data}
\label{appendix_camera_shots}

Shots used for the FNO modelling the plasma evolution as captured by the \textit{rbb camera} looking at the central solenoid: \\

\textbf{Training shots:}\\
30255, 30256, 30257, 30258, 30259, 30260, 30261, 30262, 30263,
       30264, 30265, 30266, 30267, 30269, 30270, 30290, 30291, 30298,
       30299, 30301, 30302, 30303, 30304, 30305, 30306, 30338, 30339,
       30342, 30343, 30345, 30349, 30350, 30351, 30352, 30353, 30354,
       30355, 30356, 30357, 30358, 30359, 30360, 30416, 30417, 30418,
       30419, 30420, 30421, 30422, 30423 \\
       
\textbf{Test shots:}\\   
30424, 30425, 30426, 30427, 30428 \\

Shots used for the FNO modelling the plasma evolution as captured by the \textit{rba camera} looking at the divertor: \\

\textbf{Training shots:}\\
30302, 30353, 30336, 30266, 30343, 30252, 30306, 30256, 30316,
       30322, 30262, 30356, 30257, 30317, 30323, 30263, 30290, 30352,
       30337, 30253, 30313, 30359, 30258, 30338, 30339, 30269, 30358,
       30259, 30319, 30299, 30298, 30270, 30355, 30345, 30314, 30260,
       30351, 30264, 30350, 30301, 30325, 30265, 30311, 30360, 30305\\
       
\textbf{Test shots:}\\       
 30354, 30255, 30321 \\

\end{document}